\documentclass[preprint,12pt]{elsarticle}

\usepackage{xspace}

\usepackage{multirow}
\usepackage{longtable}
\usepackage{makecell}

\usepackage{subcaption}
\usepackage{xspace}
\usepackage[edges]{forest}
\usepackage{xcolor}
\usepackage{amssymb} 
\usepackage{float}

\newcommand{\ignore}[1]{}

\newcommand{\ment}{\mu}

\newcommand{\abtbuy}{\textsf{Abt-Buy}\xspace}
\newcommand{\fodor}{\textsf{Fodor-Zagat}\xspace}
\newcommand{\company}{\textsf{Company}\xspace}
\newcommand{\wdc}{\textsf{WDC}\xspace}
\newcommand{\walmartamazon}{\textsf{Walmart-Amz}\xspace}
\newcommand{\itunesamazon}{\textsf{iTunes-Amz}\xspace}

\newcommand{\deepmatcher}{DeepMatcher\xspace}
\newcommand{\ditto}{DITTO\xspace}
\newcommand{\emtransformer}{EM Transformer\xspace}
\newcommand{\hiergat}{HierGAT\xspace}
\newcommand{\hiermatcher}{HierMatch\xspace}
\newcommand{\rotom}{RoTom\xspace}

\definecolor{cbBlue}{RGB}{0, 114, 178}
\definecolor{cbRed}{RGB}{213, 94, 0}
\definecolor{cbGreen}{RGB}{0, 158, 115}
\definecolor{cbPurple}{RGB}{204, 121, 167}
\definecolor{cbLightBlue}{RGB}{86, 180, 233}
\definecolor{cbOrange}{RGB}{230, 159, 0}

\renewcommand{\paragraph}[1]{\vspace{2mm} \noindent {\bf #1.}}

\newcommand{\takeaway}{\vspace{2mm}\noindent {\em Takeaways:}\xspace}

\definecolor{darkred}{rgb}{0.55, 0.0, 0.0}

\definecolor{darkblue}{rgb}{0.0, 0.0, 0.55} 

\usepackage{longtable}
\usepackage{enumitem}
\usepackage{booktabs} 

\usepackage{hyperref}

\hypersetup{
    colorlinks=true,
    linkcolor=red,
    citecolor=green,
    urlcolor=blue,
    debug=true
}

\usepackage{tikz} 
\usepackage{xcolor} 

\definecolor{cbf-blue}{HTML}{377EB8}    
\definecolor{cbf-orange}{HTML}{E69F00} 
\definecolor{cbf-green}{HTML}{4DAF4A}  
\definecolor{cbf-red}{HTML}{E41A1C}    
\definecolor{cbf-purple}{HTML}{984EA3} 
\definecolor{cbf-brown}{HTML}{A65628}  
\definecolor{cbf-pink}{HTML}{F781BF}   
\definecolor{cbf-gray}{HTML}{999999}   

\definecolor{ReviewerOneColor}{RGB}{240, 110, 0}   
\definecolor{ReviewerTwoColor}{RGB}{0, 139, 109}   
\definecolor{ReviewerThreeColor}{RGB}{45, 85, 205}
\definecolor{MetaReviewerColor}{RGB}{142, 69, 133} 

\usepackage{amssymb}
\usepackage{amsmath}

\journal{Data \& Knowledge Engineering}

\begin{document}

\begin{frontmatter}



\title{Heterogeneity in Entity Matching: A Survey and Experimental Analysis}

\author[uwoca]{Mohammad Hossein Moslemi\corref{cor1}}
\ead{mohammad.moslemi@uwo.ca}

\author[utsa]{Amir Mousavi}
\ead{seyedamir.mousavi@my.utsa.edu}

\author[uwoca]{Behshid Behkamal}
\ead{behshid.behkamal@uwo.ca}

\author[uwoca]{Mostafa Milani}
\ead{mostafa.milani@uwo.ca}
\cortext[cor1]{Corresponding author}

\affiliation[uwoca]{organization={University of Western Ontario},
            city={London},
            state={Ontario},
            country={Canada}}

\affiliation[utsa]{organization={University of Texas at San Antonio},
            city={San Antonio},
            state={Texas},
            country={USA}}




\begin{abstract}
Entity matching (EM) is a fundamental task in data integration and analytics, essential for identifying records that refer to the same real-world entity across diverse sources. In practice, datasets often differ widely in structure, format, schema, and semantics, creating substantial challenges for EM. We refer to this setting as \emph{Heterogeneous EM (HEM)}. 

This survey offers a unified perspective on HEM by introducing a taxonomy, grounded in prior work, that distinguishes two primary categories--\emph{representation} and \emph{semantic heterogeneity}--and their subtypes. The taxonomy provides a systematic lens for understanding how variations in data form and meaning shape the complexity of matching tasks. We then connect this framework to the \emph{FAIR principles}--\emph{Findability}, \emph{Accessibility}, \emph{Interoperability}, and \emph{Reusability}--demonstrating how they both reveal the challenges of HEM and suggest strategies for mitigating them.

Building on this foundation, we critically review recent EM methods, examining their ability to address different heterogeneity types, and conduct targeted experiments on state-of-the-art models to evaluate their robustness and adaptability under semantic heterogeneity. Our analysis uncovers persistent limitations in current approaches and points to promising directions for future research, including multimodal matching, human-in-the-loop workflows, deeper integration with large language models and knowledge graphs, and fairness-aware evaluation in heterogeneous settings.
\end{abstract}




\end{frontmatter}

\section{Introduction}\label{sec:intro}

{Entity Matching (EM) has long been a fundamental component of data integration, cleaning, and analytics pipelines \cite{doan2012principles,elmagarmid2007duplicate,christophides2020entity}. Although recent advances—especially in deep learning and AI—have accelerated progress \cite{christophides2020entity,dong2015big,li2020deep}, EM systems continue to struggle in real-world deployments. High-performing models trained on clean, benchmark datasets often fail to generalize when exposed to messy, noisy, and heterogeneous data found in practice \cite{Mudgal2018,magellan}. These failures are not incidental; they stem from a pervasive and under-addressed challenge: \textit{data heterogeneity}. Differences in formats (e.g., dates, units), schemas (e.g., attribute names, nesting), terminology (e.g., synonyms, language), and data quality (e.g., missing or inconsistent values) introduce mismatches in structure, semantics, and quality between development and deployment settings. Such heterogeneity undermines blocking, feature extraction, and similarity computation, leading to degraded performance across the entire EM pipeline. Even recent deep learning–based methods, which perform well on standard benchmarks, suffer sharp drops in accuracy when applied across domains with varying schema or semantics \cite{Mudgal2018,Li2020,gembench}.}

{In practice, data heterogeneity manifests in many intertwined ways, reinforcing the challenges outlined above. For example, the same product may appear as “Apple iPhone 14 (Blue)” in one source, “IPH14-BLU” in another, and only as an image with minimal text in a third—illustrating \emph{representation heterogeneity}. Clinical datasets often express the same concept using terms such as “Hypertension,” “High blood pressure,” or “HTN,” revealing \emph{semantic heterogeneity}. Two datasets may encode addresses differently, with one storing the full address in a single field while another splits it across multiple attributes, exemplifying \emph{structural heterogeneity}. Context also varies: job titles like “Senior” or “Manager” can carry different meanings across organizations or languages, leading to \emph{contextual heterogeneity}. Multilingual and multimodal environments introduce additional variation: the same city may appear as “München,” “Munich,” or “Munique,” and entities may be represented as tables, JSON records, knowledge-graph triples, or images. These diverse patterns illustrate the breadth and complexity of heterogeneity that EM systems must contend with in real-world deployments.}

While heterogeneity in data is well-recognized across many domains--including information retrieval~\cite{singhal2001modern}, geospatial systems~\cite{goodchild2007gis}, the Internet of Things~\cite{atzori2010internet}, and big data analytics~\cite{borges2013survey}--it poses particularly acute and evolving challenges in EM. Historically, heterogeneity in EM has been handled under themes such as schema matching~\cite{rahm2001survey}, 
duplicate record detection~\cite{elmagarmid2007duplicate}, and semantic integration~\cite{doan2005semanticintegration}, 
typically focusing on specific dimensions like attribute alignment mismatches or lexical variation. However, the modern data landscape--characterized by large-scale, semi-structured and unstructured sources from the web, data lakes, IoT devices, and enterprise systems--introduces more complex and compounded forms of heterogeneity. These include variation in data formats (e.g., JSON, XML, relational), schemas, semantics, language, granularity, and data quality. In this context, mismatches in data models and semantic assumptions severely complicate schema alignment, feature extraction, and record linkage~\cite{bernstein2011data}. These challenges call for EM methods that are explicitly robust to heterogeneity, and for a systematic understanding of the many forms that heterogeneity can now take in practice.

To address this challenge, we argue that categorizing and systematically studying data heterogeneity is essential for advancing the design, evaluation, and deployment of EM systems. A principled taxonomy of heterogeneity is useful for organizing prior work and also enables several concrete benefits in the design and assessment of EM systems. First, it guides the development of \textit{targeted model architectures}, allowing practitioners to align method design with the expected types of heterogeneity. Prior work has demonstrated that different model families excel under specific types of heterogeneity--for example, transformer-based architectures have shown robustness to semantic variation such as synonyms and abbreviations~\cite{dou2022empowering,brunner2020entity,Thirumuruganathan2021}, while graph-based methods are effective at capturing structural mismatches across schemas~\cite{hiergat,37-ED-GNN,38-sui2022trigger,40-GraphER}. Building on this foundation, our paper provides new experimental evidence supporting the need for heterogeneity-aware modeling. 

Second, such a taxonomy enables \textit{component-level stress testing}, in which researchers can evaluate how EM methods respond to controlled semantic or structural mismatch at different stages of the pipeline--from blocking to similarity computation to final classification. Third, it supports the construction of \textit{heterogeneity-aware benchmarks} and perturbation-based evaluation frameworks. Finally, an explicit understanding of heterogeneity contributes to \textit{uncertainty quantification} and \textit{robustness analysis}, which are increasingly critical for deploying EM systems in high-stakes domains such as healthcare, scientific data integration, and finance.

This paper presents a survey of recent methods in entity matching, with a specific focus on how they address the challenges introduced by data heterogeneity. Unlike prior surveys that cover traditional EM techniques~\cite{rahm2001survey}, deep learning approaches~\cite{barlaug2021neural,li2020deep,li2020survey}, blocking strategies~\cite{papadakis2020blocking}, or benchmarking frameworks~\cite{kopcke2010frameworks,govind2019entity,wang2021machamp}, our work takes a fundamentally different perspective by placing \textit{heterogeneity} at the center of analysis. We develop a hierarchical taxonomy that characterizes common forms of representation and semantic heterogeneity in EM, and we use this taxonomy to organize and critique recent EM models. Complementing the survey, we conduct targeted experiments that evaluate the robustness of state-of-the-art models under controlled semantic heterogeneity conditions. To our knowledge, this is the first work to both systematically classify heterogeneity types in EM and empirically analyze how these variations affect model behavior. Our goal is to establish heterogeneity as a first-class concern in EM research and to provide a foundation for more robust, generalizable, and transparent EM systems.

One of the most significant recent shifts in the EM landscape is the increasing influence of large language models (LLMs) and generative AI. These models offer new capabilities that are particularly relevant for addressing semantic heterogeneity, a core challenge in modern EM. Pretrained models such as BERT and GPT have demonstrated strong abilities to capture lexical and contextual variation through transfer learning, reducing the need for hand-crafted features or schema-specific engineering~\cite{Li2020}. More recently, prompting and instruction tuning have enabled the use of foundation models for zero- or few-shot entity resolution~\cite{elazar2023prompting}, making it possible to generalize across domains without extensive retraining. These trends suggest that foundation models are poised to play a growing role in heterogeneity-aware EM--a theme we revisit in detail later in the survey.

The study of heterogeneity in EM also has significant implications for data governance and interoperability, particularly in the context of the \textit{FAIR} principles for scientific data management--ensuring that data is Findable, Accessible, Interoperable, and Reusable~\cite{wilkinson2016fair}. Heterogeneity presents direct challenges to achieving FAIR compliance, especially when integrating records across fragmented, inconsistent, or mismatched sources. Conversely, EM methods that are explicitly designed to handle such heterogeneity can act as critical enablers of FAIRification by enhancing schema alignment, disambiguation, and record linkage. Throughout this paper, we emphasize the mutual relationship between EM and FAIR, and show how heterogeneity-aware EM systems contribute to building more trustworthy, transparent, and reusable data infrastructures.

Our paper makes the following contributions:
\begin{itemize}[leftmargin=15pt]
    \item We present a hierarchical taxonomy of data heterogeneity in EM—adapted from established distinctions in related areas, distinguishing between \textit{representation heterogeneity} (e.g., format, schema) and \textit{semantic heterogeneity} (e.g., language, granularity, quality).
    \item We systematically survey and categorize recent EM methods through the lens of this taxonomy, revealing how different model classes--rule-based, neural, and graph-based--address (or fail to address) specific forms of heterogeneity, and identifying patterns and gaps in current research.
    \item We develop and release a benchmark for evaluating semantic heterogeneity in EM, which we use to stress-test state-of-the-art models under controlled variations. Our experiments expose failure modes, highlight robustness differences between architectures, and demonstrate that existing benchmarks often mask these limitations.
    \item We synthesize the insights from both the survey and experiments into practical recommendations for designing heterogeneity-aware EM systems, and we outline directions for future research, including evaluation protocols, benchmark design, and architectural innovations.
\end{itemize}

The rest of the paper is organized as follows. In Section~\ref{sec:hierarchy}, we introduce our taxonomy of data heterogeneity in EM. Section~\ref{sec:survey} surveys recent EM methods and analyzes their capabilities and limitations with respect to different types of heterogeneity. Section~\ref{sec:fair} discusses the relationship between EM and the FAIR data principles, highlighting how heterogeneity-aware EM methods can support FAIRification. Section~\ref{sec:exp} presents our experimental framework and results. Section~\ref{sec:future} concludes by outlining future directions, including the role of large language models in handling heterogeneity.

\section{A Framework for Classifying HEM} \label{sec:hierarchy}

In this section, we first formalize the problem setting of heterogeneous entity matching (HEM) in a general and modality-agnostic way. This provides a unified foundation for describing where heterogeneity arises and how it affects the core EM task. We then introduce our taxonomy of heterogeneity types, which builds on this formalization and organizes representation and semantic heterogeneity into a coherent hierarchy.

\subsection{{The HEM Problem}} \label{sec:formal}

{To discuss HEM more formally, assume $E = \{e_1, \ldots, e_{|E|}\}$ denotes the set of real-world entities in an application domain. A \emph{mention} is any observable representation of an entity in $E$ within a dataset $D$, such as a relational record, a textual span (word or phrase), an image crop, or a node in a graph. For a dataset $D$, we denote its set of mentions by $\ment(D) = \{m_1, \ldots, m_{|D|}\}$. Each mention corresponds to an underlying entity through an unknown mapping $g_D : \ment(D) \rightarrow E$. Given two datasets $D$ and $D'$, the goal of EM is to determine when two mentions in the two datasets refer to the same real-world entity. Formally, EM seeks a binary matching function $f : \ment(D) \times \ment(D') \rightarrow \{0,1\}$, where $f(m,m') = 1$ if and only if $g_{D}(m) = g_{D'}(m')$. When $D = D'$, EM reduces to recovering the latent equivalence classes in $D$ induced by $g_D$.}

{Heterogeneous EM (HEM) refers to the EM task when the datasets containing the mentions exhibit heterogeneity. \emph{Between-dataset heterogeneity} arises when $D$ and $D'$ differ in how they represent, encode, or interpret information---for example, differences in schema design (attribute names, types, or structure), terminology or linguistic usage, granularity (e.g., ``USA'' vs.\ ``California''), or modality (text vs.\ images). \emph{Within-dataset heterogeneity} arises when mentions inside a single dataset $D$ vary along these same dimensions, such as mixed attribute formats, inconsistent value representations, or multilingual text within the same source. Both forms of heterogeneity complicate the decision of whether mentions refer to the same real-world entity. }

{We frame our discussion around \emph{Entity Matching (EM)}, while remaining
mindful that the underlying task appears across research communities under
different but closely related terminology. Common alternatives include
\emph{Entity Resolution (ER)}, \emph{Record Linkage}, \emph{Duplicate Detection},
\emph{Record Matching}, and \emph{Identity Resolution}
\cite{elmagarmid2007duplicate,christen2012data}. ER is often defined as a broader
pipeline involving blocking, pairwise comparison, clustering, and inconsistency
resolution~\cite{kopcke2010evaluation}, whereas EM typically
refers more narrowly to the pairwise matching or similarity assessment stage
within this pipeline~\cite{li2020deep}. Related correspondence problems also
arise in adjacent settings, such as \emph{Entity Linking} in natural language
processing~\cite{shen2015entity} and \emph{Entity Alignment} in knowledge graphs
\cite{sun2017cross}. We adopt EM as our central term because our focus is on the
matching function itself and on how representation, schema, and semantic
heterogeneity affect this component across diverse data modalities and schema
environments.}

\begin{figure}[htb]
\centering
\resizebox{\textwidth}{!}{%
\begin{forest}
for tree={
    grow=south,
    parent anchor=south,
    child anchor=north,
    anchor=center,
    calign=center,
    inner xsep=2pt,
    inner ysep=2pt,
    s sep=4pt,
    font=\footnotesize,
    edge={darkgray, line width=0.4pt},
    l sep=1pt,
    edge path={
        \noexpand\path [draw, \forestoption{edge}]
        (!u.parent anchor) -- (.child anchor)\forestoption{edge label};
    },
    align=center,
    level distance=12pt,
    sibling distance=2pt,
    level=0{
        fill=cbPurple!20,
        draw=cbPurple!80,
        line width=0.1pt,
        rounded corners=2pt
    }{},
    if level=1{
        fill=cbBlue!20,
        draw=cbBlue!80,
        line width=0.1pt
    }{},
    if level=2{
        fill=cbRed!20,
        draw=cbRed!80,
        line width=0.1pt
    }{},
    if level=3{
        fill=cbGreen!20,
        draw=cbGreen!80,
        line width=0.5pt,
        rounded corners=2pt
    }{},
    if level=4{
        fill=cbPurple!20,
        draw=cbPurple!80,
        line width=0.5pt,
        rounded corners=2pt,
        font=\scriptsize
    }{},
},
level 4/.style={
    level distance=-0pt
},
[Heterogeneity in EM, fill=cbPurple!20, draw=cbPurple!80
    [Representation
        [Multimodality]
        [Format]
        [Structural\\ (Schema)]
    ]
    [Semantic
        [Terminology \& \\ Language]
        [Contextual \\ Variability]
        [Granularity \& \\ Resolution]
        [Temporal \\ Variability]
        [Data \\ Quality]
    ]
]
\end{forest}
}
\caption{Taxonomy of heterogeneity in entity matching (HEM), including representation- and semantic-level variation.}
\label{fig:hierarchy}
\end{figure}
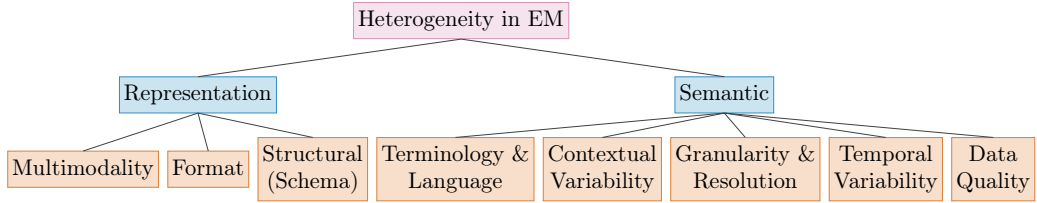

To systematically analyze heterogeneity in HEM, we categorize it into two main types: \emph{representation heterogeneity} and \emph{semantic heterogeneity}. Representation heterogeneity refers to differences in how data is structured or encoded across sources---for example, variations in modalities (text vs.\ images), file formats (JSON vs.\ XML), or schema organization~\cite{rahm2001survey,madhavan2001generic}. In contrast, semantic heterogeneity arises when data carries different meanings or interpretations despite structural alignment, often due to differences in terminology, context, granularity, or data quality. This distinction builds on foundational work in data integration~\cite{rahm2001survey, batini1986comparative, madhavan2001generic} and offers a practical framework for identifying and addressing the diverse sources of variation that affect EM pipelines. Figure~\ref{fig:hierarchy} presents our taxonomy, which organizes both categories into subtypes commonly observed in real-world EM scenarios. We elaborate on each category in the sections that follow.

{Similar high-level distinctions between structural (syntactic) and semantic heterogeneity have been discussed in related areas such as schema matching~\cite{rahm2001schema}, semantic integration~\cite{doan2005semantic}, ontology alignment~\cite{euzenat2013ontology}, and federated databases~\cite{sheth1990federated}. 
However, to our knowledge, no prior work has systematically adapted these ideas to EM or used them as an organizing framework for surveying EM methods under heterogeneity. 
Our goal is not to introduce a new theoretical taxonomy, but to employ this framework as a practical lens for structuring, comparing, and assessing recent EM approaches across diverse forms of heterogeneity.}

\subsection{Representation Heterogeneity} \label{sec:repH}

Representation heterogeneity encompasses the structural and syntactic differences that occur when datasets use different modalities, formats, or schema designs to describe entities. These differences can disrupt every stage of the EM pipeline--from feature extraction to similarity computation--by introducing incompatibilities in how records are organized or encoded. For example, one dataset might store product data as nested JSON objects, while another uses flat CSV files. Even when datasets describe the same entities, format mismatches and inconsistent attribute organization can hinder alignment. Addressing representation heterogeneity typically requires schema matching, format normalization, or modality-specific processing techniques. We break this category into three subtypes:

\begin{itemize}[leftmargin=0.35cm]

\item {{\em Multimodality:} Datasets may include diverse data types—such as text, images, and videos—for describing entities. For instance, e-commerce records may pair textual descriptions with product photos \cite{khan2021mmbert,liu2020deepfashion}. Aligning entities across modalities requires models that can jointly embed or compare representations across heterogeneous data sources \cite{xu2020layoutlm,moon2018multimodal,chen2020imagebert}.}

\item {\em Format Heterogeneity:} Data may be stored in different syntactic formats, such as JSON, XML, or CSV for text, or JPEG vs. PNG for images~\cite{noy2000algorithm}. Although the semantics may be consistent, structural variation can hinder parsing and alignment.

\item {\em Structural (Schema) Heterogeneity:} This refers to differences in attribute naming, hierarchy, and table structure~\cite{madhavan2001generic}. For example, one dataset may use ``price'' while another uses ``cost'', or may nest address fields differently. Schema matching and ontology-based alignment are common approaches for addressing this form~\cite{doan2012principles}.

\end{itemize}

While we survey methods for format and structural heterogeneity in Section~\ref{sec:survey}, we give multimodality more attention here due to its unique modeling challenges. Multimodal EM typically involves two steps: (1) multimodal entity recognition, identifying entities across modalities, and (2) multimodal linking, associating those entities to a shared identity.

Several recent works explore solutions in this space. Yu et al.~\cite{yu2020improving} introduced a multimodal transformer for aligning visual and textual data in social media. Moon et al.~\cite{moon2018multimodal} and Adjali et al.~\cite{adjali2020building} enhance entity disambiguation using image-text pairs. Gan et al.~\cite{gan2021multimodal} proposed the M3EL dataset for benchmarking visual-textual matching. Recent architectural advances include MIMIC~\cite{wang2022multimodal}, which models multi-grained interactions, and MAF~\cite{xu2022maf}, which enables cross-modal alignment through flexible attention mechanisms. Together, these works show the promise of multimodal transformers and contrastive objectives in handling complex cross-modal EM tasks.

\subsection{Semantic Heterogeneity} \label{sec:semantic}

Semantic heterogeneity arises when data shares structure or format but diverges in meaning or interpretation. This is a core challenge in EM because it undermines similarity measures and alignment logic. We group common sources of semantic heterogeneity as follows:

\begin{itemize}[leftmargin=0.35cm]

\item {\em Language and Terminology Differences:} Different datasets may use alternate terms or languages for the same concept. For instance, ``mobile phone'' vs. ``cellular device'' or ``prix'' (French) vs. ``price''. Techniques such as synonym expansion, translation, and vocabulary alignment help address this~\cite{euzenat2013ontology,rahm2001survey}.

\item {\em Contextual Variability:} The same term may have different meanings depending on context (e.g., ``apple'' as fruit vs. company). Handling this requires context-aware models such as BERT~\cite{devlin2019bert} or ELMo~\cite{peters-etal-2018-deep}, which embed tokens based on usage.

\item {\em Granularity and Resolution:} Datasets may differ in how detailed their records are. One might record locations at the country level, another at the neighborhood level. Aggregation/disaggregation techniques and ontology alignment are common remedies~\cite{halevy2001answering}.

\item {\em Temporal Variability:} Semantics can shift over time or differ due to timing. For example, product prices or job roles may change, making record alignment time-sensitive~\cite{doan2002learning,snodgrass1999developing}. EM systems must consider temporal validity or versioning.

\item {\em Data Quality:} Incomplete, noisy, or inconsistent records introduce semantic ambiguity. Typos, outdated values, or missing fields disrupt both training and inference. Addressing this often involves data cleaning, imputation, or robust training methods~\cite{rahm2000data,pirhadi2024otclean}.

\end{itemize}

Early database work recognized these challenges in federated systems~\cite{sheth1990federated}, and Semantic Web research later addressed ontology alignment~\cite{euzenat2013ontology}. Today, modern EM methods incorporate contextual modeling, domain adaptation, and knowledge resources to address semantic heterogeneity--topics we explore in depth in Section~\ref{sec:survey}.

\section{Review of HEM and Related Research Areas}
\label{sec:survey}

To systematically review existing EM methods in relation to heterogeneity, we adopted a multi-step process. We first identified a set of research areas that directly intersect with heterogeneity in EM, including \emph{schema and structural heterogeneity}, \emph{representation learning}, \emph{deep and graph-based models}, \emph{knowledge graphs and ontologies}, \emph{transfer learning and domain adaptation}, \emph{active and interactive learning}, \emph{self-supervised and evolutionary methods}, and \emph{LLMs}. While our survey emphasizes recent progress in HEM, we also include influential earlier works where necessary to provide conceptual grounding and illustrate how research in these areas has evolved.

We then collected and reviewed peer-reviewed articles from leading data management, AI, and ML venues that engage with heterogeneity in any of these areas. We excluded theses, posters, and papers that do not explicitly address heterogeneity. From this broader set of publications, we identified a core group of studies that directly target HEM, with particular emphasis on semantic, structural, and format heterogeneity, given their frequent overlap in practical EM settings.

\subsection{{Schema Heterogeneity and EM Approaches}}
\label{sec:schema-heterogeneity}

{Schema Matching and EM are distinct but related tasks. Schema matching aims to identify correspondences between attributes or structural elements of two schemas, while EM operates at the level of entity instances. Schema matching is especially relevant to schema heterogeneity as a common practical obstacle for EM: when two datasets organize or name attributes differently, some form of schema alignment is typically required before pairwise matching can be reliably performed. For this reason, schema matching is often treated as a preliminary step in traditional EM pipelines, particularly in data integration settings~\cite{rahm2001survey}.}

In modern EM, however, a growing body of work seeks to \emph{avoid} explicit schema alignment altogether by developing EM models that operate directly on heterogeneous or partially aligned schemas. Below, we review EM approaches designed to handle schema or structural variation without requiring a separate schema matching stage. These methods address \emph{schema-level representation heterogeneity} in our taxonomy (Figure~\ref{fig:hierarchy}).

Addressing this need, \emph{HERA}~\cite{54-HERA} proposes a paradigm for entity resolution that bypasses schema alignment. Instead of first integrating schemas, HERA operates directly on heterogeneous records, reducing information loss from premature schema integration. It uses a compare-and-merge process to iteratively build ``super records,'' combining both instance-based and schema-based similarity signals. An indexing structure supports efficient candidate generation and similarity computation. Empirical results show that HERA significantly outperforms state-of-the-art methods, particularly when schema heterogeneity is high and schema mapping is lossy.

Similarly, \emph{GraphER}~\cite{40-GraphER} avoids explicit schema alignment through a token-centric model built on Graph Convolutional Networks (GCNs). It constructs an Entity Record Graph (ER-Graph) that encodes relationships between records, attributes, and tokens. Through a two-layer GCN, GraphER jointly learns structural and semantic token embeddings, enabling fine-grained token-level comparisons without fixed attribute alignment. Evaluations on standard benchmarks show that GraphER consistently outperforms baseline models, especially in cases with diverse or sparse schemas.

Finally, \emph{Machamp}~\cite{wang2021machamp} introduces a benchmark for Generalized Entity Matching (GEM) that further highlights these challenges. Unlike earlier EM benchmarks that assume structured data with aligned schemas, Machamp includes structured, semi-structured (e.g., JSON), and unstructured (e.g., text) data. It covers seven real-world scenarios that reflect schema mismatches, such as matching relational with semi-structured or textual data. By repurposing and transforming existing datasets, Machamp evaluates model robustness under schema and semantic variation. Experiments reveal substantial performance drops for deep models like BERT and \ditto in heterogeneous settings, underscoring the need for more schema-agnostic techniques.

\subsection{Representation Learning and Semantic Embeddings}
\label{sec:representation_learning}

Representation learning has become foundational to modern EM~\cite{li2020deep,hiergat}, especially with the rise of deep neural models. Unlike static feature extraction, representation learning automatically identifies latent patterns that are most predictive for matching, enabling comparisons across diverse formats, schemas, and domains. Semantic embeddings—such as Word2Vec, GloVe, and BERT—encode the contextual meaning of tokens, attributes, and records, making them particularly effective for resolving semantic heterogeneity, including synonymy, polysemy, and language variation. Most recent EM systems use some form of learned representation to handle heterogeneity in both structure and meaning.

An early transformer-based approach is \emph{\emtransformer}~\cite{brunner2020entity}, which evaluates four transformer architectures—BERT, RoBERTa, XLNet, and DistilBERT—on noisy and textual EM datasets. Framed as sequence-pair classification, these models outperform classical baselines like \deepmatcher and Magellan. Their ability to handle long, unstructured records and adapt through fine-tuning highlights the strength of transformers for managing both schema and semantic heterogeneity.

Building on this direction, \emph{\ditto}~\cite{li2020deep} fine-tunes Transformer-based models such as BERT and RoBERTa for EM tasks. Treating EM as sequence-pair classification, \ditto uses contextual embeddings to capture language structure and relational cues. Its performance is further improved by (1) domain knowledge injection to highlight key tokens, (2) TF-IDF summarization to handle long records, and (3) data augmentation to enhance robustness. These design choices reduce dependency on large labeled datasets and enable generalization across schemas without requiring attribute alignment.

Extending this line of work, \emph{\hiergat}~\cite{hiergat} introduces a Hierarchical Graph Attention Transformer that models entities, attributes, and tokens through a layered graph structure. It combines self-attention and graph attention to learn multi-level contextual embeddings. The model addresses challenges such as polysemy, attribute salience, and noisy input by capturing both semantic and relational dependencies. \hiergat demonstrates strong performance on ``dirty'' datasets, showing robustness to data quality issues.

Several recent models explicitly target multi-task and contrastive representation learning for EM. \emph{Unicorn}~\cite{10.1145/3588938} trains a unified Transformer encoder jointly across EM, schema matching, entity linking, and ontology alignment, transferring knowledge across tasks. \emph{Sudowoodo}~\cite{10184556} adopts a contrastive self-supervised pretext task to learn record embeddings without labels, which are later fine-tuned for EM and other integration tasks. Both works emphasize generalizability across domains and low-resource settings, offering pathways to reduce manual supervision in EM.

Complementing these architectures,~\cite{10.14778/3598581.3598594} provide an empirical comparison of twelve off-the-shelf embeddings—including FastText, SBERT, and several BERT variants—across blocking and matching tasks. Surprisingly, they find that cosine similarity over frozen embeddings often rivals fine-tuned deep models, depending on dataset properties. These insights help practitioners select efficient and effective embedding strategies for heterogeneous matching scenarios.

\subsection{Deep Learning and Graph Neural Networks}
\label{sec:deep_learning_gnns}

Deep learning methods, including Graph Neural Networks (GNNs), have become central to recent advances in EM, offering scalable ways to handle complex heterogeneity. As summarized in surveys~\cite{barlaug2021neural,li2020deep,li2020survey}, these methods automatically learn relevant features across diverse modalities and data structures, reducing the need for manual feature engineering. Transformer-based models capture semantic and structural patterns in text, tables, and mixed data, while attention mechanisms enable fine-grained schema alignment. GNNs further extend this by modeling relational dependencies between records, attributes, and entities, directly addressing structural and relational heterogeneity.

\emph{Seq2SeqMatcher}~\cite{7-Seq2SeqMatcher} treats EM as a token-level sequence-to-sequence problem. Its align–compare–aggregate architecture handles schema and format heterogeneity by resolving attribute mismatches and managing noisy data.

Graph-based models have shown particular promise in heterogeneous settings. \emph{LinKG}~\cite{12-LinKG} offers a scalable framework for linking heterogeneous entity graphs. It combines LSTM-based encoding for textual data, locality-sensitive hashing for scalability, and heterogeneous graph attention networks to resolve ambiguous links. Its deployment in Microsoft Academic Search highlights its practical effectiveness.

\emph{HierMatcher}~\cite{6-HierMatcher} introduces a hierarchical network that models entities at token, attribute, and entity levels. It combines cross-attribute token alignment, attribute-aware attention, and entity-level aggregation, addressing heterogeneity due to non-aligned attributes and noisy or missing data.

\emph{R-SupCon}~\cite{peeters2022supervised} introduces a supervised contrastive pretraining strategy for transformers. The model pulls together records referring to the same product and pushes apart unrelated ones. To handle missing product IDs, it employs a source-aware sampling strategy. After contrastive training, the encoder is fine-tuned with labeled pairs, achieving state-of-the-art F1 scores on several e-commerce datasets. This approach exemplifies deep metric learning applied to EM.

\emph{GTA}~\cite{dou2022empowering} integrates graph contrastive learning with Transformers. It constructs hybrid graphs for dual-level matching and multi-granularity interaction, achieving high accuracy by jointly leveraging semantic embeddings and relational structure.

\emph{RELATER}~\cite{84-RELATER} applies graph-based reasoning to handle dynamic relationships and temporal heterogeneity. It propagates positive evidence and applies temporal constraints as negative signals, refining clusters in datasets with evolving attributes. While it does not explicitly use GNNs, its reasoning mechanisms align with GNN principles.

\emph{ED-GNN}~\cite{37-ED-GNN} directly applies GNNs—such as GraphSAGE, R-GCN, and MAGNN—for medical entity disambiguation. It targets terminology and context variation through relational modeling, showing strong performance in domain-specific EM.

\emph{EMBA}~\cite{zhang2024emba} presents a multi-task BERT-based architecture that predicts both record matches and shared entity IDs. It incorporates an attention-over-attention layer to emphasize token interactions critical to each task, improving both classification and resolution performance.

\subsection{{Knowledge Graphs and Ontologies for EM and Heterogeneous EM}}
\label{sec:kg_for_em}

{Knowledge Graphs (KGs) and ontologies have long been used in data integration and linkage~\cite{8-OGKS,bornemann2023matching}, and they provide structured, semantically rich representations of entities, attributes, and relationships that can support EM under both semantic and structural heterogeneity. By encoding synonymy, polysemy, hierarchical relations, and domain constraints, KGs enable EM systems to reconcile terminology mismatches, disambiguate ambiguous attributes, and exploit contextual signals not explicitly present in raw records. When incorporated into EM pipelines, KGs can support attribute-level and entity-level matching, improving robustness in noisy or heterogeneous environments.}

{KGs also help mitigate representation heterogeneity by offering schema-independent cues. For example, mappings between concept hierarchies or ontological types allow EM systems to compare records even when their schemas differ or when attribute names are misaligned. Unstructured or semi-structured sources---such as text corpora, enterprise metadata, or domain-specific ontologies---further complement structured KGs by providing latent semantic information that can be integrated through embedding models.}

{Several contributions illustrate how KGs can directly enhance EM. Ontological Graph Keys (\emph{OGKs})~\cite{8-OGKS} extend classical graph-key approaches by leveraging external ontologies to detect syntactically divergent but semantically equivalent subgraphs. Their scalable budgeted-Chase–based algorithms allow EM systems to reconcile attribute-level inconsistencies under semantic heterogeneity. Temporal KGs provide another important signal for EM: Bornemann et al.~\cite{bornemann2023matching} align entities with evolving roles using time-stamped constraints, enabling EM systems to incorporate temporal semantics when matching entities whose descriptions change over time.}

{Integrating external knowledge resources—including ontologies, domain schemas, and large KGs such as DBpedia and Wikidata—provides powerful semantic context that can improve both the accuracy and explainability of EM, particularly in settings marked by terminology variation, contextual ambiguity, or schema drift.}

\subsection{{Ontology Matching and KG Alignment Under Heterogeneity}}
\label{sec:ontology_alignment}

{Ontology matching and knowledge graph (KG) alignment are well-established areas~\cite{logmap,aml,paris} that address heterogeneity at the schema, concept, and entity levels. Although distinct from EM, these areas confront many of the same challenges—terminology variation, language differences, granularity mismatches, contextual ambiguity, and structural divergence. Techniques developed for ontology and KG alignment therefore offer valuable insights for building heterogeneity-aware EM systems.}

{Traditional ontology matching methods such as LogMap~\cite{logmap}, PARIS~\cite{paris}, and AgreementMakerLight (AML)~\cite{aml} focus on aligning classes, properties, and schema elements across ontologies using combinations of lexical cues, structural constraints, and logical reasoning. These systems explicitly target representation heterogeneity by reconciling divergent schema structures and terminologies across domains. PARIS, in particular, demonstrates how instance-, schema-, and lexical-level evidence can be integrated through probabilistic reasoning to address multi-level heterogeneity.}

{Recent KG-alignment methods extend these ideas to entity-level alignment in large-scale, often multilingual knowledge graphs. Early embedding-based models such as MTransE~\cite{mtranse} and BootEA~\cite{bootea} learn shared latent spaces that align structurally similar entities across heterogeneous KGs. Subsequent approaches such as RDGCN~\cite{rdgcn} leverage Graph Neural Networks to integrate structural, textual, and relational contexts, while CrossKG~\cite{60-CrossKG} enriches attribute information using attribute triples, character-level embeddings, and transitivity rules. Methods such as RREA~\cite{rrea} and MultiKE~\cite{multike} further incorporate multi-view or relation-aware representations to capture semantic variation, language differences, and schema divergence.}

{\emph{CollectiveEA}~\cite{collectiveea} combines structural, semantic, and lexical signals to align entities across heterogeneous KGs. It extracts graph-neighborhood features, textual cues, and string similarities, and performs collective alignment through a stable matching formulation that considers interdependencies between entities. CollectiveEA exemplifies how multi-source evidence can be integrated to address substantial structural and semantic heterogeneity in large KGs.}

{Although ontology and KG alignment target different tasks from EM, they share key goals: reconciling mismatched representations, resolving terminology variation, and establishing semantic equivalence across heterogeneous sources. Techniques from this literature—including cross-lingual embeddings, structure-aware reasoning, and multi-view alignment—offer promising directions for future heterogeneity-aware EM systems.}

\subsection{{Benchmarks for HEM}}
\label{sec:hem_benchmarks}

{Benchmarks play an essential role in evaluating EM systems under different forms of heterogeneity~\cite{magellan,wdc,gembench,matchgpt,mtranse}, yet many widely used datasets focus primarily on clean, schema-aligned tables with limited variation in semantics, granularity, or representation. For HEM, benchmarks must expose EM models to realistic sources of heterogeneity—including terminology variation, schema drift, inconsistent attribute granularity, noisy or missing values, multilingual data, and domain shifts. Several recent benchmark suites and evaluation frameworks explicitly address these challenges.}

{The \emph{Magellan family}~\cite{magellan,autoblock} introduced a set of diverse EM tasks spanning structured, web-derived, and enterprise datasets. Although not originally designed for heterogeneity, many Magellan datasets contain natural representation variability, schema inconsistencies, and string-level noise, making them suitable for evaluating blocking, similarity-based matching, and supervised learners under non-uniform data conditions. Subsequent extensions such as AutoBlock~\cite{autoblock} provide large, heterogeneous blocking datasets with realistic attribute misalignments.}

{\emph{WDC Web Table Matching}~\cite{wdc} and \emph{WDC Product Matching} datasets contain highly heterogeneous web-extracted product data with significant noise, inconsistent taxonomies, missing values, and differing attribute vocabularies. These datasets directly test robustness to terminology heterogeneity, representation drift, and unstructured data integration, and are widely used in recent deep EM papers such as DeepMatcher and Ditto.}

{\emph{GEM and GEMBench}~\cite{gembench} were specifically designed to evaluate generalization in EM models across domains, schemas, and modalities. The datasets span multiple domains with different schemas, attribute distributions, and linguistic styles, exposing deep EM methods to cross-domain and cross-schema heterogeneity. GEMBench additionally provides systematic splits for in-domain, out-of-domain, and zero-shot evaluation, aligning directly with representation, contextual, and semantic variability.}

{More recently, LLM-oriented EM benchmarks such as \emph{PromptEM} and \emph{MatchGPT}~\cite{matchgpt} include natural-language–rich attributes, schema inconsistencies, and diverse domains. These datasets are constructed to test reasoning-based matching and robustness to contextual and semantic heterogeneity. They also highlight new sources of error from long free-text attributes, entity descriptions, and domain-shift scenarios.}

{Finally, multilingual EM datasets—such as the multilingual WDC collections and cross-lingual KG alignment datasets (e.g., DBP15K)~\cite{mtranse}—provide natural test beds for evaluating heterogeneity induced by language differences, polysemy, and culturally specific schemas. Although DBP15K predates many recent benchmarks, it remains a standard evaluation dataset for cross-lingual and cross-schema heterogeneity.}

{The growing ecosystem of heterogeneous EM benchmarks reveals that models must handle not only noise and missingness, but also schema drift, semantic inconsistency, domain shift, and multilingual variation. These benchmarks provide essential test beds for evaluating heterogeneity-aware EM methods and highlight the need for robust generalization beyond narrow, dataset-specific conditions.}

\subsection{{Instance Coreference Resolution and HEM}}
\label{sec:coreference}

{Instance coreference resolution (CR) is the task of identifying textual or semi-structured mentions that refer to the same real-world entity \cite{hobbs1978resolving,soon2001machine,lee2017end,joshi2020spanbert}. Although traditionally studied in natural language processing, CR addresses many of the same heterogeneity challenges that arise in EM: variability in surface forms, contextual ambiguity, differences in granularity, missing or partial descriptions, and cross-lingual variation. As such, CR provides a rich and mature body of techniques that can inform heterogeneity-aware EM.}

{Classical CR systems~\cite{hobbs1978resolving,soon2001machine} rely on
string-based, syntactic, and rule-driven cues to determine whether two mentions are coreferent. These systems must resolve substantial representation heterogeneity, as the same entity may be expressed through names, aliases, pronouns, definite descriptions, or abbreviations. Modern neural CR methods
significantly extend this capability. The end-to-end neural coreference model of Lee et al.~\cite{lee2017end} jointly learns mention detection and coreference scoring using contextual embeddings, enabling the model to infer entity equivalence from latent semantic signals rather than explicit surface overlap. SpanBERT-based models~\cite{joshi2020spanbert} further improve
robustness by capturing fine-grained semantic similarity across mentions with rich contextualized representations. These neural models show strong ability to reconcile heterogeneity in terminology, context, and syntax—issues that also arise prominently in EM.}

{Beyond unstructured text, recent work extends CR to semi-structured and multimodal data. For example, resolving product or organization mentions in web tables, listings, XML records, and documents enriched with metadata requires integrating textual cues with structured context, schema information, or visual features~\cite{kummerer2023layoutcoref,agarwal2021weakly}.
Multimodal CR systems combine text, layout, and image signals to identify cross-mention equivalence in documents, illustrating how heterogeneous data sources can be jointly leveraged when surface similarity is insufficient. Such approaches demonstrate that CR methods increasingly operate in settings where heterogeneity is similar in nature to EM: fragmented or noisy representations, schema variation across sources, differing attribute granularity, and contextual shifts between mentions.}

{Despite targeting different tasks, CR and EM share core objectives: resolving whether two heterogeneous representations correspond to the same underlying entity. CR's long-standing emphasis on modeling contextual semantics, handling ambiguity, integrating multiple modalities, and performing global
consistency reasoning offers valuable methodological insight for EM under heterogeneity. Advances in CR—particularly in representation learning, contextual modeling, and cross-document reasoning—therefore provide promising directions for designing future HEM systems capable of robust performance across diverse and highly variable data sources.}

\subsection{Transfer Learning and Domain Adaptation}
\label{sec:transfer_learning}

Transfer learning and domain adaptation have become powerful tools for tackling HEM~\cite{1-AutoEM-zhao2019auto,2-AdaMEL-jin2021deep,trabelsi2022dame,10.14778/3565816.3565836,tu2022domain}, particularly in scenarios with limited labeled data or significant domain shifts. Transfer learning enables models to reuse knowledge learned from large, general-purpose datasets by fine-tuning them on smaller, domain-specific EM tasks, helping align terminology, language, and semantic patterns across datasets. Domain adaptation complements this by explicitly addressing discrepancies between source and target domains—such as differences in data schemas, vocabulary, or structure—using techniques like adversarial training, domain-specific embeddings, and distribution alignment. Together, these methods reduce the need for task-specific supervision while improving generalization across heterogeneous sources.

\emph{Auto-EM}~\cite{1-AutoEM-zhao2019auto} introduces a transfer learning framework for EM by leveraging deep models pre-trained on large-scale knowledge bases. It uses a hierarchical neural network to pre-train entity-specific models (e.g., for locations or organizations) using rich synonyms and contextual information. These models can be fine-tuned or directly applied to new tasks, reducing reliance on labeled data. Auto-EM effectively aligns semantically similar entities and adapts across types, demonstrating strong performance on diverse EM benchmarks.

\emph{AdaMEL}~\cite{2-AdaMEL-jin2021deep} presents a deep transfer learning method for multi-source entity linkage. It employs attribute-level self-attention to capture the importance of individual features, while domain adaptation mechanisms allow generalization across different distributions. AdaMEL integrates labeled data from multiple domains, enhancing accuracy and robustness. It effectively addresses both semantic heterogeneity (e.g., terminology shifts) and representation heterogeneity (e.g., format or attribute variation).

\emph{DAME}~\cite{trabelsi2022dame} tackles domain shift by modeling EM as a mixture-of-experts framework. Each domain expert specializes in a particular source, and a shared global model aggregates their knowledge. DAME uses adversarial training and attention mechanisms to bridge domains and performs well even in zero-shot settings. It demonstrates robustness to schema and terminology differences, outperforming models like Ditto and DeepMatcher across benchmarks.

\emph{PromptEM}~\cite{10.14778/3565816.3565836} adopts a prompt-based approach, framing each record pair as a fill-in-the-blank question for a pre-trained language model. With only a handful of labeled examples, it uses self-training to bootstrap performance under domain shift. PromptEM is particularly well-suited for low-resource settings, where labeled data is scarce or new domains are introduced frequently.

\emph{DADER}~\cite{tu2022domain} presents a modular domain adaptation framework for EM consisting of: (1) a Feature Extractor for vectorizing entity pairs, (2) a Matcher for predicting links, and (3) a Feature Aligner to minimize distributional gaps between domains. The aligner can be implemented using discrepancy-based methods (e.g., Maximum Mean Discrepancy), adversarial techniques (e.g., Gradient Reversal), or reconstruction-based approaches (e.g., autoencoders). DADER shows consistent improvements in both in-domain and cross-domain settings by learning domain-invariant representations that address both semantic and structural heterogeneity.

\subsection{Active Learning and Interactive Methods}
\label{sec:active_learning}

Active learning has long been explored in EM and related data integration tasks~\cite{82-ALMSER,80-DIAL,10.14778/2876473.2876474,10.14778/3523210.3523226,81-DAEM}, and it is particularly relevant to HEM due to the increased need for labeled data in complex, heterogeneous environments. Unlike standard EM settings where modest supervision often suffices, HEM demands targeted labeling to handle diversity in schemas, formats, and semantics. Active learning addresses this by selecting the most informative or uncertain record pairs—using strategies like uncertainty sampling or committee-based selection—to maximize model performance with minimal annotation cost.

Interactive methods complement this by incorporating user feedback into the matching loop. Users can validate matches, correct errors, or guide the system on ambiguous cases. This interactivity helps overcome context-dependent or domain-specific heterogeneity, while also allowing systems to adapt dynamically to evolving schemas and datasets. Together, active learning and interaction provide scalable and human-in-the-loop strategies for robust EM under heterogeneity.

Early work such as~\cite{10.14778/2876473.2876474} frames EM as a progressive labeling process, where an oracle labels record pairs on demand. Their algorithms optimize recall under a fixed query budget, creating an adaptive feedback loop. Extending this idea,~\cite{10.14778/3523210.3523226} develop a system for querying EM results on specific data subsets without executing a full pipeline, enabling low-latency user-driven exploration through dynamically constructed indices.

\emph{JedAI 2.0}~\cite{15-JedAI} offers an end-to-end, interactive entity resolution platform. Its GUI allows users to configure workflows, visualize matches, and refine strategies. JedAI supports schema-free and loosely structured data, improving usability and performance for HEM in practical scenarios.

\emph{ALMSER}~\cite{82-ALMSER} introduces a graph-based active learning framework for multi-source EM. It constructs correspondence graphs to identify informative record pairs and uses graph propagation to augment training data. ALMSER improves matching performance across multiple sources, effectively addressing structural and terminological heterogeneity.

\emph{DIAL}~\cite{80-DIAL} proposes a scalable active learning approach using an Index-By-Committee framework. It jointly optimizes blocking recall and matching precision, using pre-trained transformers to compute semantic embeddings. DIAL scales to large Cartesian product spaces and performs well on multilingual datasets, addressing both semantic and format heterogeneity.

\emph{CollaborER}~\cite{74-CollaborEM} introduces a self-supervised framework that generates pseudo-labels and trains matchers collaboratively. It integrates graph- and sentence-level features, outperforming existing unsupervised baselines and rivaling supervised models—while addressing both semantic and structural heterogeneity.

Beyond active learning, \emph{DAEM}~\cite{81-DAEM} combines adversarial active learning with dynamic blocking. It fills missing textual values using a neural model, selects informative samples for annotation, and uses adversarial examples to improve robustness. DAEM adapts well to heterogeneous schemas and noisy data.

\subsection{{Progressive and Incremental EM and Resolution}}
\label{sec:progressive_er}

{
Progressive ER methods aim to return high-quality matches early by prioritizing
candidate pairs under time or budget constraints. These approaches differ from
active learning because they do not rely on human feedback; instead, they
incrementally refine similarity scores or ranking functions as additional
evidence becomes available. This makes them highly relevant to HEM, where
schema drift, missing attributes, datatype inconsistencies, and heterogeneous
formats degrade blocking quality and enlarge the candidate space. Classical
work such as progressive duplicate detection~\cite{peukert2010progressive} and
large-scale systems like BigDansing~\cite{khayyat2015bigdansing} demonstrate
how adaptive ordering can improve efficiency and robustness in heterogeneous
environments.
}

{
Subsequent “pay-as-you-go’’ approaches~\cite{maskat2015payg} extend these ideas
by adaptively refining similarity signals when schema drift or missing
attributes reduce the reliability of individual features. More recent
progressive frameworks explicitly model uncertainty arising from heterogeneous
or incomplete signals. PERC~\cite{nguyen2020perc} dynamically reorders
comparisons as similarity evidence changes, providing strong performance when
attribute-level cues are noisy or inconsistent. Together, these methods align
naturally with HEM because representation and semantic heterogeneity make full
Cartesian matching impractical and amplify uncertainty in similarity
estimation. Progressive ER therefore offers a principled mechanism for coping
with heterogeneity-induced ambiguity and computational cost.
}

\subsection{{Self-Supervised and Pseudo-Label EM}}
\label{sec:self_supervised}

{
Self-supervised EM methods learn record representations or matching functions
without requiring labeled pairs, making them well suited for heterogeneous
settings where supervision is scarce or non-transferable. Heterogeneity across
schemas, formats, and vocabularies often limits the applicability of labeled
data, and self-supervision offers a way to extract domain-agnostic structure
from raw records. \emph{EmbDI}~\cite{61-EmbDI} exemplifies this direction by
constructing training corpora from random walks over token–attribute–tuple
graphs to capture structural and semantic consistency across heterogeneous
tables. More recent approaches such as \emph{CollaborER}~\cite{74-CollaborEM}
leverage graph- and sentence-level signals to generate pseudo-labels and train
matchers collaboratively, reducing reliance on human supervision while
addressing both semantic and structural heterogeneity.
}

{
Contrastive pretraining methods further strengthen robustness to heterogeneity
by learning similarity functions from augmented or multi-view representations.
Related contrastive and autoencoder-based approaches
(e.g.,~\cite{cocoE,contrastiveER}) learn embeddings resilient to missing values,
terminology variation, and schema mismatch. \emph{Sudowoodo}~\cite{10184556}
extends this direction by using contrastive learning to model semantic
similarity across noisy, structurally inconsistent textual attributes. These
self-supervised strategies directly address core HEM challenges by reducing
reliance on labeled data and producing representations that generalize across
heterogeneous sources.
}

\subsection{{Evolutionary, Meta-Heuristic, and Hybrid Approaches}}
\label{sec:evolutionary}

{
Evolutionary computation (EC) and meta-heuristic search methods have long been used to optimize complex EM pipelines~\cite{whang2010evolvingrules,zhao2019evolutionaryEMsurvey}, particularly in settings where heterogeneity renders manually designed rules or fixed classifier parameters ineffective. EC refers to a family of population-based algorithms inspired by biological evolution—including Genetic Algorithms (GA), Genetic Programming (GP), and evolutionary strategies—which explore large combinatorial spaces of matching rules, similarity functions, and blocking configurations. By encoding matching configurations (e.g., attribute selections, similarity metrics, weight vectors) as chromosomes, EC-based methods iteratively evolve high-performing solutions under objectives such as F1 score, precision–recall balance, and coverage, enabling adaptive tuning across heterogeneous schemas, formats, and semantic variations.
}

{
Early work by~\cite{whang2010evolvingrules} showed how matching rules can evolve over time to adjust to dynamic or heterogeneous environments. Genetic programming approaches such as ERGP~\cite{sun2014ergp} evolve composite similarity functions by combining heterogeneous attribute-level comparisons. Beyond individual rules, evolutionary search has been applied to full EM pipelines:~\cite{vermaas2014ontology} optimize attribute-weight configurations in ontology-based product matching to account for schema and terminology drift, and~\cite{maskat2015payg} use GA optimization to tune blocking strategies, thresholds, and similarity metrics in a pay-as-you-go framework.
}

{
Recent advances extend EC to multi-objective and hybrid optimization settings. For example, multi-objective evolutionary algorithms have been used to jointly optimize accuracy, interpretability, and computational cost, or to evolve interpretable rule sets that complement deep neural matchers~\cite{zhao2019evolutionaryEMsurvey}. These hybrid EC–ML systems demonstrate that evolutionary optimization can enhance the robustness, explainability, and resource-awareness of learning-based EM pipelines in heterogeneous data environments.
}

{
More broadly, the EM literature has recently seen the rise of \emph{hybrid architectures} that combine heterogeneous forms of evidence—rules, similarity features, blocking signals, schema information, deep embeddings, or even LLM outputs—into unified EM systems. Examples include hybrid symbolic–neural EM frameworks such as DeepER and DeeperFlow~\cite{gong2022deeperflow}, hybrid blocking methods that integrate token-based and embedding-based signals (e.g., DeepBlocker~\cite{deepblocker}), and systems that blend schema-based alignment with neural models for instance matching (e.g., HERA~\cite{hera2020}). Although distinct from evolutionary computation, hybrid EM systems share the goal of combining diverse matching signals and modeling paradigms to overcome the limitations of any single technique, and thus naturally relate to EC approaches within the broader landscape of heterogeneity-aware EM.
}

{
In comparison to purely neural EM approaches, evolutionary and hybrid methods
exhibit complementary strengths and limitations under different forms of
heterogeneity. Evolutionary and meta-heuristic techniques are particularly
effective in settings dominated by schema, structural, or granularity
heterogeneity, where explicit control over attribute selection, similarity
functions, and blocking strategies provides flexibility and interpretability.
Hybrid architectures further benefit from combining symbolic rules, schema
signals, and learned representations, making them more robust to schema drift,
missing attributes, and representation mismatch. In contrast, end-to-end
neural approaches typically excel under large-scale semantic and linguistic
heterogeneity, especially when sufficient labeled data are available, but may
be more sensitive to distribution shift, data quality issues, and changes in
schema or record structure. This comparison highlights that evolutionary and
hybrid methods are not substitutes for neural models, but complementary tools
that are often preferable in heterogeneous, low-supervision, or rapidly
evolving data environments.
}

\subsection{Large Language Models}
\label{sec:llms_foundation_models}

LLMs have recently emerged as powerful tools for EM~\cite{chatgpt-em-peeters,leveraging-llms-em,finetuning-llms-em,match-compare-select,10.14778/3529337.3529356,10.1007/s00778-023-00824-x}, offering strong capabilities for handling heterogeneity. Their ability to generalize across tasks and domains makes them highly adaptable to diverse data types and formats. With zero-shot and few-shot learning, LLMs can operate effectively in low-supervision settings—a critical advantage in HEM scenarios where labeled data is often limited. LLMs leverage pre-trained knowledge to align records across differing schemas and semantic contexts, enabling robust performance on complex matching tasks.

Beyond text, LLMs can reason over multimodal data by integrating heterogeneous input types such as structured tables, numerical fields, and free text. Retrieval-augmented methods allow dynamic access to external knowledge sources, helping the model adapt to domain-specific vocabulary or evolving context. LLMs can also synthesize training data or provide natural-language explanations for match decisions, enhancing both performance and interpretability. These characteristics make them particularly suited for the diverse challenges of heterogeneity in EM.

Early work such as~\cite{10.14778/3529337.3529356} analyzes the internal behavior of BERT-based matchers, studying attention stability, token influence, and sensitivity to input order, with a primary focus on interpretability. An extended study by~\cite{10.1007/s00778-023-00824-x} evaluates additional datasets and perturbations, revealing how domain shift, sequence length, and training data size affect transformer-based EM.

More recent work directly applies LLMs to EM tasks. \emph{MatchGPT}~\cite{chatgpt-em-peeters} evaluates ChatGPT (gpt3.5-turbo-0301) in zero-shot and in-context settings, showing that competitive performance is possible with carefully designed prompts and rules. \emph{BoostER}~\cite{leveraging-llms-em} improves the cost-efficiency of LLM-driven EM by selecting questions that minimize uncertainty using entropy-based heuristics, balancing annotation quality and API cost.

\emph{FT-LLM}~\cite{finetuning-llms-em} investigates fine-tuning strategies for LLMs in EM, showing that structured explanations and example selection significantly boost in-domain accuracy, especially for smaller models. However, challenges remain in cross-domain generalization. \emph{COMEM}~\cite{match-compare-select} proposes a three-pronged framework—matching, comparing, and selecting—that leverages global record context. The “selecting’’ module, which integrates broader dataset-level semantics, is particularly effective in complex HEM scenarios.

Collectively, these studies highlight the versatility of LLMs in EM and their promise for addressing the multifaceted challenges of HEM—particularly through prompt engineering, adaptation, interpretability, and integration with external knowledge sources.

\subsection{{Relating Surveyed Approaches to the HEM Taxonomy}}
\label{sec:mapping-taxonomy}

{Figure~\ref{fig:hierarchy} organizes heterogeneity into eight second-level categories spanning representation and semantic differences. To make this taxonomy more operational, we briefly summarize how each topic reviewed in this section targets specific heterogeneity types. This mapping also clarifies more about why these methodological areas are relevant to HEM.}

\begin{itemize}[leftmargin=0.5cm]

\item {{\em Schema and Structural Heterogeneity.}
Methods in Section~\ref{sec:schema-heterogeneity} (e.g., HERA, Machamp, GraphER) primarily address \emph{schema/structural} heterogeneity: mismatched attribute names, missing fields, varying nesting layouts, and inconsistent relational structure. They also interact with \emph{datatype/format} heterogeneity (e.g., JSON vs.\ tables) and, to a lesser extent, \emph{granularity} heterogeneity when attributes differ in resolution or abstraction across datasets.}

\item {{\em Representation Learning and Semantic Embeddings.}
Section~\ref{sec:representation_learning} targets \emph{terminology/vocabulary} heterogeneity through contextual embeddings (e.g., DITTO, HierGAT). These models also mitigate \emph{contextual semantics} by encoding relational and positional cues; handle \emph{within-dataset representation} heterogeneity through noise-tolerant embedding spaces; and partially support \emph{datatype/format} heterogeneity by operating over linearized or text-derived representations.}

\item {{\em Knowledge Graphs and Ontologies for EM.}
Section~\ref{sec:kg_for_em} leverages ontologies and KGs to resolve \emph{terminology} and \emph{contextual} heterogeneity via hierarchical types, semantic relations, and synonym mappings. KGs also reduce \emph{schema/structural} and \emph{granularity} heterogeneity by offering schema-independent, logically grounded representations; and support limited \emph{linguistic} heterogeneity through multilingual knowledge bases.}

\item {{\em Ontology Matching and KG Alignment.}
Section~\ref{sec:ontology_alignment} primarily addresses \emph{schema/structural} and \emph{terminology} heterogeneity by aligning classes, relations, and identifiers across heterogeneous ontologies. Modern KG alignment methods additionally target \emph{contextual semantics} through structure-aware embeddings and handle \emph{linguistic} heterogeneity through cross-lingual mappings and multilingual encoders.}

\item {{\em Benchmarks for HEM.}
Section~\ref{sec:hem_benchmarks} presents datasets that instantiate multiple forms of heterogeneity simultaneously. These include \emph{representation} heterogeneity (schema drift, datatype inconsistencies, granularity mismatches, multimodal attributes), \emph{semantic} heterogeneity (terminology variation, contextual ambiguity), and \emph{linguistic} heterogeneity (multilingual product descriptions and KG labels). Such benchmarks allow controlled evaluation across several dimensions of HEM.}

\item {{\em Instance Coreference Resolution.}
Section~\ref{sec:coreference} corresponds mainly to \emph{terminology}, \emph{contextual}, and \emph{linguistic} heterogeneity: CR systems must reconcile aliases, abbreviations, and context-dependent references across documents and domains. CR also touches on \emph{granularity} heterogeneity (e.g., entity vs.\ sub-entity mentions) and mild \emph{within-dataset representation} heterogeneity when mentions vary in completeness or surface form.}

\item {{\em Transfer Learning and Domain Adaptation.}
Section~\ref{sec:transfer_learning} addresses \emph{terminology}, \emph{contextual}, and \emph{linguistic} heterogeneity by adapting models across domains with differing vocabularies, styles, and label distributions. Domain adaptation also mitigates \emph{datatype/format} shifts and \emph{within-dataset representation} heterogeneity by aligning feature distributions or learning domain-invariant representations.}

\item {{\em Active Learning and Interactive Methods.}
Section~\ref{sec:active_learning} primarily helps resolve \emph{terminology} and \emph{contextual} heterogeneity by allowing models to query users on ambiguous or domain-specific pairs. Interactive correction also reduces the effects of \emph{granularity} heterogeneity (e.g., attribute grouping differences) and \emph{within-dataset representation} heterogeneity (inconsistent fields, missing values).}

\item {{\em Progressive and Incremental EM.}
Section~\ref{sec:progressive_er} relates primarily to \emph{within-dataset representation} and \emph{schema/structural} heterogeneity. Progressive ER methods reorder or prioritize comparisons when attribute overlap is low, schemas differ across sources, or similarity signals are unreliable. They also address \emph{datatype/format} heterogeneity by adapting to partially missing or inconsistently typed attributes, and indirectly handle \emph{contextual} heterogeneity by allocating computation to pairs with ambiguous or conflicting evidence.}

\item {{\em Self-Supervised and Pseudo-Label EM.}
Section~\ref{sec:self_supervised} directly targets \emph{terminology/vocabulary} and \emph{contextual} heterogeneity by learning representations from raw text, relational structure, or token–attribute–tuple graphs without needing aligned schemas or labeled examples. These approaches also mitigate \emph{within-dataset representation} heterogeneity (noise, missing values, inconsistent fields) and support \emph{datatype/format} heterogeneity by extracting domain-agnostic structure from mixed or semi-structured inputs. Contrastive and graph-based self-supervision additionally helps address mild \emph{schema/structural} heterogeneity when attribute layouts differ across sources.}

\item {{\em Evolutionary, Meta-Heuristic, and Hybrid Approaches.}
Section~\ref{sec:evolutionary} addresses \emph{schema/structural}, \emph{granularity}, and \emph{within-dataset representation} heterogeneity by evolving matching rules, feature subsets, or hybrid similarity operators. These systems also handle \emph{datatype/format} heterogeneity through flexible rule search and occasionally incorporate \emph{contextual} or \emph{terminology} cues when combined with neural or symbolic components.}

\item {{\em LLM-Based EM.}
Section~\ref{sec:llms_foundation_models} naturally handles \emph{terminology}, \emph{contextual}, and \emph{linguistic} heterogeneity through pretrained semantic knowledge and in-context reasoning. LLMs also partially address \emph{datatype/format} heterogeneity by interpreting semi-structured inputs (e.g., JSON, tables) as text, and can mitigate \emph{within-dataset representation} heterogeneity through robust contextualization of noisy attributes.}

\end{itemize}

{This mapping shows that each methodological area in Section~3 addresses a distinct subset of heterogeneity challenges and collectively spans all eight second-level dimensions in our taxonomy. It also clarifies how the taxonomy guides the organization and interpretation of the surveyed literature.}

\section{Entity Matching and the FAIR Principles}
\label{sec:fair}

The FAIR principles promote data practices that make information
\emph{Findable}, \emph{Accessible}, \emph{Interoperable}, and
\emph{Reusable}~\cite{FAIR-Nature2016}. These guidelines are widely embraced
across scientific and industrial domains to support data sharing,
reproducibility, and large-scale data integration. Central to this vision is
the ability to identify, link, and reuse entities across datasets that differ
in structure, representation, and semantics—precisely the setting addressed
by HEM.

As discussed throughout Section~\ref{sec:survey}, representation and semantic
heterogeneity simultaneously obstruct FAIRification and motivate the design
of robust EM techniques. Schema mismatches, inconsistent representations,
terminology variation, and granularity differences all complicate the
realization of FAIR goals. Conversely, EM systems that are explicitly designed
to operate under heterogeneity play a critical role in enabling FAIR-compliant
data infrastructures.

{
To provide a more structured synthesis of this relationship, we explicitly
connect the EM research areas reviewed in Section~\ref{sec:survey} to the individual FAIR
principles. Table~\ref{tab:fair-em-structured} summarizes how representative
methods from each surveyed area contribute to Findability, Accessibility,
Interoperability, and Reusability. Rows correspond directly to the major EM
areas discussed in Section~\ref{sec:survey}, and each cell lists concrete methods (with
references) that most directly support a given FAIR dimension. The table is
intended to highlight traceable technical connections rather than provide
exhaustive coverage.
}

\begin{table*}[t]
\centering
\resizebox{\textwidth}{!}{%
\begin{tabular}{p{4.5cm} l l l l}
\toprule
\textbf{EM Area} 
& \textbf{Findable} 
& \textbf{Accessible} 
& \textbf{Interoperable} 
& \textbf{Reusable} \\
\midrule

\parbox[t]{4.5cm}{Schema \& structural\\ heterogeneity (\S\ref{sec:schema-heterogeneity})}
& HERA~\cite{54-HERA}
& GraphER~\cite{40-GraphER}
& --
& Machamp~\cite{wang2021machamp} \\
\midrule

\parbox[t]{4.5cm}{Repr. learning\\ \& sem emb (\S\ref{sec:representation_learning})}
& Ditto~\cite{li2020deep}
& Ditto~\cite{li2020deep}
& HierGAT~\cite{hiergat}
& \parbox[t]{4.6cm}{Sudowoodo~\cite{10184556}\\ Unicorn~\cite{10.1145/3588938}} \\
\midrule

\parbox[t]{4.5cm}{DL \& GNNs (\S\ref{sec:deep_learning_gnns})}
& --
& Seq2SeqMatcher~\cite{7-Seq2SeqMatcher}
& \parbox[t]{4.6cm}{LinKG~\cite{12-LinKG}\\ HierMatcher~\cite{6-HierMatcher}}
& \parbox[t]{4.6cm}{R-SupCon~\cite{peeters2022supervised}\\ GTA~\cite{dou2022empowering}} \\
\midrule

\parbox[t]{4.5cm}{KGs \& ontologies \\ for EM (\S\ref{sec:kg_for_em})}
& --
& --
& OGKs~\cite{8-OGKS}
& Bornemann et al.~\cite{bornemann2023matching} \\
\midrule

\parbox[t]{4.5cm}{Ontology matching \&\\ KG alignment (\S\ref{sec:ontology_alignment})}
& --
& --
& \parbox[t]{4.6cm}{LogMap~\cite{logmap}\\ AML~\cite{aml}\\ PARIS~\cite{paris}}
& \parbox[t]{4.6cm}{MTransE~\cite{mtranse}\\ BootEA~\cite{bootea}\\ RDGCN~\cite{rdgcn}\\ CrossKG~\cite{60-CrossKG}\\ RREA~\cite{rrea}\\ MultiKE~\cite{multike}\\ CollectiveEA~\cite{collectiveea}} \\
\midrule

\parbox[t]{4.5cm}{Self-supervised \&\\ pseudo-label EM (\S\ref{sec:self_supervised})}
& --
& --
& CollaborER~\cite{74-CollaborEM}
& \parbox[t]{4.6cm}{EmbDI~\cite{61-EmbDI}\\ Sudowoodo~\cite{10184556}} \\
\midrule

\parbox[t]{4.5cm}{Evolutionary \&\\ hybrid approaches (\S\ref{sec:evolutionary})}
& --
& --
& --
& \parbox[t]{4.6cm}{ERGP~\cite{sun2014ergp}\\ DeepER/DeeperFlow~\cite{gong2022deeperflow}\\ DeepBlocker~\cite{deepblocker}} \\
\midrule

\parbox[t]{4.5cm}{LLMs (\S\ref{sec:llms_foundation_models})}
& --
& PromptEM~\cite{10.14778/3565816.3565836}
& MatchGPT~\cite{chatgpt-em-peeters}
& COMEM~\cite{match-compare-select} \\

\bottomrule
\end{tabular}%
}
\caption{Structured mapping between EM research areas reviewed in Section~\ref{sec:survey}
and the FAIR principles they most directly support. Each cell lists
representative methods and references, highlighting concrete technical
mechanisms rather than exhaustive coverage.}
\label{tab:fair-em-structured}
\end{table*}

\begin{itemize}[leftmargin=0.35cm]

\item \emph{Findability.}
Findability requires that entities and metadata be consistently indexed and
retrievable by humans and machines using persistent identifiers and
well-defined representations. In heterogeneous settings, duplicated or
fragmented entity descriptions hinder reliable indexing and discovery.
{
EM approaches addressing schema and representation heterogeneity
(Sections~\ref{sec:schema-heterogeneity} and~\ref{sec:representation_learning})
directly support Findability by constructing normalized or schema-independent
entity representations. For example, HERA~\cite{54-HERA} incrementally builds
super-records across heterogeneous schemas, while contextual embedding models
such as Ditto~\cite{li2020deep} produce canonical textual representations that
enable consistent indexing across sources, as summarized in
Table~\ref{tab:fair-em-structured}.
}

\item \emph{Accessibility.}
Accessibility emphasizes that data and metadata should be retrievable through
well-defined, machine-readable protocols. Structural and format heterogeneity
directly obstruct accessibility when schemas are undocumented, inconsistent,
or incompatible across sources.
{
Schema-agnostic EM methods reviewed in
Sections~\ref{sec:schema-heterogeneity} and~\ref{sec:deep_learning_gnns}—such as
GraphER~\cite{40-GraphER} and Ditto~\cite{li2020deep}—enable record comparison
without requiring strict prior schema alignment. By operating over loosely
structured or heterogeneous inputs, these approaches operationalize FAIR
Accessibility in settings where explicit schema harmonization is infeasible
(Table~\ref{tab:fair-em-structured}).
}

\item \emph{Interoperability.}
Interoperability aims to ensure that datasets can be combined and interpreted
within a shared semantic framework, typically relying on common vocabularies,
ontologies, or data models. Semantic heterogeneity—arising from terminology
variation, granularity mismatches, and contextual ambiguity—poses a primary
barrier to this goal.
{
Ontology- and knowledge-aware EM methods reviewed in
Sections~\ref{sec:kg_for_em} and~\ref{sec:ontology_alignment}, such as
OGKs~\cite{8-OGKS}, PARIS~\cite{paris}, and RDGCN~\cite{rdgcn}, explicitly encode
semantic relations and hierarchical structure to reconcile mismatched meanings
across datasets. Deep hierarchical models such as HierGAT~\cite{hiergat}
further integrate semantic and structural signals, supporting interoperability
at both the schema and instance levels, as reflected in
Table~\ref{tab:fair-em-structured}.
}

\item \emph{Reusability.}
Reusability focuses on enabling future use of data through semantic clarity,
quality assurance, and trustworthiness. Unresolved heterogeneity in entity
identity propagates ambiguity to downstream applications, reducing confidence
in reuse.
{
Self-supervised and pseudo-label EM methods
(Section~\ref{sec:self_supervised}), such as Sudowoodo~\cite{10184556} and
EmbDI~\cite{61-EmbDI}, learn robust representations that generalize across
heterogeneous datasets with limited supervision. Evolutionary and hybrid EM
approaches (Section~\ref{sec:evolutionary}) further enhance reusability by
combining neural, symbolic, and rule-based signals to produce more interpretable
and adaptable matching outcomes (Table~\ref{tab:fair-em-structured}).
}

\end{itemize}

{
Viewed through the FAIR lens, heterogeneity is not merely an obstacle for
entity matching but a defining constraint that shapes data reuse at scale.
Conversely, the EM techniques surveyed in Section~\ref{sec:survey}—particularly
schema-agnostic models, semantic embeddings, knowledge-aware approaches,
self-supervised methods, and hybrid pipelines—constitute concrete technical
enablers of FAIRification. By making these connections explicit,
Table~\ref{tab:fair-em-structured} clarifies how advances in
heterogeneity-aware EM directly support the construction of findable,
accessible, interoperable, and reusable data infrastructures.
}

\section{Experimental Analysis}
\label{sec:exp}

This section evaluates recent EM methods under different forms of \emph{semantic heterogeneity}, including synonym variation (Sections~\ref{sec:syn-exp} and~\ref{sec:syn-exp-random}), data granularity differences (Section~\ref{sec:ex-hier}), and dirty or noisy data (Section~\ref{sec:dirty-exp}). These experiments target three key types of semantic heterogeneity: terminology and language, granularity and resolution, and data quality. Results are summarized in Section~\ref{sec:results}. While prior work has studied EM under noise (e.g.,~\cite{li2020deep}), this is the first focused evaluation across these semantic dimensions using recent models.

\subsection{Experimental Setting}
\label{sec:expMycode}
We first describe the setup and infrastructure used in our experiments. Additional implementation details are available in our repository~\cite{HeterogeneityEMSurvey}.

\subsubsection{Datasets and Preparation} \label{sec:prepare}

We use six widely studied datasets--\abtbuy, \company, \fodor, \wdc, \walmartamazon, and \itunesamazon--summarized in Table~\ref{tab:datasets}. These datasets span diverse characteristics: small to large sizes (from hundreds to hundreds of thousands of records), structured and textual attributes, and hierarchical fields (e.g., ``category'' and ``brand'' in \walmartamazon). This diversity allows us to apply all experimental perturbations described in the following sections. Including additional datasets or both clean and noisy variants would expand the study without altering the core findings.

\begin{table}[H]
    \centering
    {
    \setlength{\tabcolsep}{2pt} 
    \begin{tabular}{lcccccc}
    \toprule
        \textbf{Dataset} & \textbf{\#Rec (tr/te)} & \textbf{\#Attr.} & \textbf{\#Pairs (tr/te)} & \textbf{\%Pos (tr/te)} \\
    \midrule
        \wdc & 24,107 + 4,500 & 5 & 8,839 + 500 & 37\% + 11\% \\
        \company & 90,129 + 22,503 & 1 & 22,560 + 5,640 & 25\% + 25\% \\
        \abtbuy & 7,659 + 1,916 & 3 & 822 + 206 & 11\% + 11\% \\
        \fodor & 757 + 189 & 6 & 88 + 22 & 12\% + 12\% \\
        \walmartamazon & 8,193 + 2,049 & 5 & 769 + 193 & 9\% + 9\% \\
        \itunesamazon & 430 + 109 & 8 & 105 + 27 & 24\% + 25\% \\
    \bottomrule
    \end{tabular}
    }
    \caption{Dataset characteristics.}
    \vspace{-2mm}
    \label{tab:datasets}
\end{table}

We prepare the selected datasets by injecting different types of semantic heterogeneity in a controlled manner. Our goal is to evaluate two key properties of EM models: {\em robustness} and {\em generalizability}. 

To test robustness, we introduce heterogeneity into the training data while keeping the test data unchanged. This simulates scenarios where models are trained on heterogeneous datasets. For generalizability, we inject heterogeneity into the test data while using unaltered training data, mimicking deployment in new environments. We consider three heterogeneity types: (1) terminology and language, (2) granularity and resolution, and (3) data quality. We describe each data perturbation below.

\begin{itemize}[leftmargin=0.35cm]
\item \emph{Synonym Injection:} To simulate semantic heterogeneity from terminology variation, we replace words in textual attributes with contextually appropriate synonyms. This is applied to datasets with rich textual content--\abtbuy, \company, and \wdc. We first extract words from the textual attributes, removing stopwords, numeric tokens, and non-alphabetic terms. We use the KeyBERT library~\cite{grootendorst2020keybert} to extract candidate keywords, filtering out product names and domain-specific terms. 

To generate synonyms, we compare WordNet~\cite{miller1995wordnet}, BERT~\cite{devlin2019bert}, and LLMs such as GPT-4 and Gemini against a small manually labeled test set. GPT-4 outperforms all others in contextual accuracy, so we adopt it to generate synonyms. Using a prompt-based approach, we ask GPT-4 to replace words with their most appropriate synonyms based on sentence context. These replacements are applied randomly to a specified proportion of candidate words. At 100\% synonym ratio, we modify approximately 440k/3.9M tokens in \wdc, 257k/616k in \abtbuy, and 84M/237M in \company (test/train).

{To rigorously distinguish between the effects of semantic heterogeneity and random lexical noise, we establish a Random Word Noise baseline. As implemented in our benchmarking scripts, this baseline isolates the identical token positions targeted by the synonym injection process. However, instead of using contextually relevant synonyms, we replace these tokens with unrelated words randomly sampled from the NLTK English word corpus. This process includes preprocessing steps to preserve structural consistency, such as handling compound terms (e.g., standardizing ``light-emitting diode'') before injection. By maintaining the exact same noise distribution and token indices as the synonym set, this baseline allows us to attribute performance drops specifically to semantic drift rather than simple vocabulary mismatch.}


    \item \emph{Hierarchical Data Distortion:} To simulate granularity-based semantic heterogeneity, we modify hierarchical attributes such as time, location, and categorization. 
{Instead of random noise, we employ domain-specific taxonomy trees to systematically alter the level of abstraction. We implement two specific perturbation mechanisms based on the attribute type:
\begin{itemize}
    \item \textbf{Categorical Generalization:} For nominal attributes (e.g., \textit{City}, \textit{Brand}, \textit{Category}), we utilize nested dictionaries to map specific entities to their semantic parents. For example, in the \walmartamazon dataset, a specific brand like ``Acer'' is mapped to ``Computers,'' which is further mapped to ``Electronics.'' The perturbation function traverses this hierarchy to replace a leaf node with a randomly selected ancestor.
    \item \textbf{Numerical and Temporal Discretization:} For continuous or high-cardinality values, we apply interval-based binning hierarchies. Exact values are replaced with range descriptors or broader timeframes. For instance, in the \itunesamazon dataset, a specific song duration (e.g., 210 seconds) is generalized to a ``Moderate'' length bucket, while release dates are abstracted to their release year or decade (e.g., ``1999'' becomes ``1990s'').
\end{itemize}
These transformations allow us to inject controlled semantic heterogeneity, simulating data integration scenarios where sources report at different levels of granularity.}
We quantify information loss using entropy, computed as the sum of individual column entropies across independent attributes, following standard information theory~\cite{shannon1948}. Datasets with suitable hierarchies--\itunesamazon, \walmartamazon, and \fodor--are selected for these experiments. The number of affected test/train cells is approximately 450/1.7k for \itunesamazon, 550/2.2k for \fodor, and 6.1k/24.5k for \walmartamazon.

    


\item \emph{Dirty Data Injection:} To simulate data quality heterogeneity, we introduce missing values, attribute noise, and label noise.
{We implement distinct perturbation logic for each category to model real-world data corruption patterns:
\begin{itemize}
    \item \textbf{Missing Values:} We employ three mechanisms defined by their dependency structures. 
    \begin{itemize}
        \item For \emph{Missing Completely at Random (MCAR)}, we uniformly sample row and column indices across the dataset to remove values, ensuring no dependency on the data content.
        \item For \emph{Missing at Random (MAR)}, the missingness probability is conditioned on the record's ground truth label. We assign a higher base probability weight to matching record pairs (0.8) compared to non-matches (0.2). This weight is then passed through an arctangent function to generate a smoothed probability ($P = \arctan(weight) / \arctan(N)$), determining whether a value is masked.
        \item For \emph{Missing Not at Random (MNAR)}, we introduce a dependency on the unobserved value itself. The probability weight is calculated by combining the class label weight with a normalized hash of the attribute's specific value (added as a factor $hash(value) \% 100 / 100$). This ensures that specific values (e.g., specific price points or high-cardinality strings) have distinct probabilities of being missing.
    \end{itemize}
    \item \textbf{Attribute Noise:} We apply type-specific distortions. 
    \begin{itemize}
        \item For \textbf{string attributes}, we simulate typographical errors using a two-step process: first, we select approximately 30\% of the string's positions and replace the characters with random ASCII letters; second, we append an additional random character to the end of the string to simulate insertion errors.
        \item For \textbf{numerical attributes}, we apply multiplicative noise by perturbing the original value with a random factor drawn uniformly from the range $[-20\%, +20\%]$ and rounding the result to the nearest integer.
    \end{itemize}
    \item \textbf{Label Noise:} To simulate annotation errors, we randomly select a subset of training and validation records determined by the noise ratio (e.g., 5--25\%) and flip their binary labels ($0 \leftrightarrow 1$), creating valid-but-incorrect supervision signals.
\end{itemize}
}
For missing values and attribute noise, we use \fodor (~1k/~4k test/train cells), \walmartamazon (~9k/~40k), and \itunesamazon (~800/~3.4k). Label noise is introduced separately by flipping a percentage of match labels.

\end{itemize}

\subsubsection{Entity Matching Models Evaluated}

We evaluate four EM models in our experiments: \deepmatcher, \ditto, \emtransformer, and \hiergat. This selection balances practical considerations and architectural diversity. While earlier sections review a broad landscape of EM methods, pilot experiments showed that many recent models exhibit similar performance trends. Including all of them would add complexity without significantly altering conclusions. Thus, we focus on models that are actively maintained, run on modern libraries, and install without legacy dependencies--criteria that many older systems no longer meet.

The selected models span distinct design paradigms. \deepmatcher~\cite{Mudgal2018} is a widely-used baseline with a relatively simple architecture that continues to perform well, especially when sufficient labeled data is available. It combines hybrid attention over tokenized inputs, pre-trained word embeddings, and optional metadata to compute similarity scores. The remaining three methods--\ditto, \emtransformer, and \hiergat--introduce architectural innovations aimed at better handling heterogeneous EM (HEM). \ditto leverages input augmentation and a Transformer encoder, \emtransformer blends rule-based and learned matching strategies, and \hiergat uses graph-based contextual modeling. Together, these models offer a diverse and representative set of approaches for evaluating robustness under data heterogeneity.

\subsection{Experimental Results}\label{sec:results}

We report matching accuracy using the Area Under the Receiver Operating Characteristic Curve (AUC), which captures how well a model distinguishes between matched and non-matched record pairs. AUC summarizes performance across all classification thresholds and is well-suited to settings where models produce continuous similarity scores. A higher AUC reflects better ranking ability and overall discrimination performance, independent of a fixed classification cutoff. We analyze AUC trends as different forms of semantic heterogeneity are introduced.

{
Unless stated otherwise, we repeat each stochastic setting (e.g., different random seeds and, when relevant, different perturbations) and report mean ROC AUC with $\pm 1$ standard deviation (shaded bands; mean $\pm$ SD). Our goal in this section is to assess robustness trends under controlled heterogeneity, not to over-interpret small gaps between similar methods. Although significance tests for AUC differences are possible, applying them across many datasets and conditions would require heavy multiple-comparison correction and can highlight trivial effects. We therefore focus on run-to-run variability and treat differences within that variability as inconclusive.}


\begin{figure*}[ht]
    \centering
    \begin{subfigure}[b]{0.30\textwidth}
        \includegraphics[width=\textwidth]{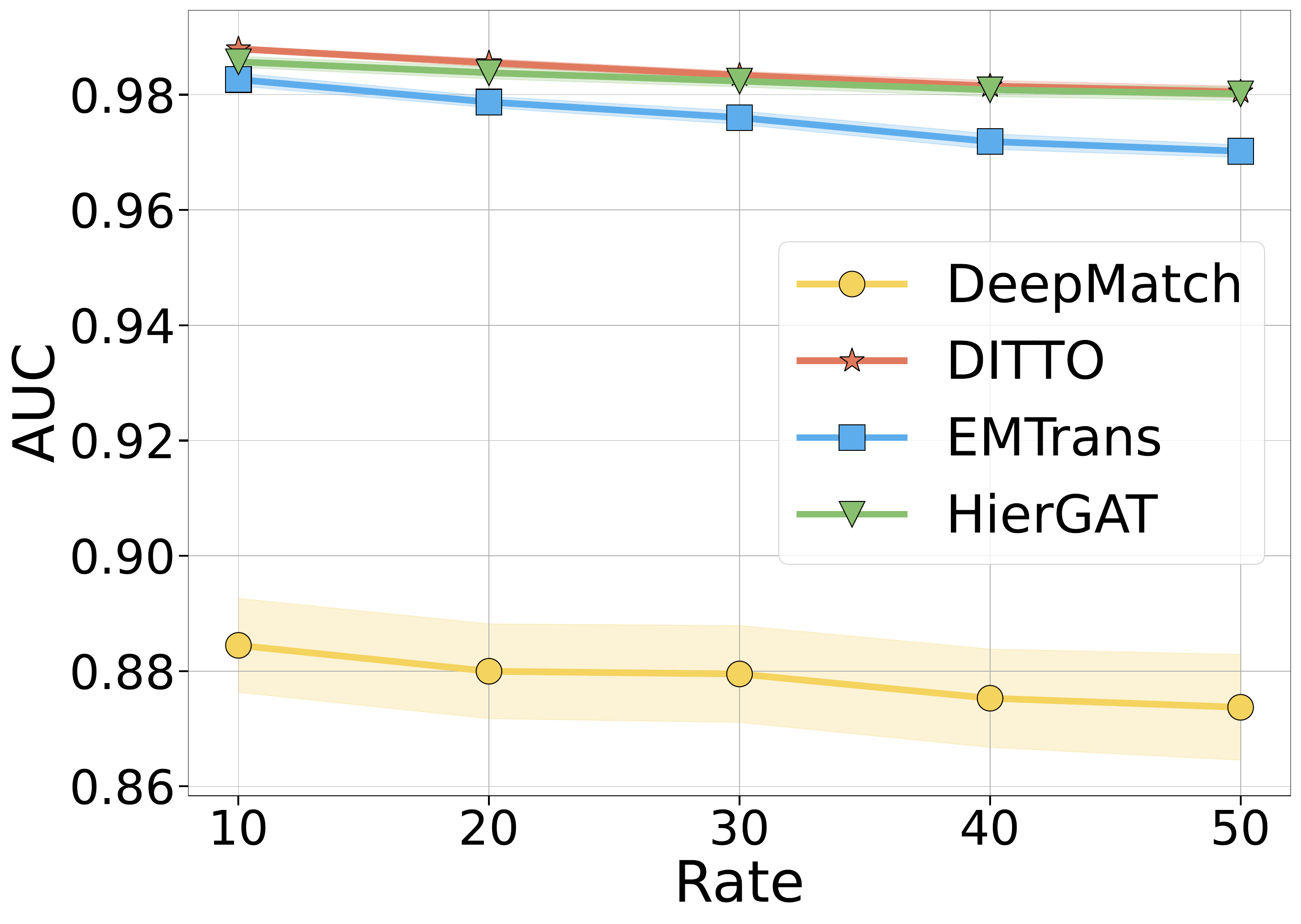}
        \caption{\abtbuy}
        \label{fig:abt_buy-syn}
    \end{subfigure}
    \hspace{2mm}
    \begin{subfigure}[b]{0.30\textwidth}
        \includegraphics[width=\textwidth]{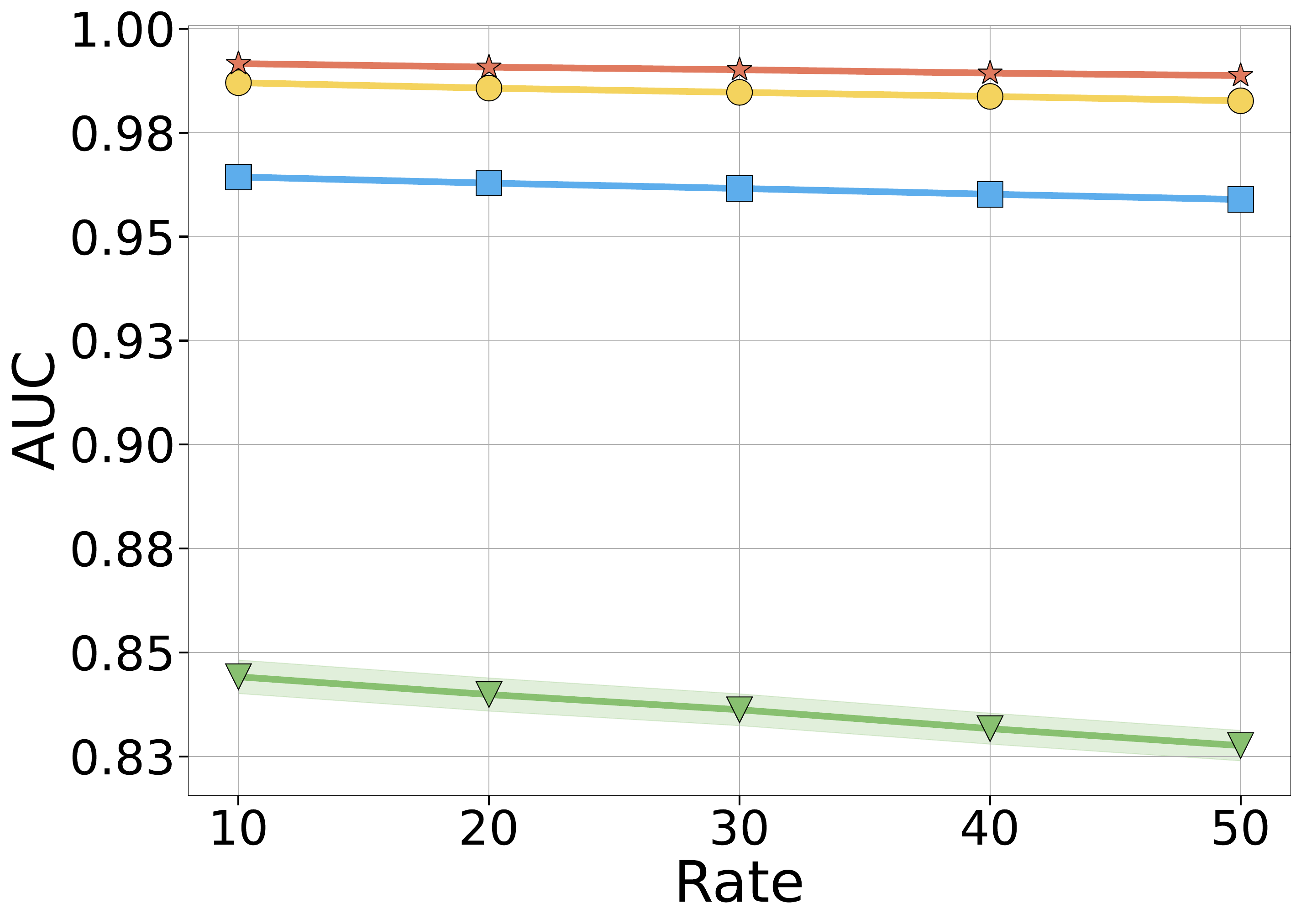}
        \caption{\company}
        \label{fig:company-syn}
    \end{subfigure}
    \hspace{2mm}
    \begin{subfigure}[b]{0.30\textwidth}
        \includegraphics[width=\textwidth]{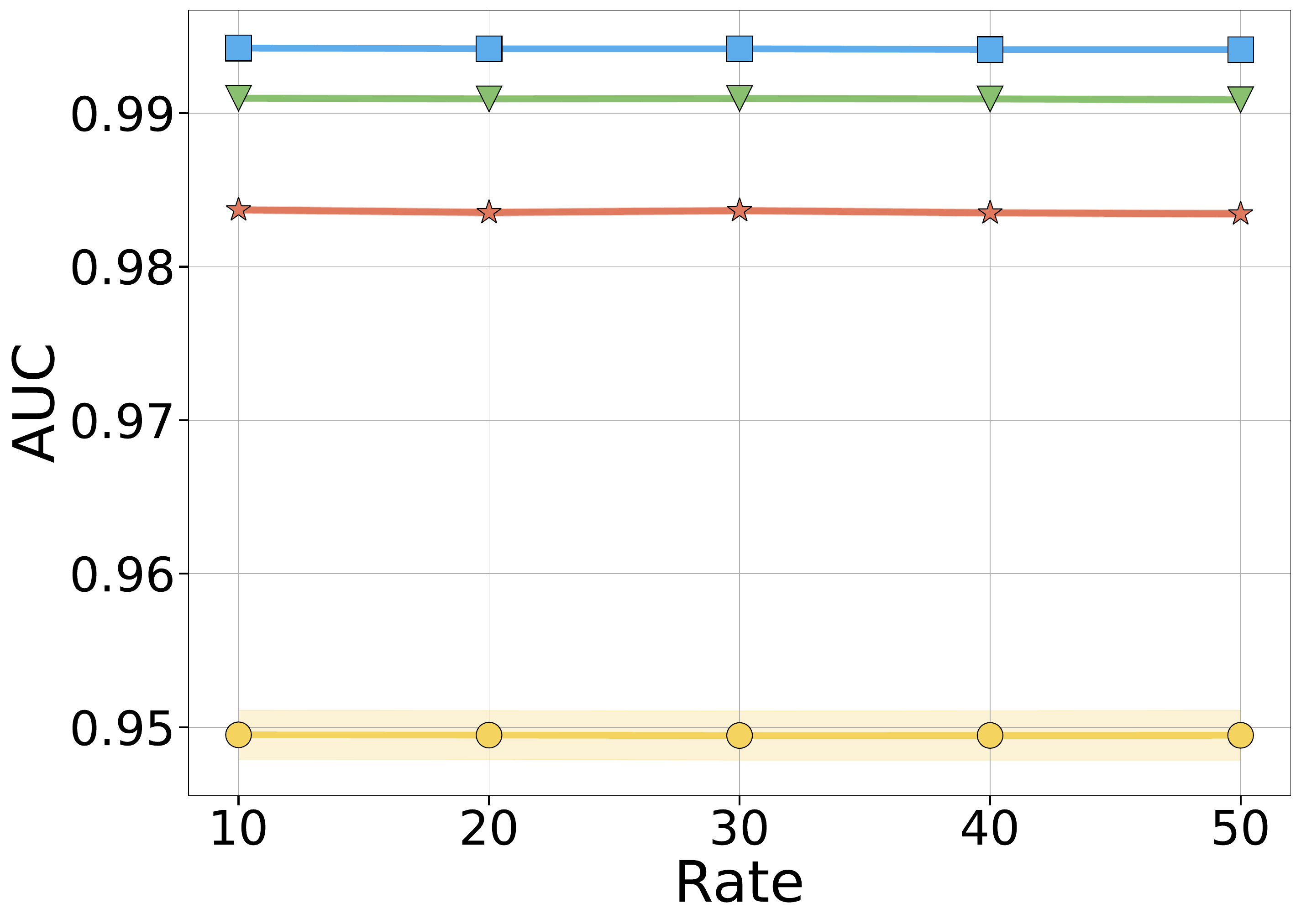}
        \caption{\wdc Product}
        \label{fig:wdc-syn}
    \end{subfigure}
    \vskip\baselineskip
    \begin{subfigure}[b]{0.30\textwidth}
        \includegraphics[width=\textwidth]{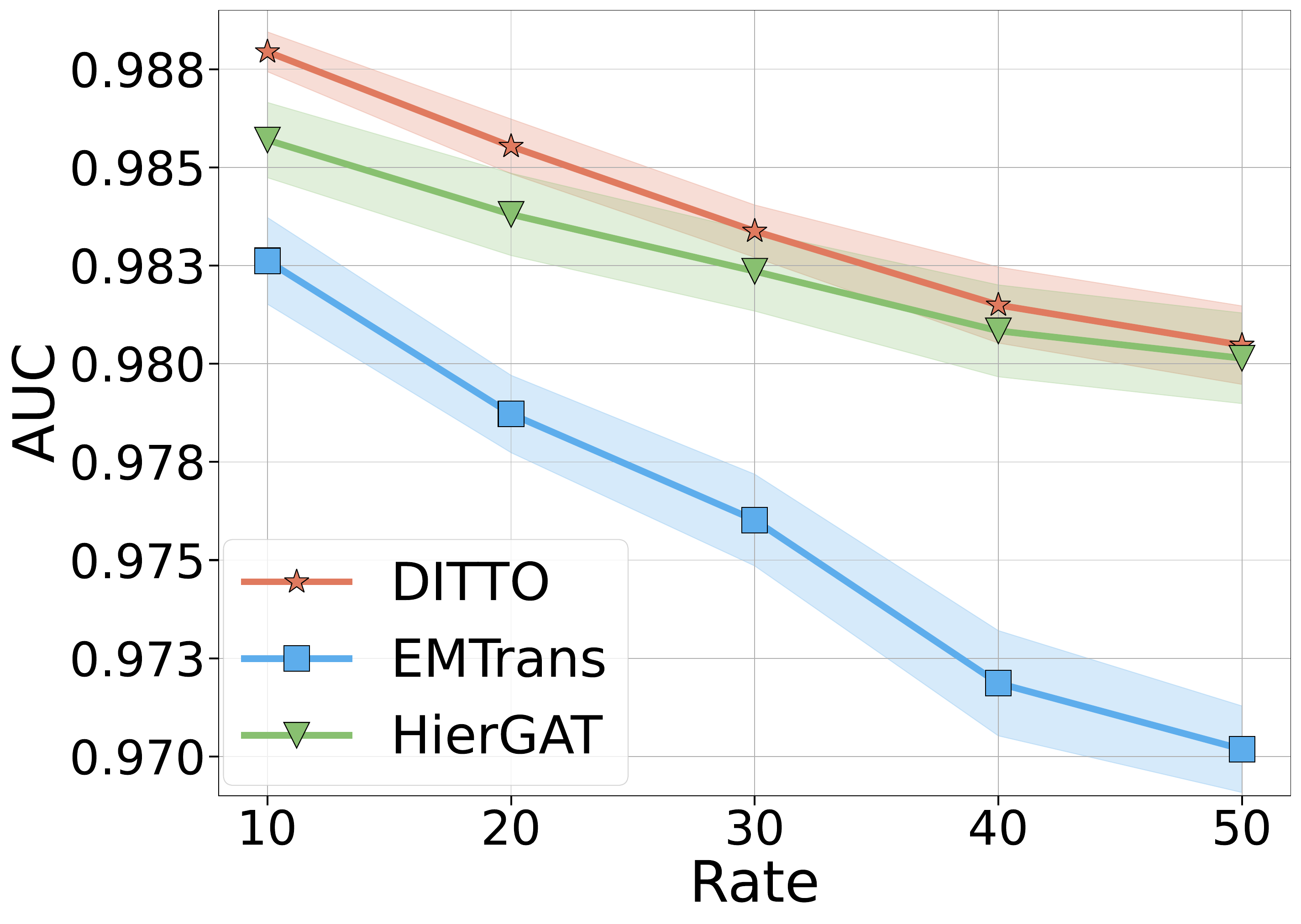}
        \caption{\abtbuy}
        \label{fig:abt_buy-syn-c}
    \end{subfigure}
    \hspace{2mm}
    \begin{subfigure}[b]{0.30\textwidth}
        \includegraphics[width=\textwidth]{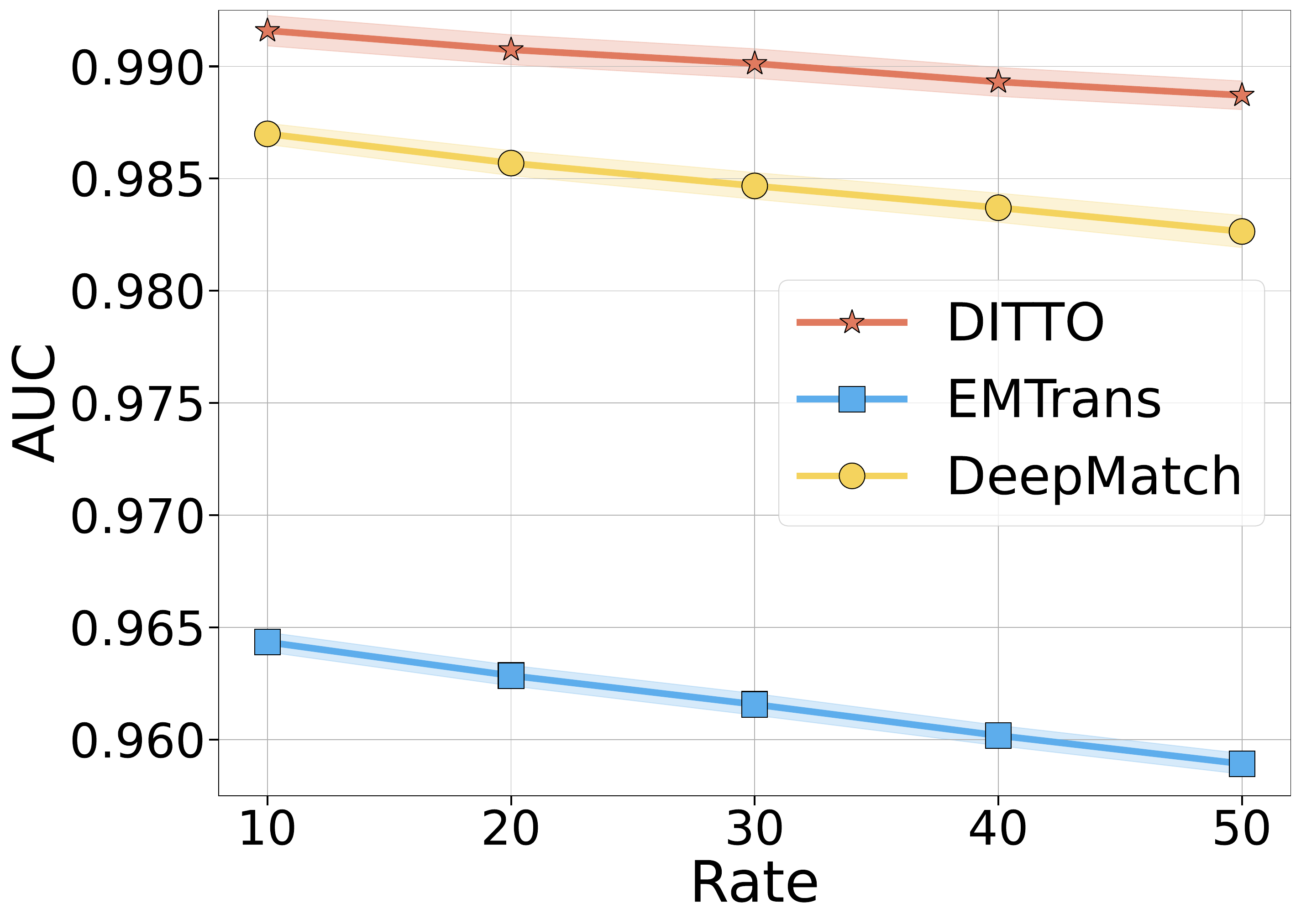}
        \caption{\company}
        \label{fig:company-syn-c}
    \end{subfigure}
    \hspace{2mm}
    \begin{subfigure}[b]{0.30\textwidth}
        \includegraphics[width=\textwidth]{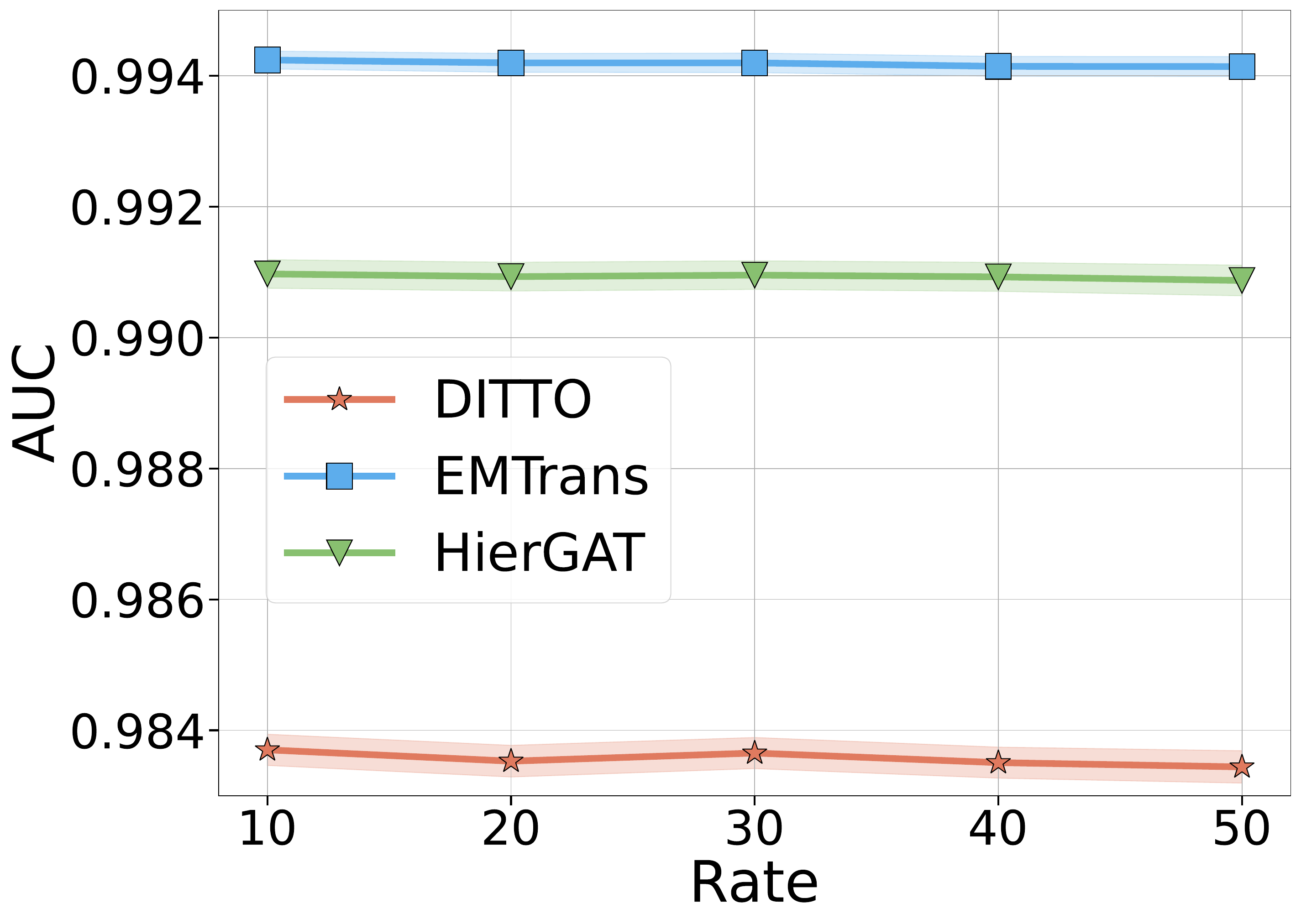}
        \caption{\wdc}
        \label{fig:wdc-syn-c}
    \end{subfigure} 
    \vspace{-3mm}
    \caption{Impact of synonym injection in test data across EM methods and datasets. Second-row figures provide detailed views of high-performing methods from the first-row figures.}
    \label{fig:syn}
\end{figure*}

\subsubsection{Language \& Terminology Heterogeneity with Synonyms}\label{sec:syn-exp}

Figure~\ref{fig:syn} shows how injecting synonyms into the test data affects model performance across different datasets and synonym replacement rates. The second row of plots zooms in on the top-performing methods from the first row.

In the \abtbuy dataset (Figure~\ref{fig:abt_buy-syn}), all models show a marked performance decline as the synonym ratio increases. \deepmatcher performs the worst, with both a low starting AUC and the steepest drop. This is due to its reliance on static embeddings, which lack contextual sensitivity. \emtransformer performs moderately better but falls behind \ditto and \hiergat. As shown in Figure~\ref{fig:abt_buy-syn-c}, \ditto maintains robustness through BERT-based fine-tuning, while \hiergat benefits from its hierarchical graph attention, capturing both local and global context.

In the \company dataset (Figure~\ref{fig:company-syn}), initial AUC scores are slightly higher, and the performance drop from synonym injection is more gradual. \hiergat performs the worst in this setting, while \deepmatcher and \ditto lead. Figure~\ref{fig:company-syn-c} suggests that \deepmatcher benefits from the dataset’s well-structured attributes (e.g., ``name'', ``address''), where static embeddings suffice. \ditto remains strong due to its semantic generalization, whereas \hiergat's attention mechanisms are less useful in strictly structured data.

For the \wdc dataset (Figure~\ref{fig:wdc-syn}), AUC remains largely stable across all models, regardless of the synonym ratio. \deepmatcher again shows the weakest performance, followed by \ditto. \emtransformer and \hiergat are the most resilient. As illustrated in Figure~\ref{fig:wdc-syn-c}, the minimal impact is likely due to redundancy in attributes such as ``title'', ``brand'', and ``price'', which give models alternative signals. \ditto’s performance is relatively lower here, likely because the structure of the dataset reduces the advantage of contextual embeddings. \emtransformer and \hiergat are more effective due to their modeling of attribute interactions and hierarchy.

\takeaway Synonym injection reduces EM performance across all models, but the severity varies by dataset. \ditto and \hiergat are generally the most robust, especially in unstructured or complex settings. \deepmatcher struggles due to its use of static embeddings but performs relatively well in highly structured datasets. These results underscore the importance of aligning model choice with dataset characteristics when dealing with semantic heterogeneity.

\begin{figure*}[ht]
    \centering
    \begin{subfigure}[b]{0.4\textwidth}
        \centering
        \includegraphics[width=\textwidth]{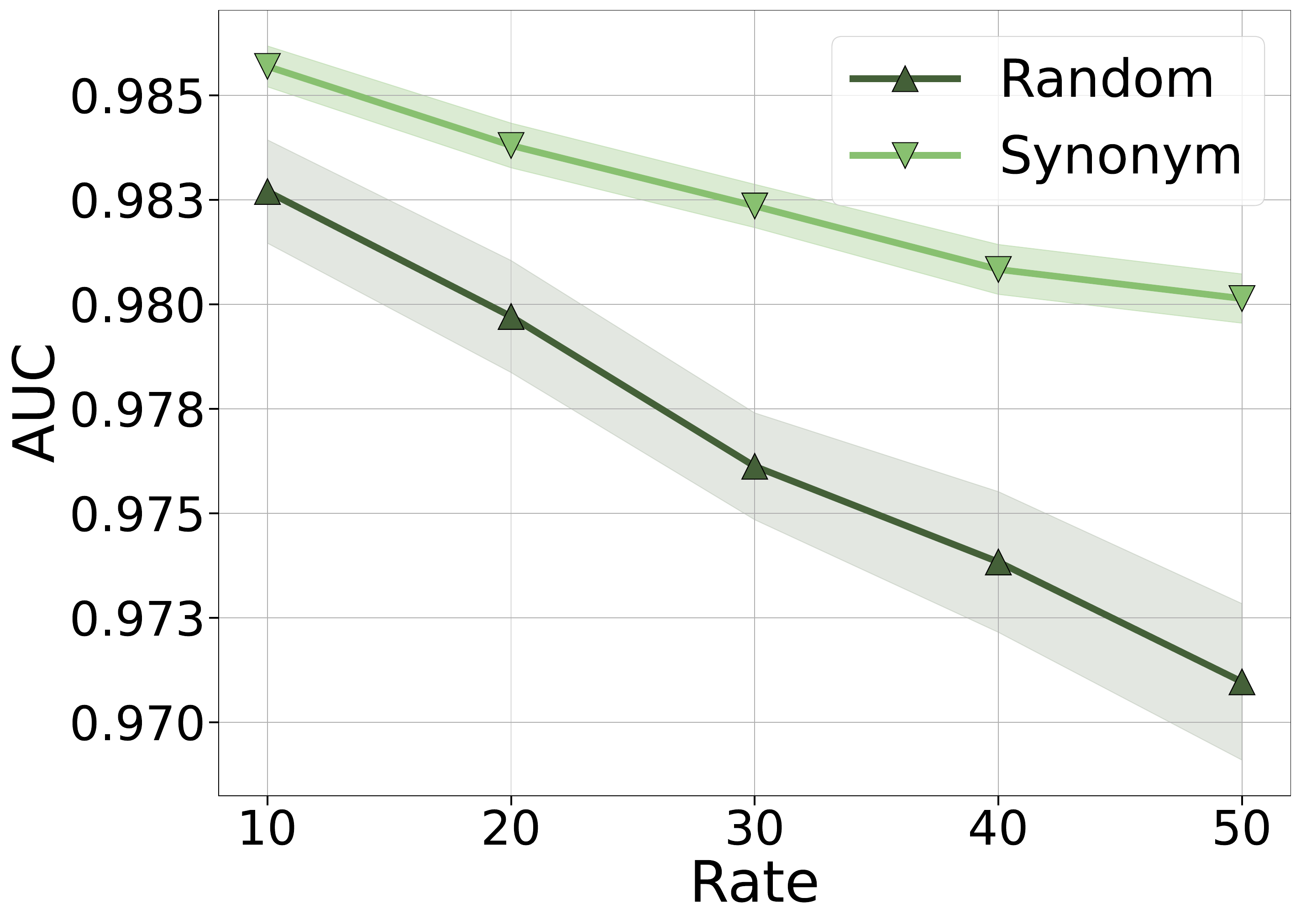}
        \caption{\hiergat}
        \label{fig:gat-random-abt}
    \end{subfigure}
    \begin{subfigure}[b]{0.4\textwidth}
        \centering
        \includegraphics[width=\textwidth]{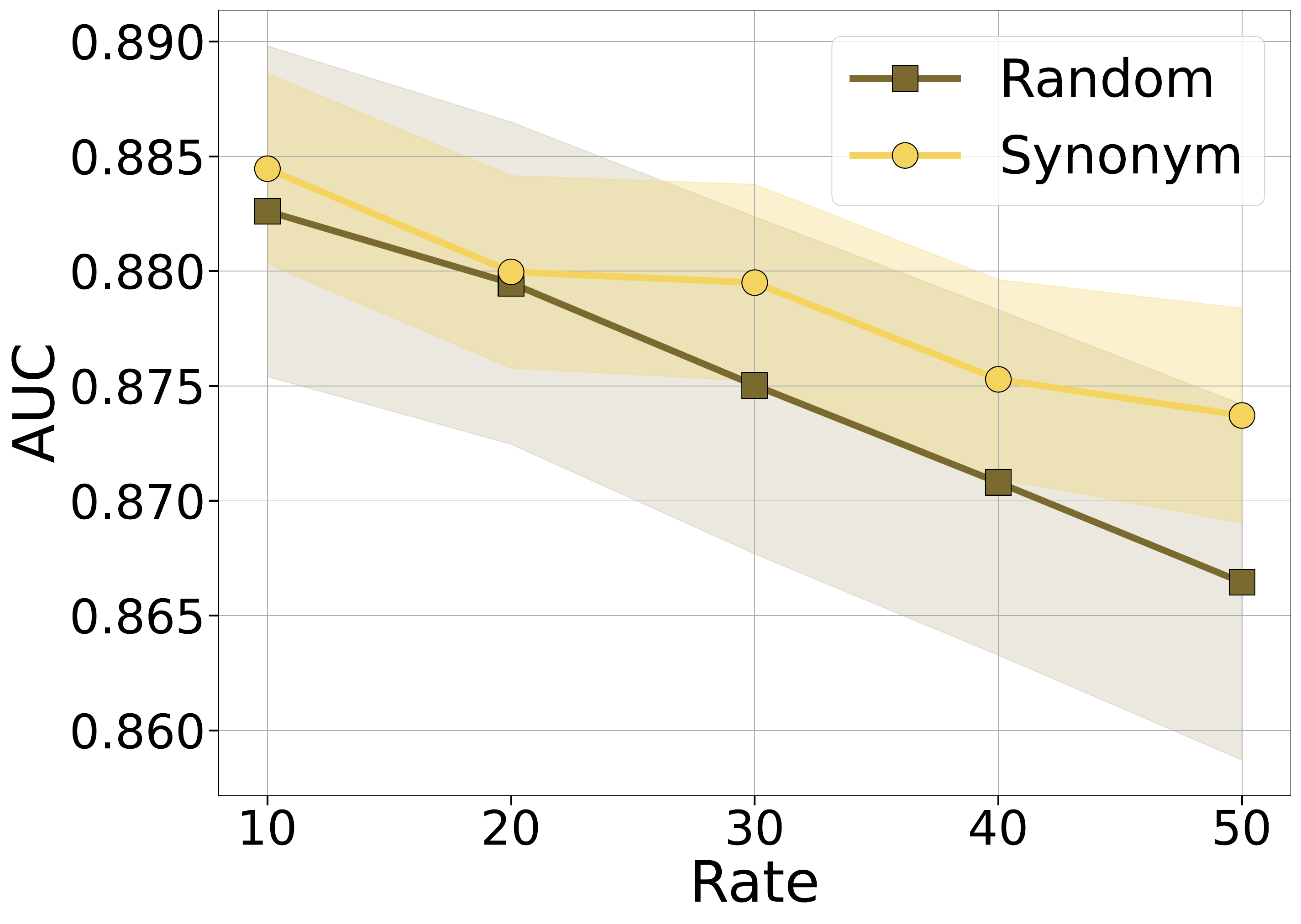}
        \caption{\deepmatcher}
        \label{fig:deepm-random-abt}
    \end{subfigure}
    \hfill
    \begin{subfigure}[b]{0.4\textwidth}
        \centering
        \includegraphics[width=\textwidth]{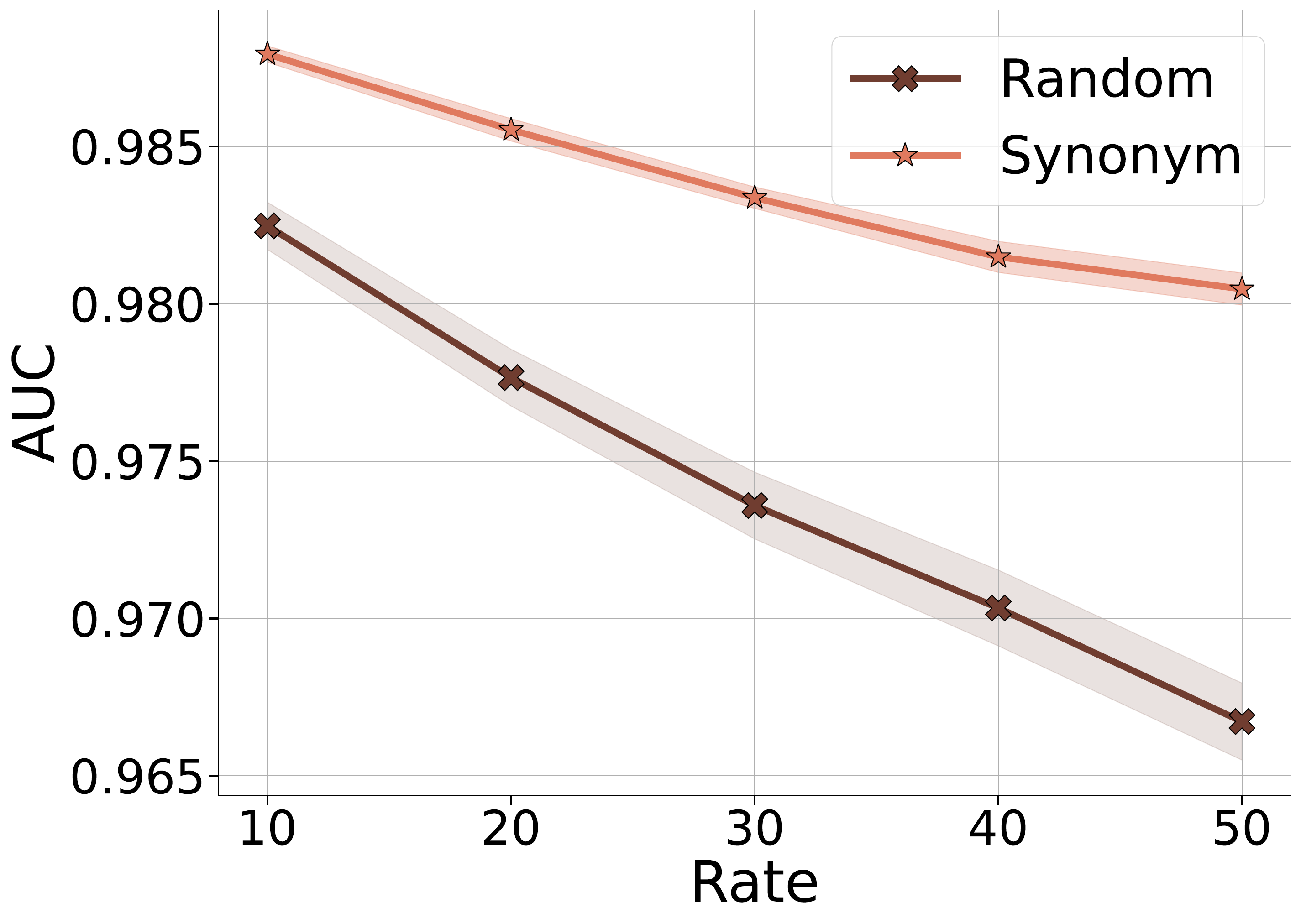}
        \caption{\ditto}
        \label{fig:ditto-random-abt}
    \end{subfigure}
    \begin{subfigure}[b]{0.4\textwidth}
        \centering
        \includegraphics[width=\textwidth]{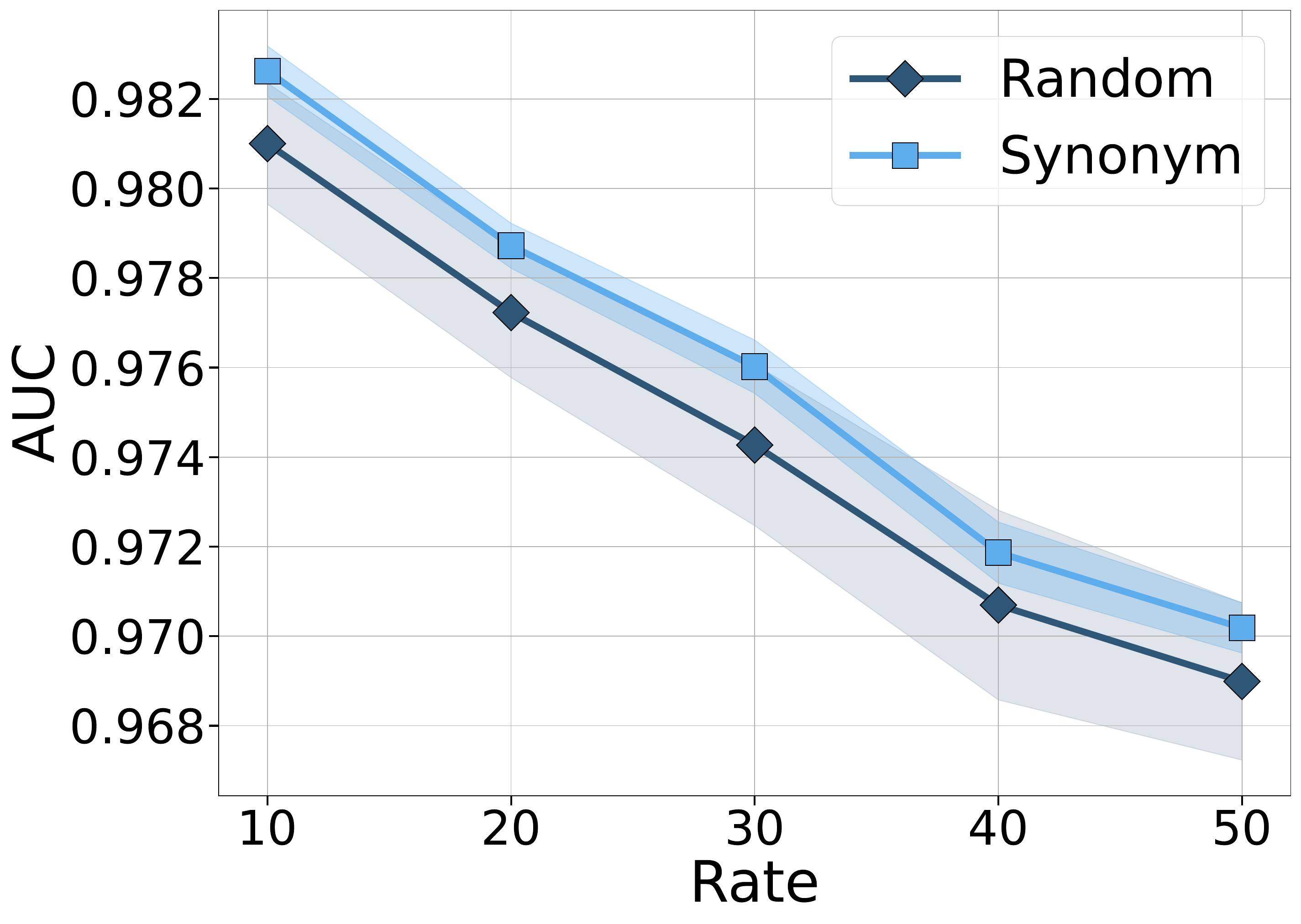}
        \caption{\emtransformer}
        \label{fig:tran-random-abt}
    \end{subfigure}
    \caption{Random word vs. synonym replacement in \abtbuy}
    \label{fig:syn-random-abt}
\end{figure*}

\begin{figure*}[ht]
    \centering
    \begin{subfigure}[b]{0.4\textwidth}
        \centering
        \includegraphics[width=\textwidth]{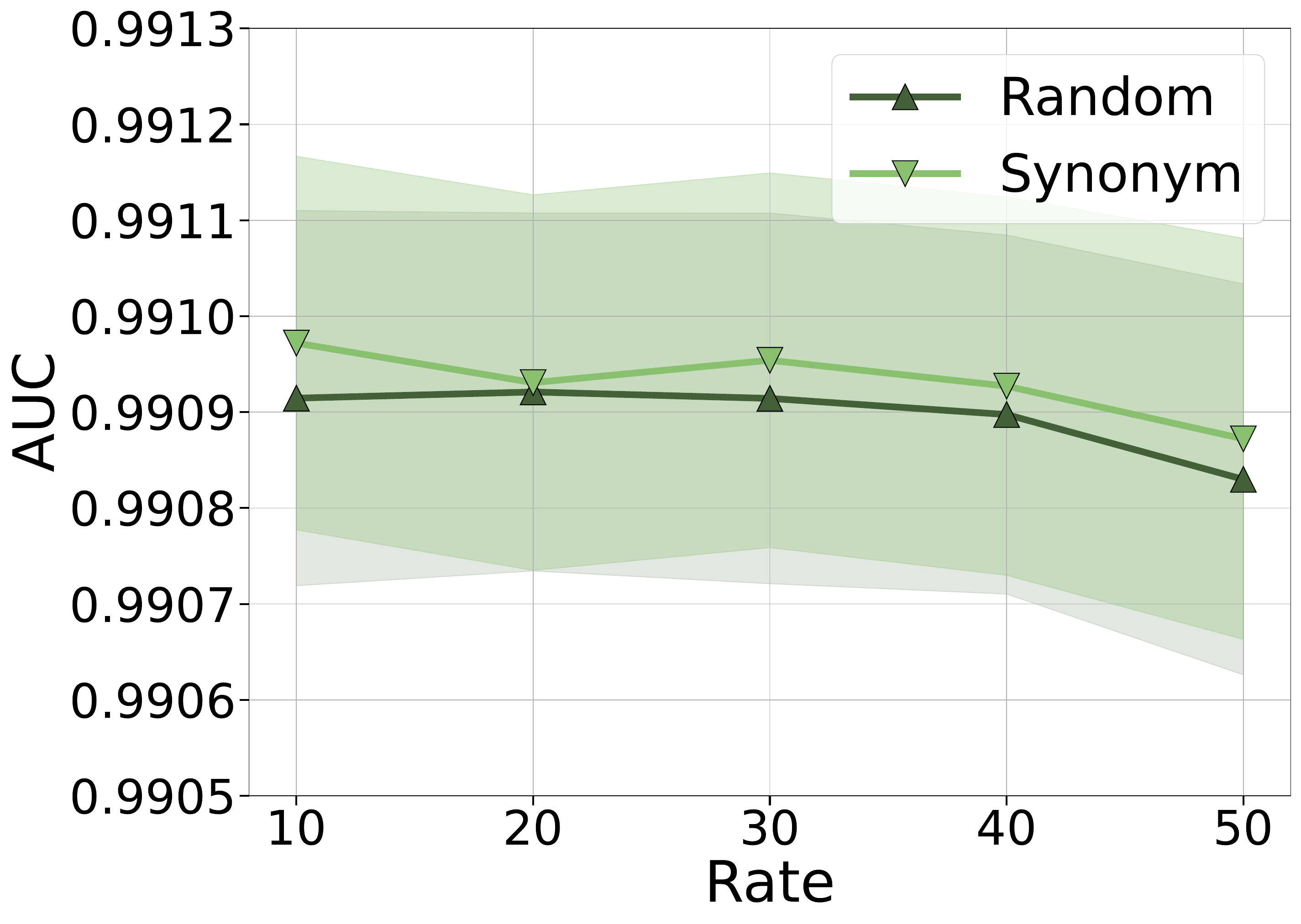}
        \caption{\hiergat}
        \label{fig:gat-random-wdc}
    \end{subfigure}
    \begin{subfigure}[b]{0.4\textwidth}
        \centering
        \includegraphics[width=\textwidth]{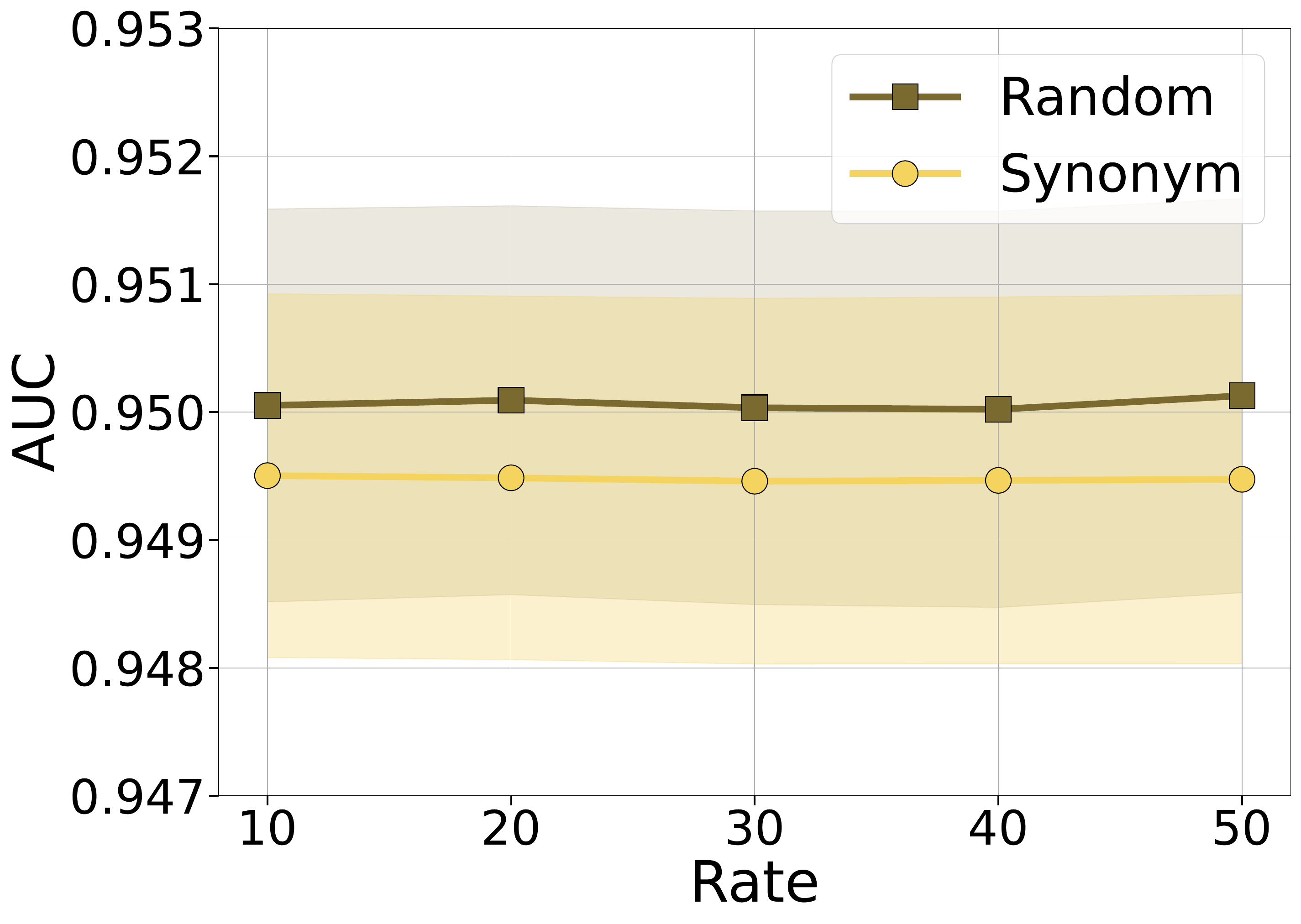}
        \caption{\deepmatcher}
        \label{fig:deepm-random-wdc}
    \end{subfigure}
    \hfill
    \begin{subfigure}[b]{0.4\textwidth}
        \centering
        \includegraphics[width=\textwidth]{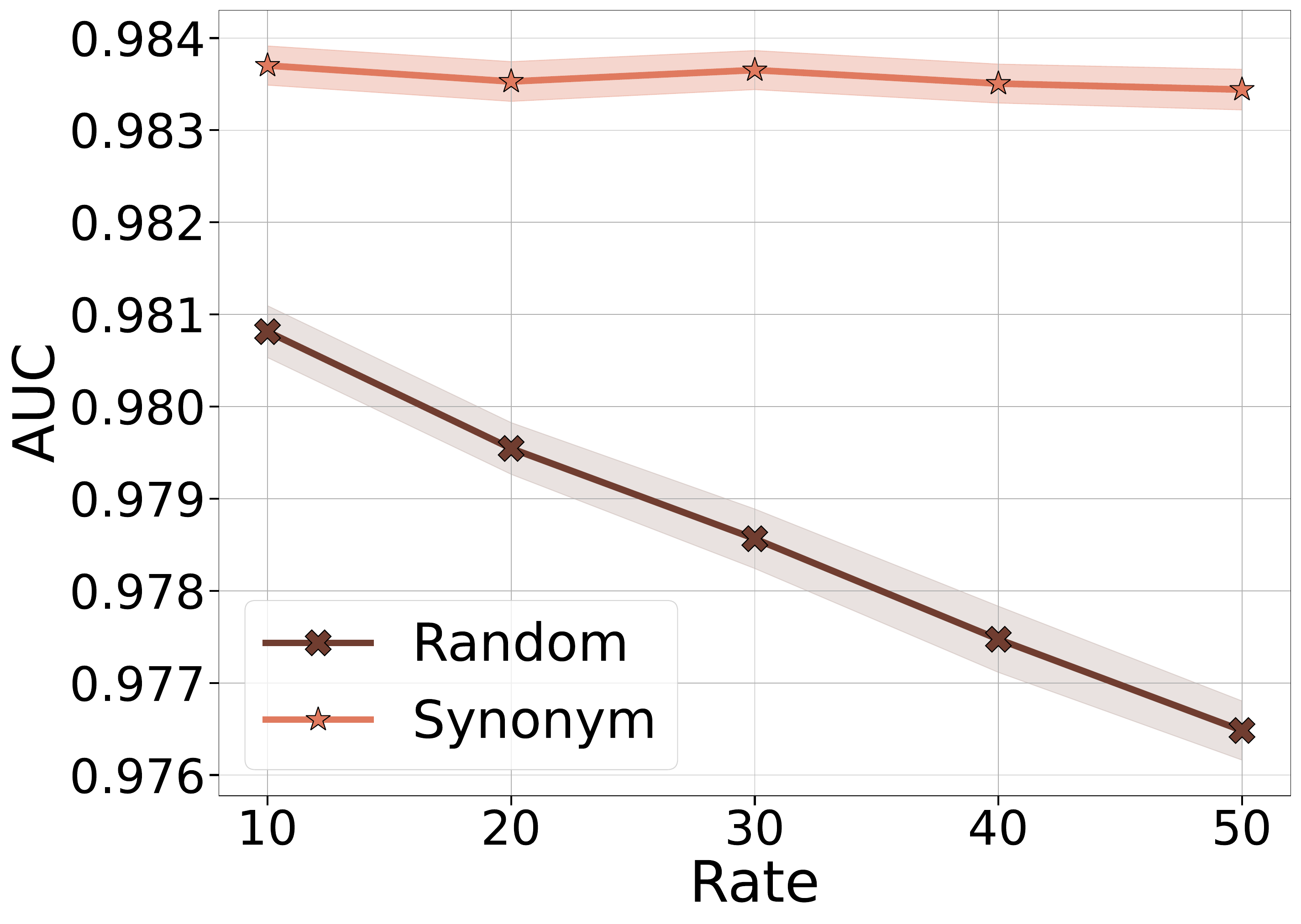}
        \caption{\ditto}
        \label{fig:ditto-random-wdc}
    \end{subfigure}
    \begin{subfigure}[b]{0.4\textwidth}
        \centering
        \includegraphics[width=\textwidth]{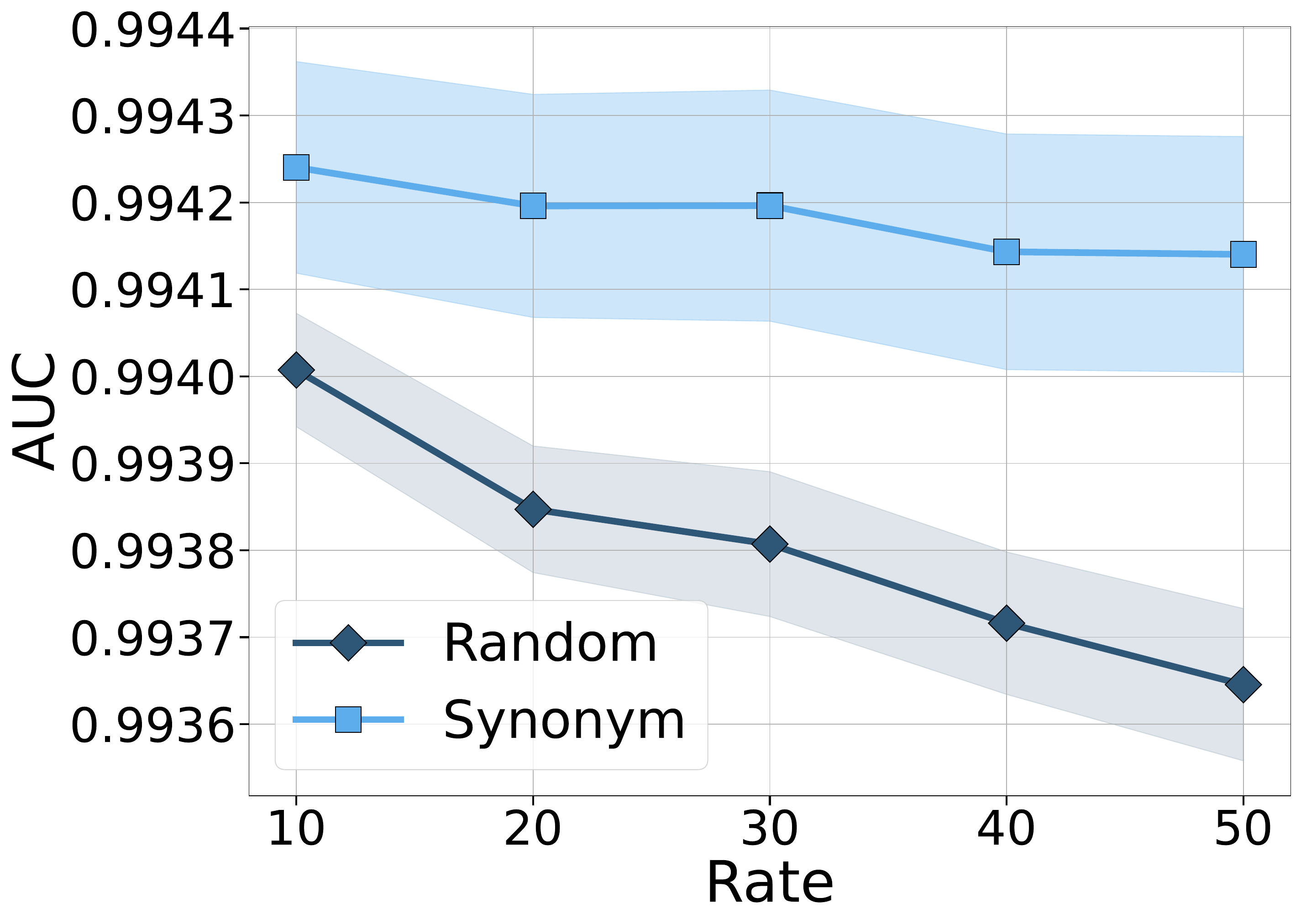}
        \caption{\emtransformer}
        \label{fig:tran-random-wdc}
    \end{subfigure}
    \caption{Random word vs. synonym replacement in \wdc}
    \label{fig:syn-random-wdc}
\end{figure*}

\subsubsection{Synonyms vs Random Words} \label{sec:syn-exp-random}

This experiment evaluates whether EM models can effectively leverage semantic relationships, such as synonymy. We compare their performance in two scenarios: one where words are replaced with contextually appropriate synonyms, and another where words are replaced with random, unrelated terms. If a model cannot exploit semantic relationships, its performance should degrade similarly in both settings.

Figures~\ref{fig:syn-random-abt} and~\ref{fig:syn-random-wdc} show AUC curves for \abtbuy and \wdc under increasing replacement rates. As expected, AUC drops in both settings as the noise increases. However, synonym replacements consistently lead to better performance than random ones, showing that most models can use semantic signals. This gap becomes more pronounced at higher replacement rates, highlighting the role of semantic understanding in robustness.

\ditto demonstrates strong performance across both datasets (Figures~\ref{fig:ditto-random-abt} and~\ref{fig:ditto-random-wdc}), maintaining a large gap between synonym and random replacements. Its fine-tuned BERT-based embeddings capture semantic relationships effectively. \hiergat and \emtransformer also show sensitivity to synonym injection (Figures~\ref{fig:gat-random-abt}, \ref{fig:tran-random-wdc}), though their performance is somewhat dataset-dependent. In contrast, \deepmatcher exhibits nearly overlapping performance curves for synonym and random replacements (Figures~\ref{fig:deepm-random-abt},~\ref{fig:deepm-random-wdc}), indicating its static embeddings fail to encode semantic similarity. Similarly, \emtransformer struggles in \abtbuy (Figure~\ref{fig:tran-random-abt}) but performs better in \wdc, reflecting its reliance on dataset structure.

\takeaway \ditto is most effective at leveraging semantic relationships, followed by \hiergat and \emtransformer. \deepmatcher shows little benefit from synonym-aware training, due to its static embedding design.

\subsubsection{Granularity \& Resolution Heterogeneity with Hierarchical Distortion} \label{sec:ex-hier}

\begin{figure*}[ht]
    \centering
    \begin{subfigure}[b]{0.30\textwidth}
        \includegraphics[width=\textwidth]{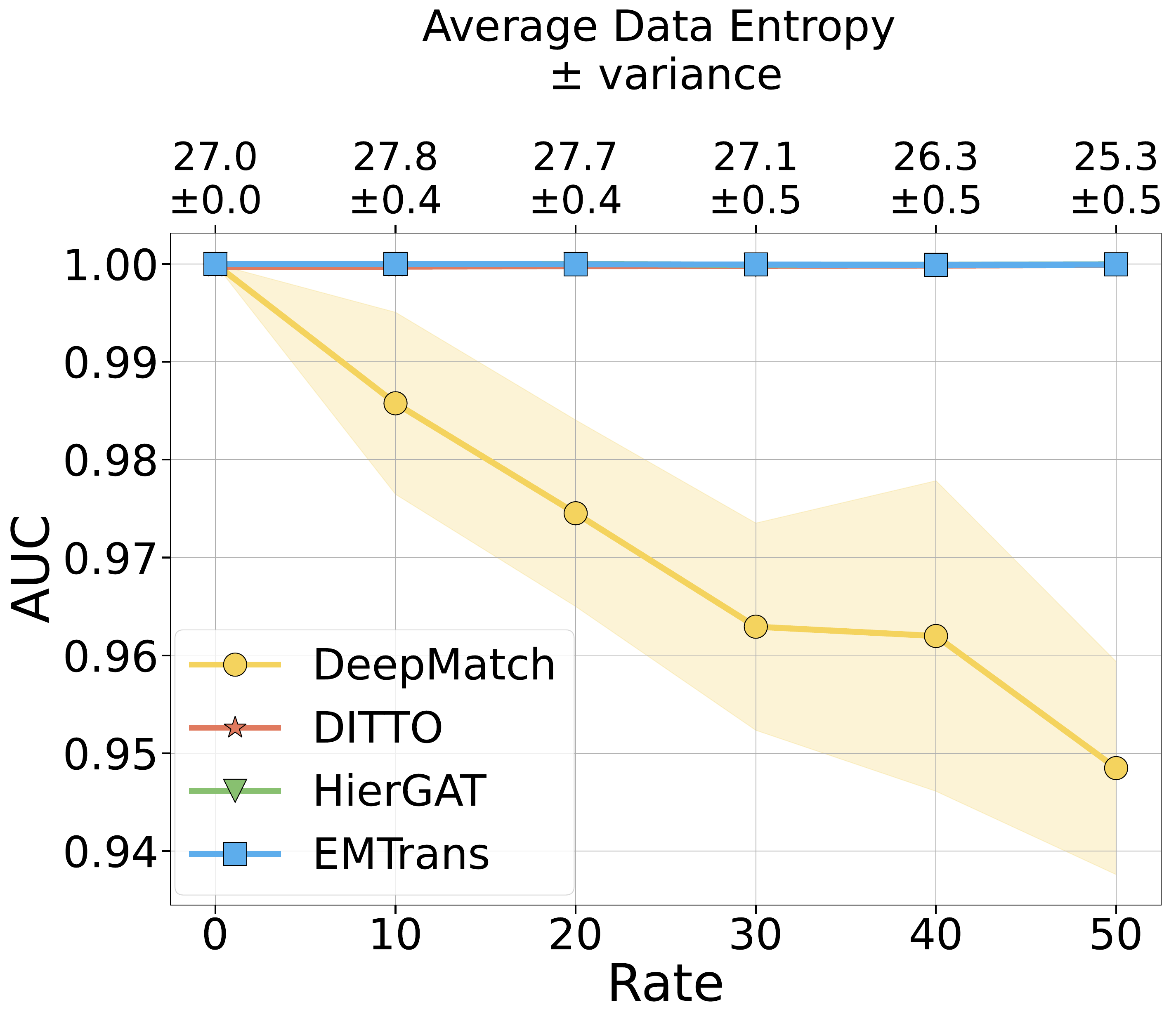}
        \caption{\fodor}
        \label{fig:hier-testFodor}
    \end{subfigure}
     \hspace{2mm}
    \begin{subfigure}[b]{0.30\textwidth}
        \includegraphics[width=\textwidth]{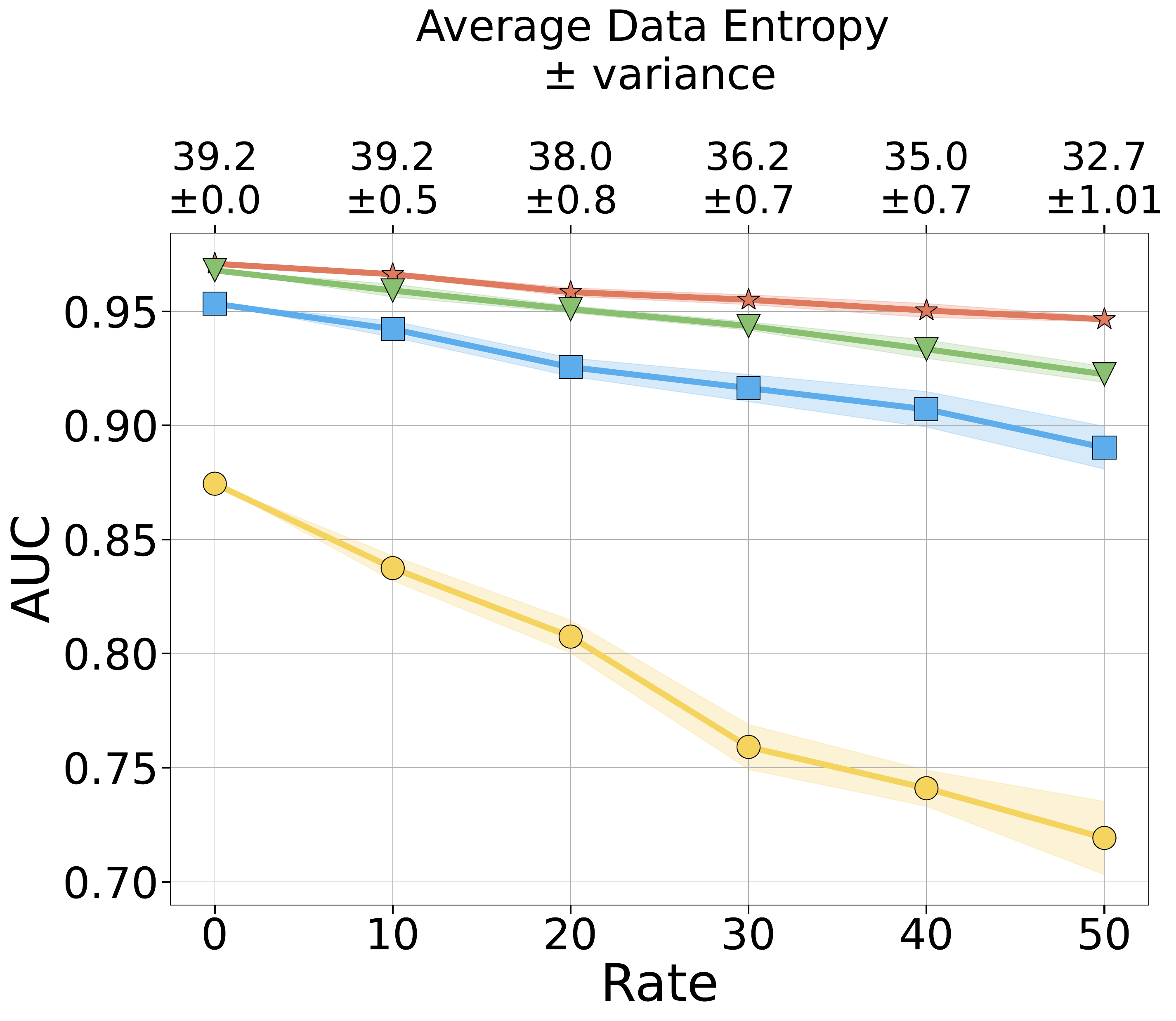}
        \caption{\walmartamazon}
    \end{subfigure}
    \hspace{2mm}
    \begin{subfigure}[b]{0.30\textwidth}
        \includegraphics[width=\textwidth]{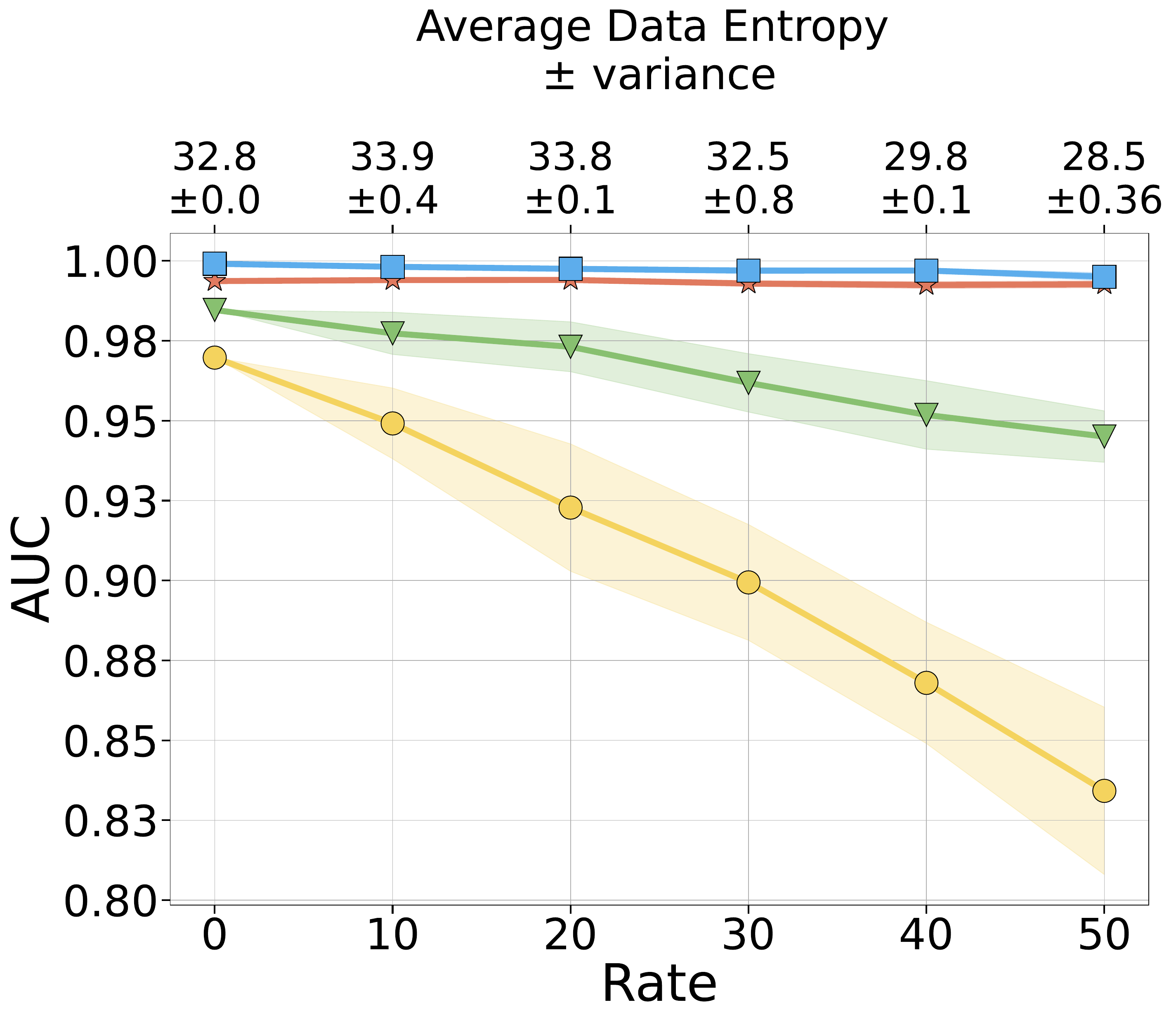}
        \caption{\itunesamazon}
    \end{subfigure}
    \caption{Performance vs hierarchical data distortion (information loss) when changing test data.}
    \label{fig:hier-test}
\end{figure*}

Figure~\ref{fig:hier-test} reports the impact of hierarchical distortion on model performance across datasets. Distortion is applied to the test data, simulating mismatches in data granularity or resolution. While distortion rate indicates how many values are changed, it does not fully capture semantic loss. We therefore also report entropy values to quantify information loss, based on column-level entropy from information theory~\cite{shannon1948}.

In general, entropy decreases with higher distortion as attribute values become more coarse. However, in Figure~\ref{fig:hier-testFodor}, entropy increases slightly at low distortion levels (27.0 to 27.8 between 0\% and 10\%), due to frequent values being replaced with less common but more general alternatives. This reflects a corner case unique to the \fodor dataset.

The relationship between distortion and entropy is non-linear, depending on hierarchy structure and attribute distributions. All models show performance degradation as distortion increases. \deepmatcher is the most sensitive, experiencing steep declines. \ditto and other transformer-based methods show greater robustness. This may be attributed to BERT’s capacity to link generalized or distorted values to their more specific counterparts via contextual embeddings, while static embeddings in \deepmatcher fail to compensate for resolution loss.

\takeaway All models degrade under resolution heterogeneity, but transformer-based models such as \ditto are more resilient. Static embedding methods like \deepmatcher are more susceptible to hierarchical distortions.

\begin{figure*}[ht]
    \centering
    \begin{subfigure}[b]{0.45\textwidth}
        \includegraphics[width=\textwidth]{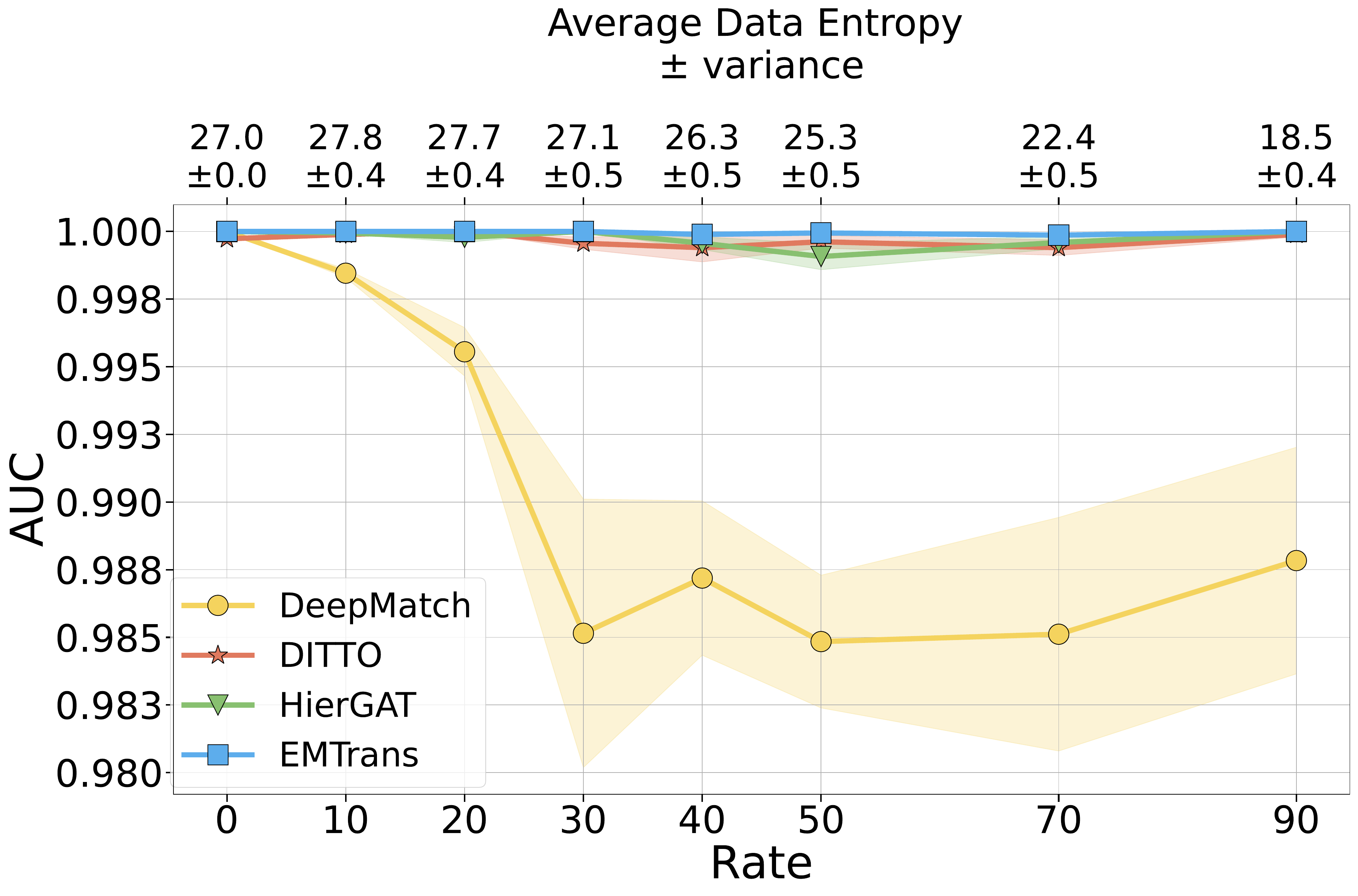}
        \caption{\fodor}
    \end{subfigure}
    \hspace{2mm}
    \begin{subfigure}[b]{0.45\textwidth}
        \includegraphics[width=\textwidth]{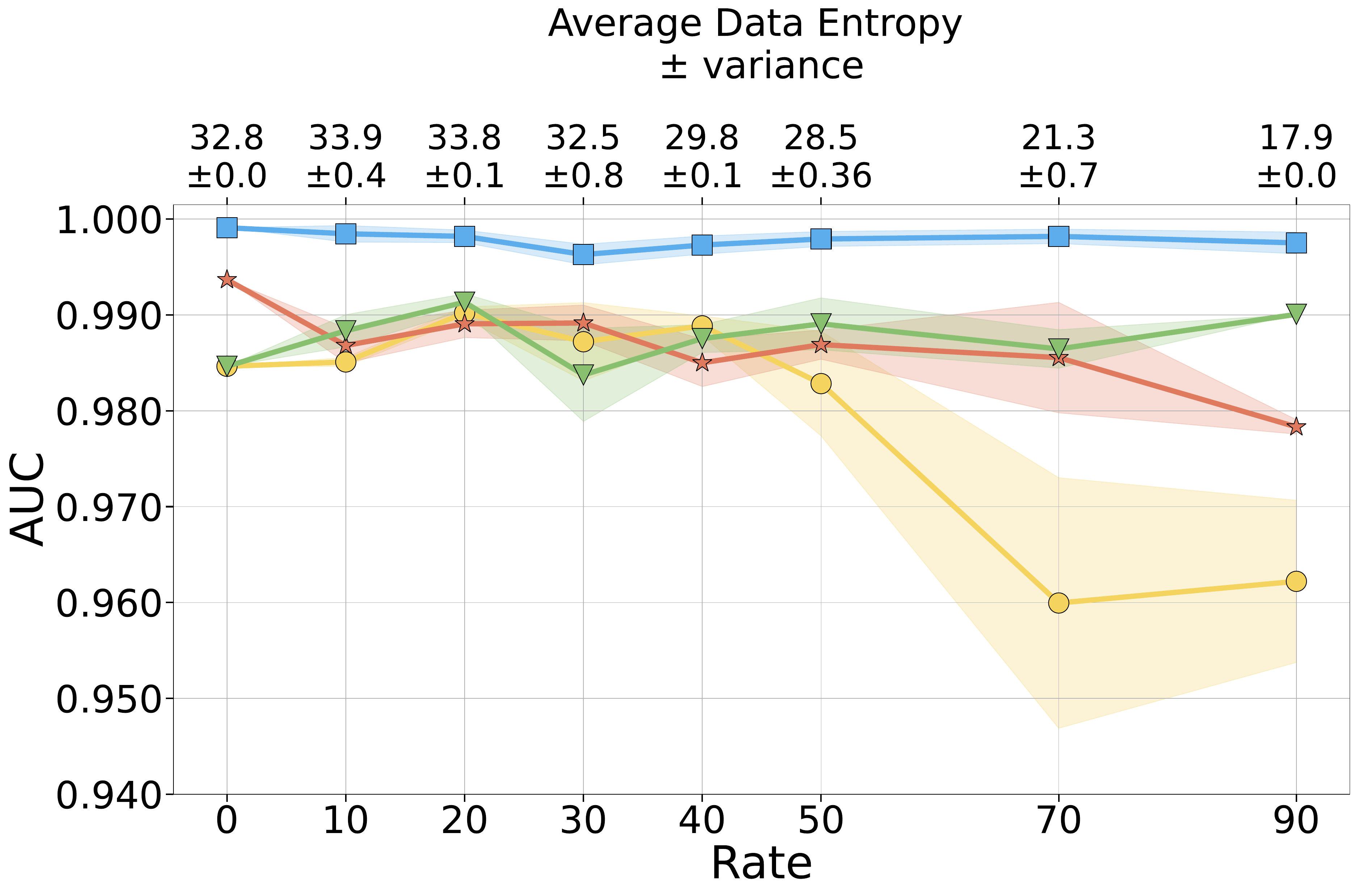}
        \caption{\itunesamazon}
    \end{subfigure}
    \caption{Performance vs hierarchical data distortion (information loss) when changing training data.}
    \label{fig:hier-train}
\end{figure*}

Figure~\ref{fig:hier-train} mirrors the previous experiment in Figure~\ref{fig:hier-test}, but applies hierarchical distortion to the \emph{training} data while keeping the test data unchanged. The AUC trends show that performance remains relatively stable across all models, particularly for advanced methods like \ditto and \emtransformer, which exhibit minimal decline despite the degraded training data.

The contrast with Figure~\ref{fig:hier-test} is notable. When the test data is distorted, critical features used for matching are obfuscated, leading to significant performance drops--often to the point where even human annotators would struggle. In contrast, when only the training data is distorted, robust models can still infer which attributes are most informative and learn effective representations. This allows them to generalize well to clean test data.

\takeaway Granularity-related (hierarchical) heterogeneity can significantly degrade EM performance, particularly when it affects the test data. However, models like \ditto and \emtransformer demonstrate strong resilience when trained on distorted data, effectively identifying key features and mitigating training noise. These results highlight that generalizing to distorted test data is more challenging than learning from heterogeneous training data, emphasizing the importance of robust architectures for handling real-world granularity shifts.

\begin{figure*}[ht]
    \centering
    \begin{subfigure}[b]{0.30\textwidth}
        \includegraphics[width=\textwidth]{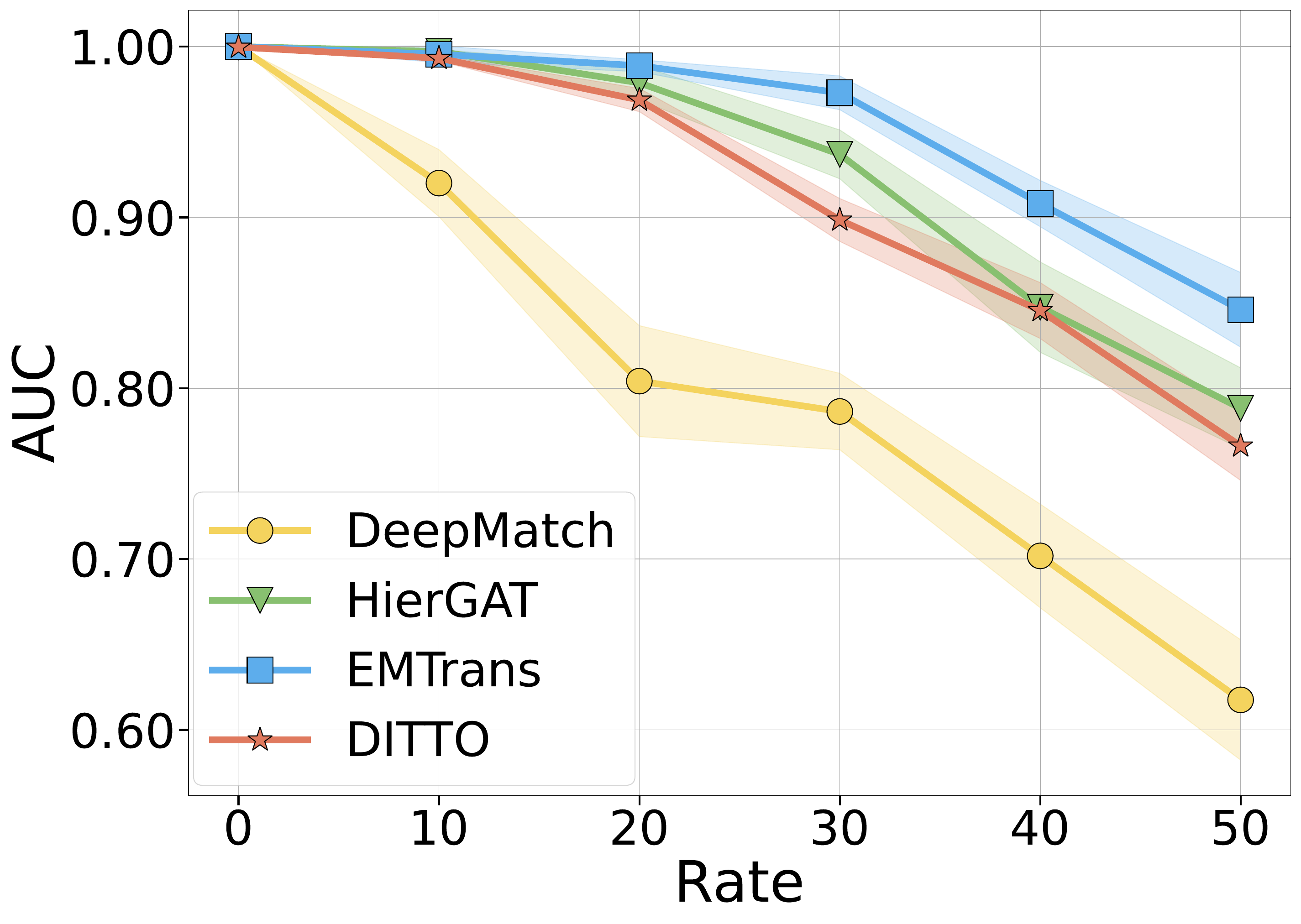}
        \caption{MCAR}
    \end{subfigure}
    \hspace{2mm}
    \begin{subfigure}[b]{0.30\textwidth}
        \includegraphics[width=\textwidth]{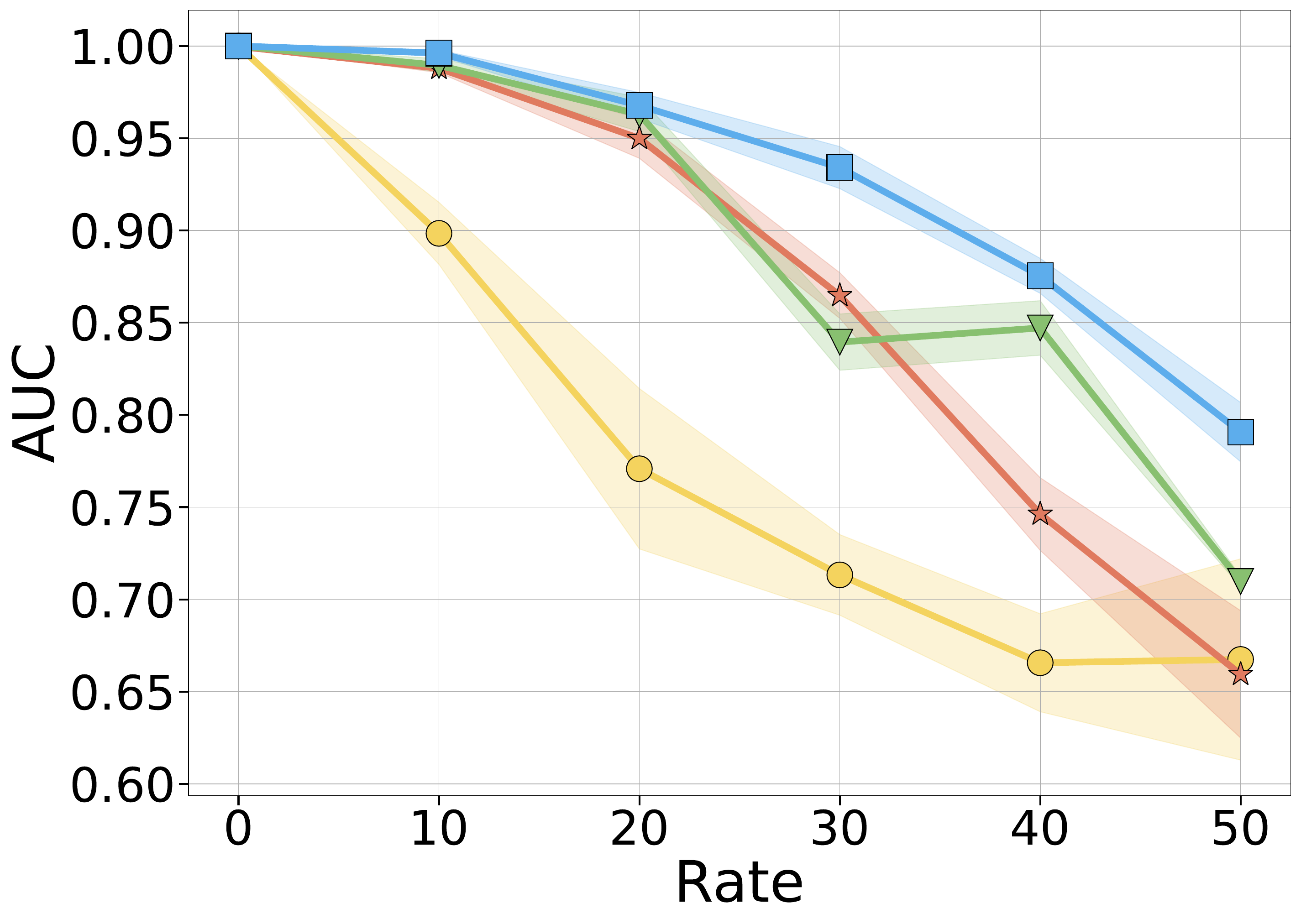}
        \caption{MAR}
        \label{fig:MARfig7}
    \end{subfigure}
    \hspace{2mm}
    \begin{subfigure}[b]{0.30\textwidth}
        \includegraphics[width=\textwidth]{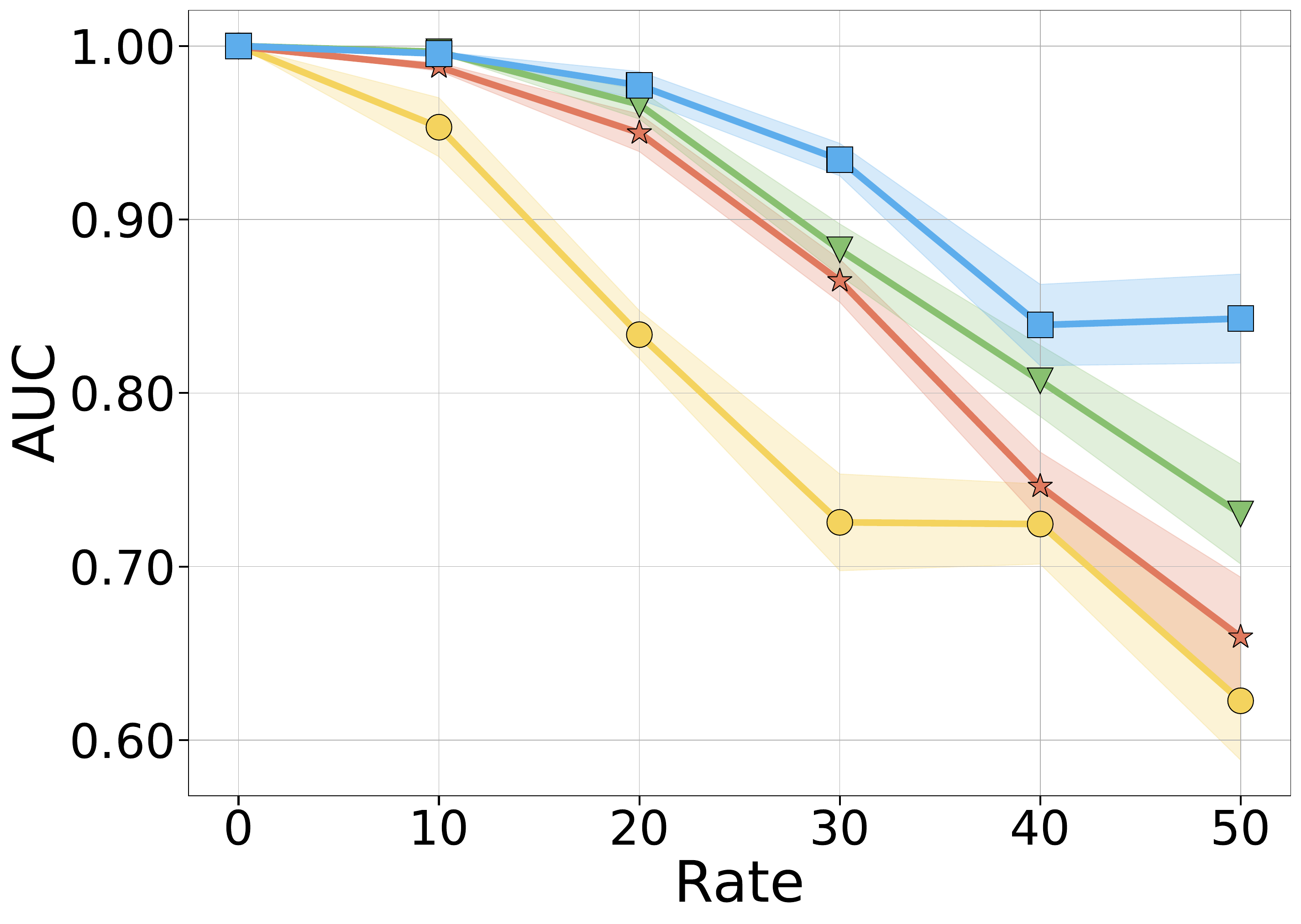}
        \caption{MNAR}
    \end{subfigure}
    \caption{Missing data in \fodor: the test data is dirty, and the training data is unchanged.}
       \label{fig:miss-test-fodor}
\end{figure*}

\begin{figure*}[h!]
    \centering
    \begin{subfigure}[b]{0.30\textwidth}
        \includegraphics[width=\textwidth]{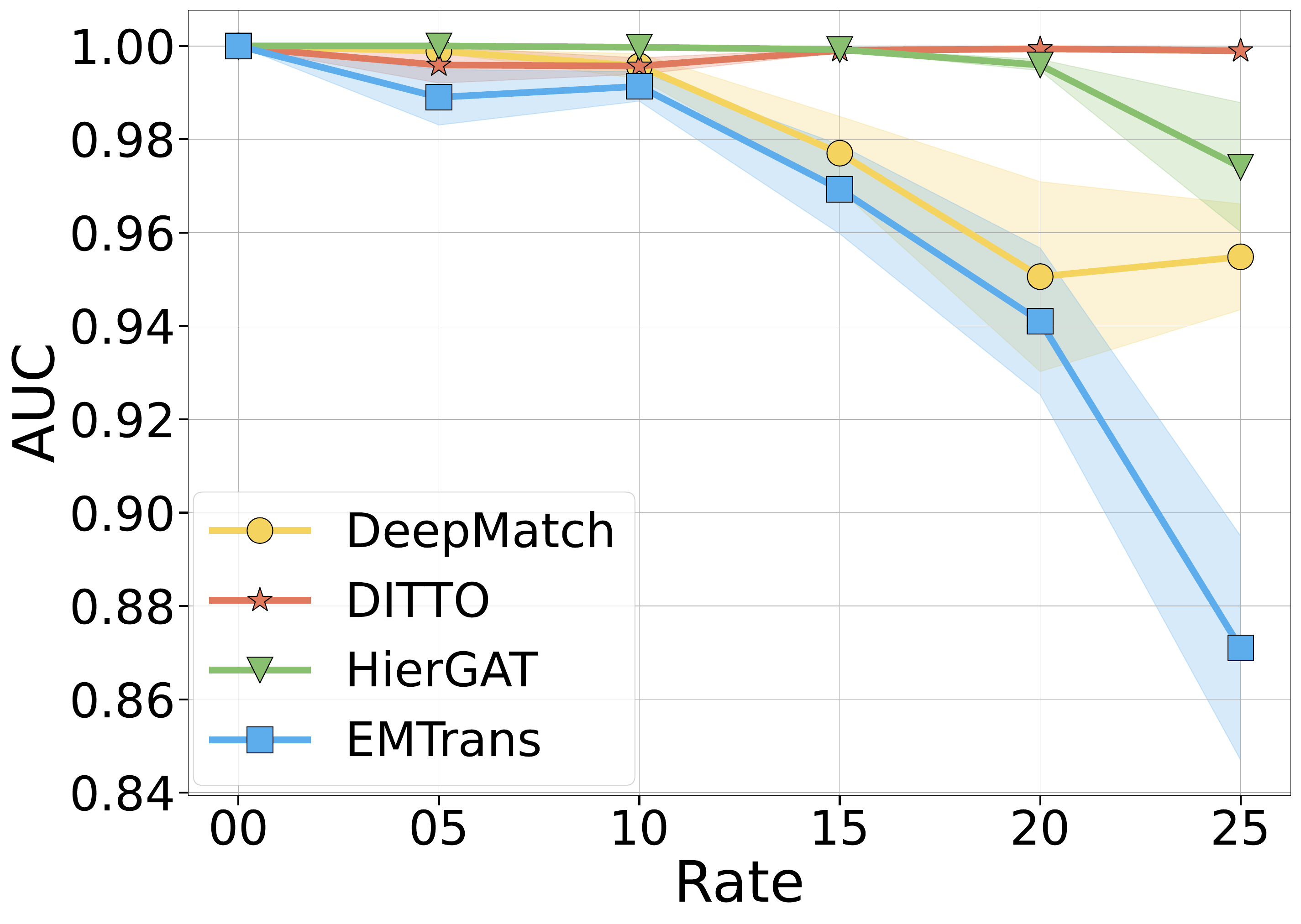}
        \caption{\fodor}
    \end{subfigure}
    \hspace{2mm}
    \begin{subfigure}[b]{0.30\textwidth}
        \includegraphics[width=\textwidth]{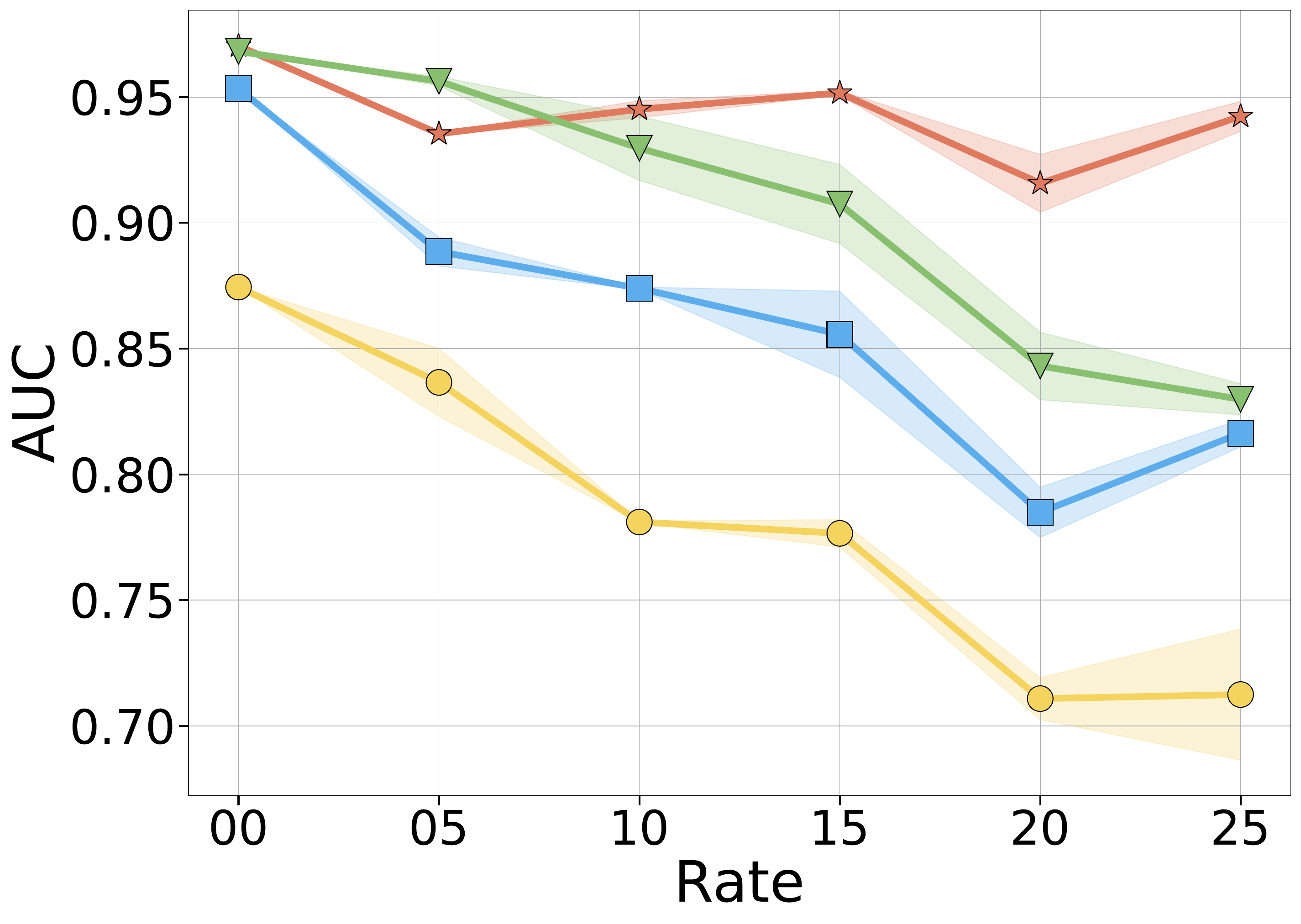}
        \caption{\walmartamazon}
    \end{subfigure}
    \hspace{2mm}
    \begin{subfigure}[b]{0.30\textwidth}
        \includegraphics[width=\textwidth]{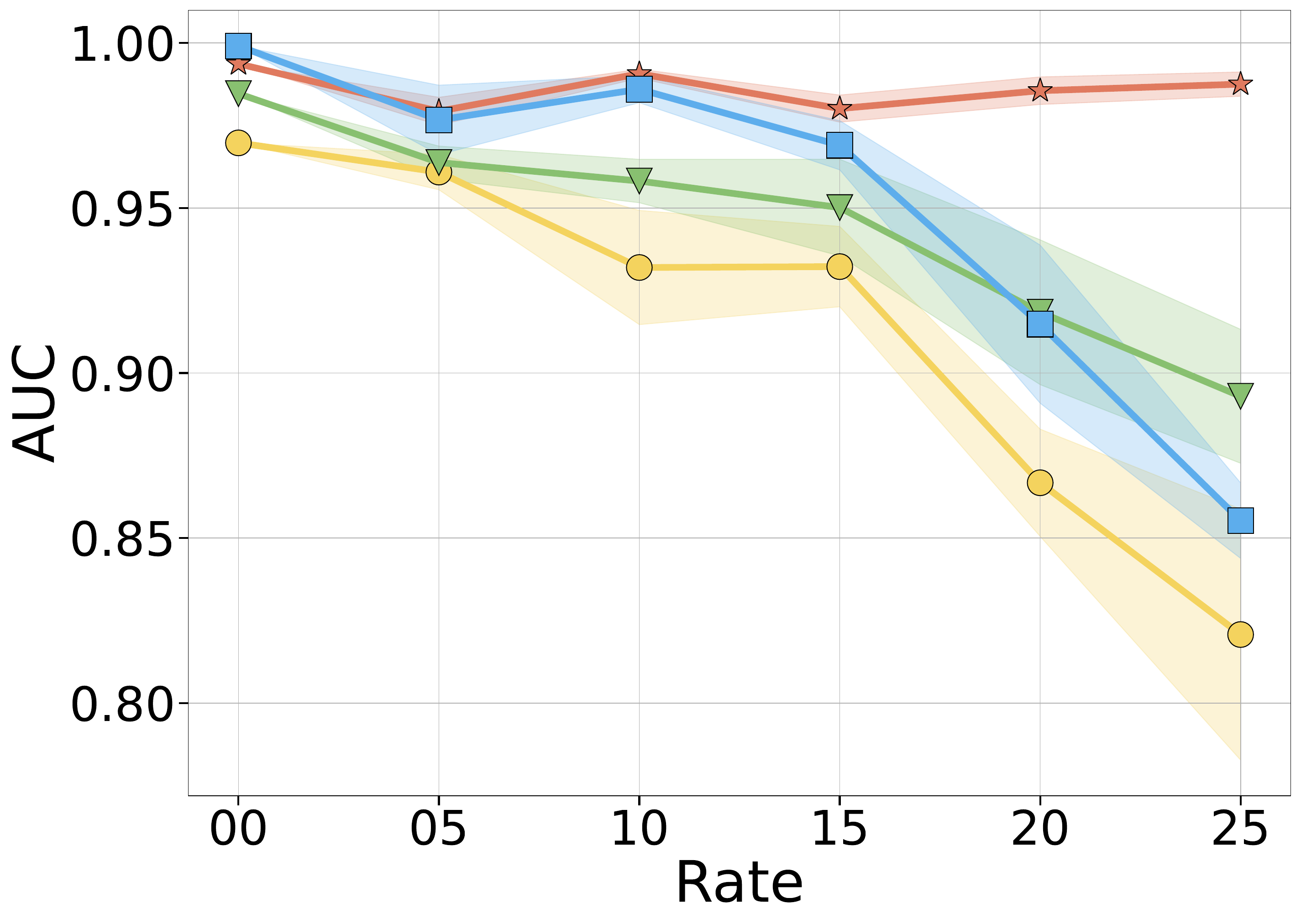}
        \caption{\itunesamazon}
    \end{subfigure}
    \caption{Label noise: the training data is dirty, and the test data is unchanged.}
    \label{fig:lbl-noise}
\end{figure*}

\subsubsection{Heterogeneity Caused by Data Quality Differences} \label{sec:dirty-exp}

We now evaluate the robustness of EM methods under semantic heterogeneity caused by data quality issues, such as missing values, attribute noise, and label noise.

Figure~\ref{fig:miss-test-fodor} shows results from injecting missing values into \fodor's test data, using standard missingness patterns: MCAR (completely at random), MAR (conditional on observed features), and MNAR (dependent on unobserved values). Similar trends were observed for \itunesamazon and \walmartamazon (figures omitted). As expected, model performance degrades as missingness increases. \deepmatcher shows the steepest decline, highlighting its vulnerability to incomplete input. In contrast, \ditto and \hiergat remain more stable, leveraging contextual and structural cues to compensate for missing information.

To assess sensitivity to noisy labels, we flipped a fraction of labels in the training set and measured performance on clean test data (Figure~\ref{fig:lbl-noise}). As label noise increases, AUC declines across all models, but to varying degrees. In \fodor, performance remains stable up to 10\% noise before dropping sharply--especially for \emtransformer and \hiergat. In \walmartamazon, all models degrade quickly, with \deepmatcher most affected. On \itunesamazon, \ditto and \hiergat show stronger resilience, whereas \emtransformer and \deepmatcher degrade rapidly.

These differences reflect architectural tradeoffs. \ditto's BERT-based architecture enables robust contextualization, helping it filter noise. \hiergat's graph-attention mechanisms capture structural dependencies, though its sensitivity varies with schema complexity. \emtransformer performs moderately well but lacks specialized noise-handling mechanisms. \deepmatcher, as a simpler model with static embeddings, fails to adapt to noisy conditions.

\begin{figure*}[ht]
    \centering
    \begin{subfigure}[b]{0.30\textwidth}
        \includegraphics[width=\textwidth]{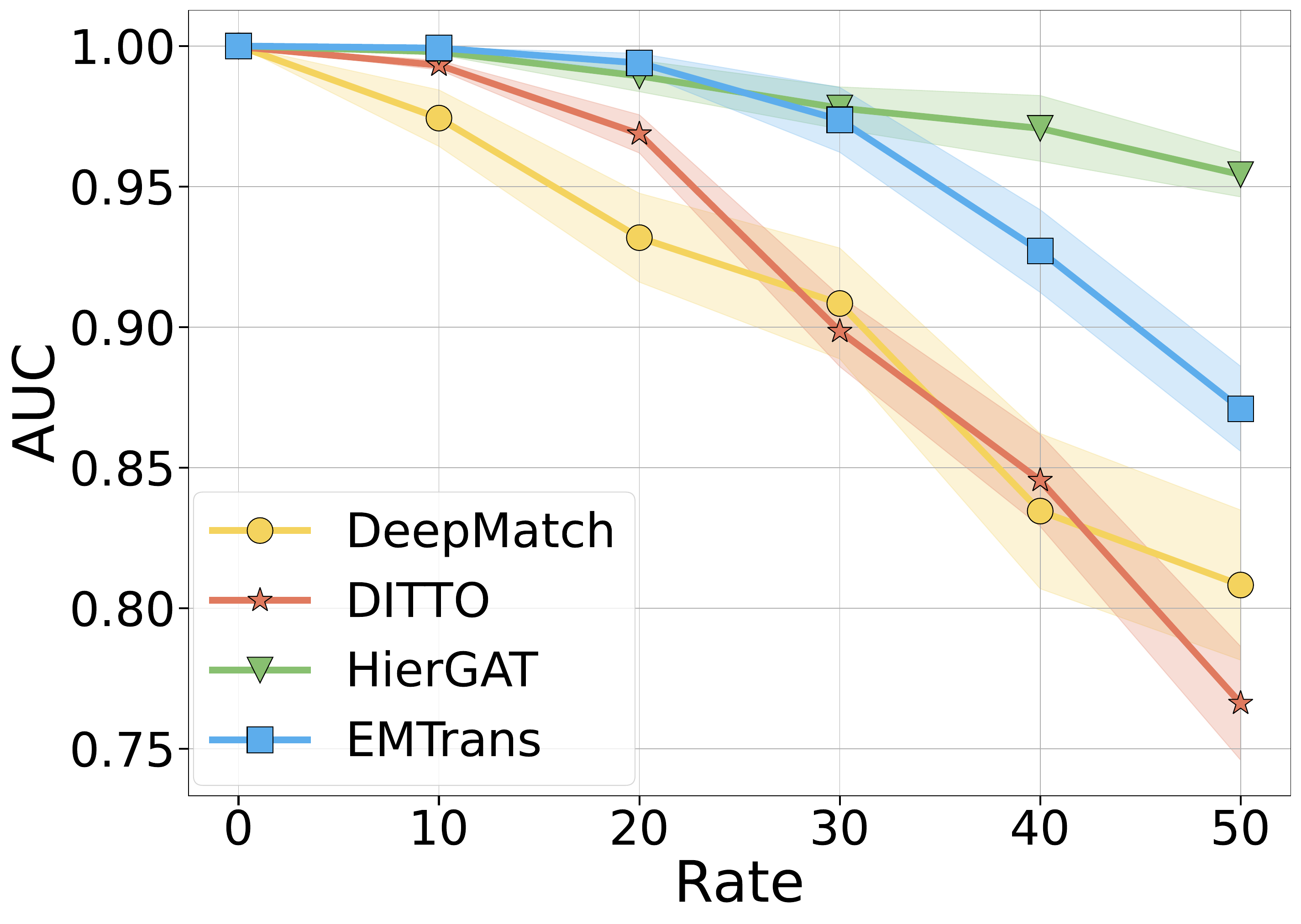}
        \caption{\fodor}
    \end{subfigure}
    \hspace{2mm}
    \begin{subfigure}[b]{0.30\textwidth}
        \includegraphics[width=\textwidth]{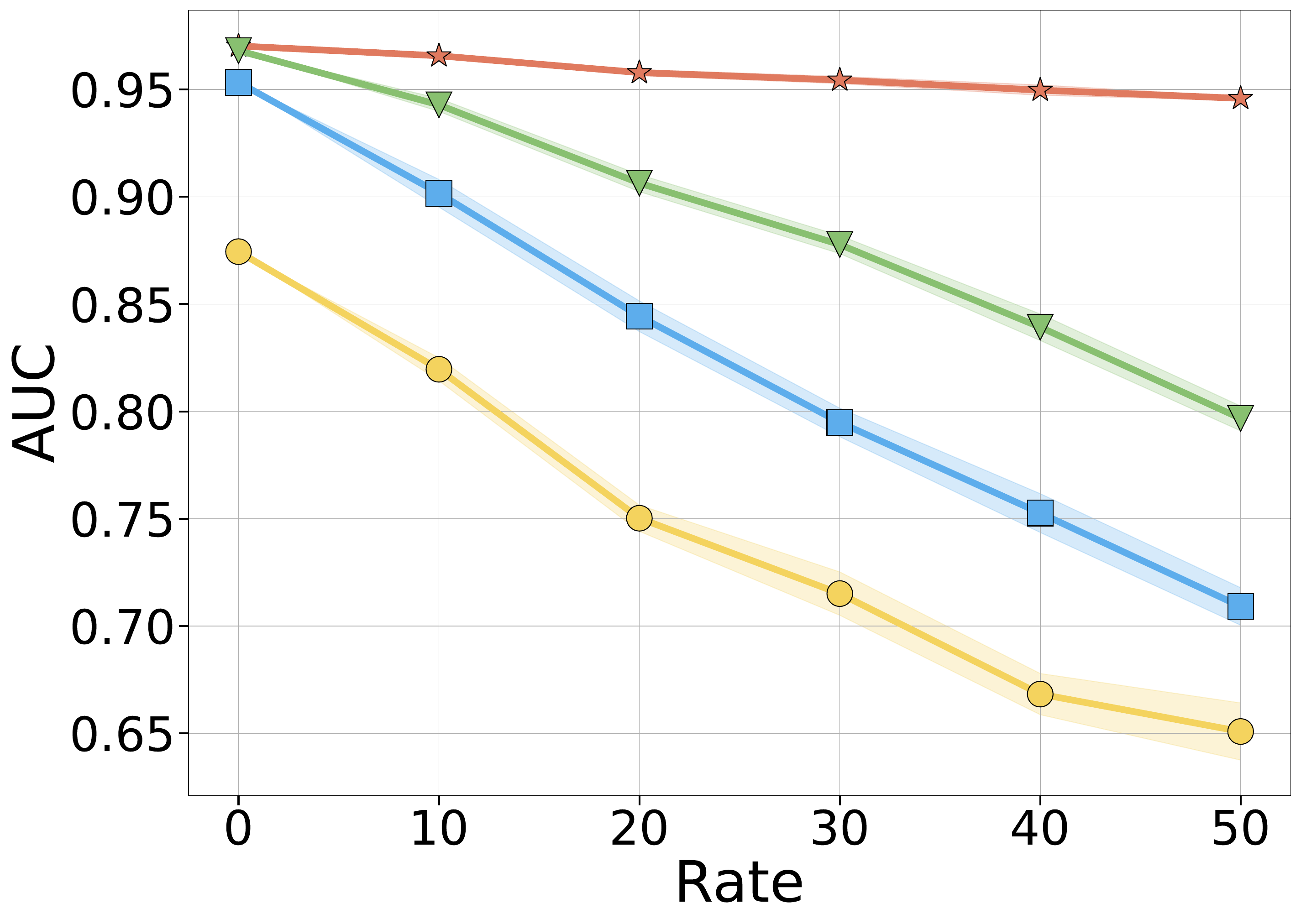}
        \caption{\walmartamazon}
    \end{subfigure}
    \hspace{2mm}
    \begin{subfigure}[b]{0.30\textwidth}
        \includegraphics[width=\textwidth]{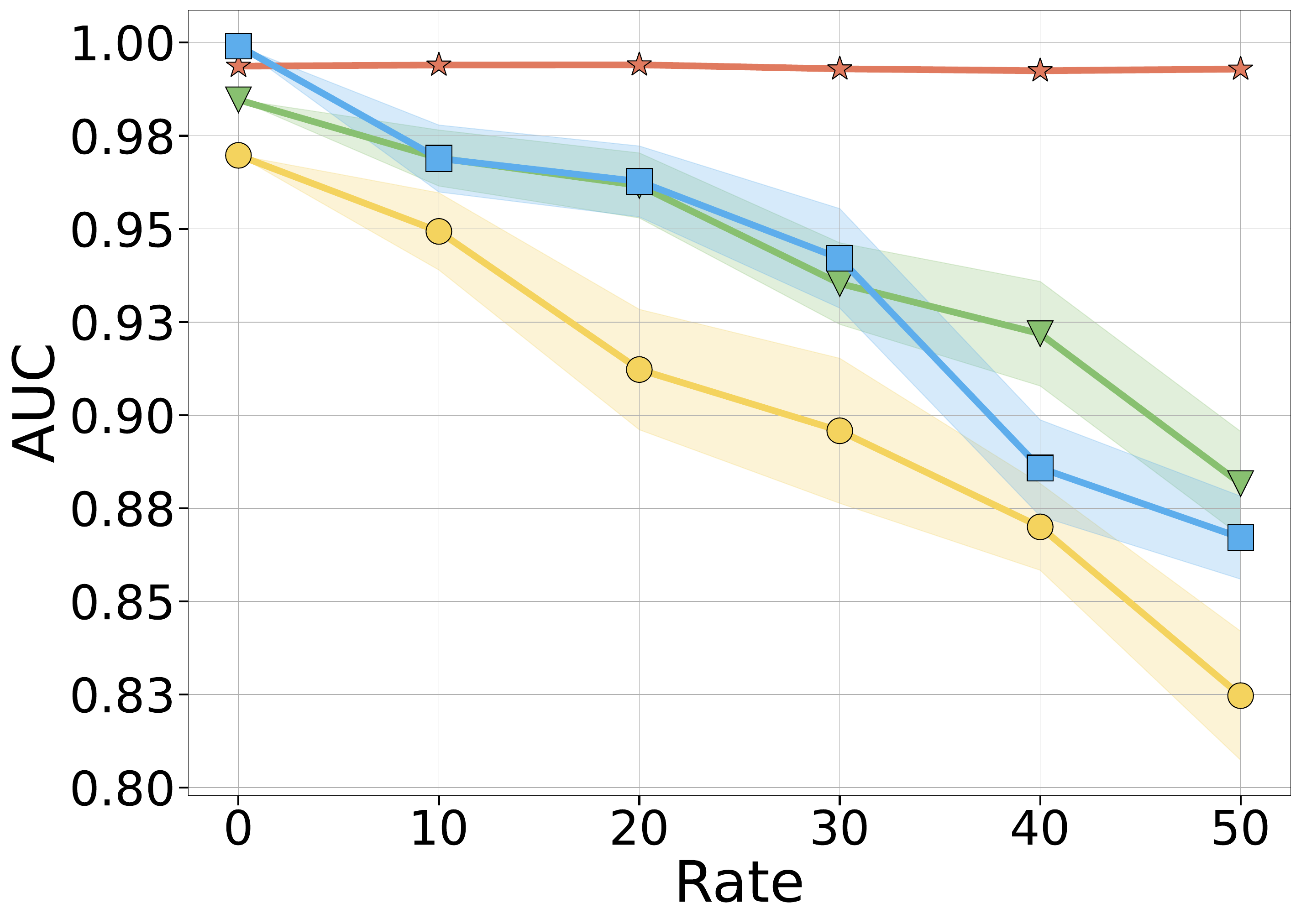}
        \caption{\itunesamazon}
    \end{subfigure}
    \caption{Attribute noise: the test data is dirty, and the training data is unchanged.}
    \label{fig:attrnoisetest}
\end{figure*}

Next, we introduced attribute noise into test data (Figure~\ref{fig:attrnoisetest}) by randomly modifying one attribute per row. As with label noise, AUC drops with increased corruption. Robustness again varies by model and dataset. In \fodor, \ditto maintains high accuracy due to its contextual embeddings, while \hiergat and \emtransformer show moderate resilience. \deepmatcher suffers steep declines.

In \walmartamazon, which contains more diverse and complex attributes, all models are more vulnerable. \deepmatcher and \emtransformer degrade the most, while \ditto and \hiergat perform comparatively better. On \itunesamazon, performance holds up at low noise levels, but degrades with higher corruption. Once again, \deepmatcher exhibits the most significant drop, while \ditto remains consistently strong.

\takeaway These experiments highlight the importance of model architecture in handling data quality heterogeneity. \ditto and \hiergat demonstrate strong resilience to missing and noisy data, thanks to their use of transformers and graph attention. Simpler models like \deepmatcher show limited robustness, especially on complex datasets like \walmartamazon. Importantly, test-time noise (heterogeneity at deployment) has a more severe impact than training-time corruption, underscoring the challenge of generalizability in real-world heterogeneous environments. To build effective EM pipelines, models must not only be robust to noise but also generalize to unseen, imperfect data.

\subsubsection{{Impact of Representation Heterogeneity}}
\label{sec:repr-exp}

{To simulate representation heterogeneity at the schema level, we start from the clean versions of \fodor, \itunesamazon, and \walmartamazon and evaluate three representative neural matchers: \deepmatcher, \hiermatcher~\cite{Peukert2020}, and \rotom~\cite{miao2021rotom}. Each model is trained once on the original column order and then evaluated under two test-time conditions: (i) the original column order and (ii) a randomly shuffled order of attributes for each record pair. For every model–dataset–condition combination we repeat evaluation 20 times with different random seeds (and, for the shuffled case, different permutations), and report the mean ROC AUC and standard deviation in Table~\ref{tab:col-shuffle}; the “No Shuffle” columns correspond to the original, unpermuted test schema.}

{
\begin{table}[H]
\centering
\setlength{\tabcolsep}{4pt}
\begin{tabular}{l l c | c c}
\toprule
\textbf{Dataset} & \textbf{Model} &
\textbf{$\text{AUC}_{\text{Normal Order}}$} &
\textbf{$\text{AUC}_{\text{Shuffle}}$} &
\textbf{$\Delta$AUC} \\
\midrule
\multirow{3}{*}{\walmartamazon}
 & \hiermatcher   & 94.34 $\pm$ 0.00 & 80.51 $\pm$ 14.09 & -13.83 \\
 & \deepmatcher   & 80.31 $\pm$ 0.91 & 72.60 $\pm$ 11.41 & -7.71 \\
 & \rotom         & 96.31 $\pm$ 0.30 & 94.88 $\pm$ 0.82 & -1.43 \\
\midrule

\multirow{3}{*}{\itunesamazon}
 & \hiermatcher   & 93.32 $\pm$ 0.00 & 74.47 $\pm$ 19.37 & -18.85 \\
 & \deepmatcher   & 98.42 $\pm$ 0.40 & 81.18 $\pm$ 11.25 & -17.24 \\
 & \rotom         & 99.22 $\pm$ 0.24 & 95.32 $\pm$ 4.38 & -3.90 \\
\midrule

\multirow{3}{*}{\fodor}
 & \hiermatcher   & 100.00 $\pm$ 0.00 & 79.98 $\pm$ 16.47 & -20.02 \\
 & \deepmatcher   & 99.92 $\pm$ 0.08 & 88.63 $\pm$ 11.73 & -11.29 \\
 & \rotom         & 99.99 $\pm$ 0.01 & 99.95 $\pm$ 0.07 & -0.04 \\
\bottomrule
\end{tabular}
\caption{Impact of test-time column-order shuffling on AUC. $\Delta$AUC indicates the change in AUC after shuffling.}
\label{tab:col-shuffle}
\end{table}}

{In Table~\ref{tab:col-shuffle}, we observe that \hiermatcher, \deepmatcher, and \rotom all attain high ROC AUC when evaluated on the original column order, but that \hiermatcher and \deepmatcher suffer substantial drops once the attributes in the test records are randomly permuted. In addition, the standard deviations of AUC in the shuffled condition are large for these two models, indicating that their predictions are highly unstable across different permutations of the same records. \rotom, in contrast, retains almost all of its performance under column shuffling, with only minor decreases in AUC and consistently low variance, showing that it is effectively robust to this type of schema heterogeneity.}

{A plausible explanation for this behavior is that \hiermatcher and \deepmatcher are designed around a fixed attribute layout: they encode each tuple as a sequence of attribute representations whose positions are implicitly tied to particular fields, and they rely on RNN/attention layers and attribute-level parameters that are not permutation invariant. When the order of columns changes at test time, the model still interprets position $i$ as ``the $i$-th training attribute'', so semantically mismatched features are compared and the learned decision boundary no longer aligns with the input. \rotom, on the other hand, linearizes records into text with explicit column-name markers and is trained together with data-augmentation operators (including column shuffling) on top of a pre-trained language model. As a result, the model learns to condition primarily on the column labels and textual content rather than their order, which naturally results the strong permutation robustness seen in the shuffled setting.}

\subsection{Key Findings, Limitations, and Implications}\label{sec:discussion}

Our experiments offer several insights into the impact of heterogeneity on EM models and practical strategies for improving robustness and generalizability.

First, all forms of semantic heterogeneity--including language and terminology differences, granularity mismatches, and data quality issues--pose substantial challenges to entity matching. While advanced models like \ditto and \hiergat demonstrate greater resilience due to their use of contextualized embeddings and attention mechanisms, simpler architectures like \deepmatcher are highly sensitive to such variations, suffering steep performance declines in noisy or semantically inconsistent settings. This underscores the importance of using models that can capture deeper semantic and structural relationships.

Second, test-time heterogeneity has a more severe effect on performance than heterogeneity during training. Most models can adapt to noisy training data by learning stable features, but generalizing to unseen heterogeneity during deployment remains difficult. This highlights the need for designing methods that prioritize transferability and robustness to distribution shifts across deployment environments.

Third, model performance varies significantly by dataset. Complex or noisy datasets such as \walmartamazon induce larger performance drops than simpler ones like \itunesamazon. Tailoring methods to the characteristics of the data--e.g., attribute richness, schema complexity, or error patterns--can improve outcomes and guide model selection.

Fourth, techniques like domain adaptation, retrieval-augmented matching, and external knowledge integration show promise for managing heterogeneity in evolving or dynamic environments. Fine-tuning pre-trained models or integrating external context can boost robustness, while mechanisms like adaptive attention and robust loss functions can mitigate the effects of label or attribute noise.

Fifth, interactive and user-in-the-loop methods remain valuable in practical settings. When heterogeneity leads to ambiguity or context-specific variation, human input can resolve edge cases that automated systems may misclassify. Coupling robust models with feedback mechanisms can significantly improve EM in real-world deployments.

{Sixth, our error analysis reveals three architectural failure modes that help explain the performance gaps observed across Figures~\ref{fig:syn}–\ref{fig:attrnoisetest}. In the synonym and synonym-vs-random experiments, \deepmatcher exhibits out-of-vocabulary failure: it relies on fixed GloVe/FastText vectors and maps many GPT-4–generated synonyms to generic unknown tokens, whereas transformer-based models with subword tokenization maintain a usable semantic signal. In the attribute-noise experiments, \hiergat suffers from graph-propagation of noise, since its message-passing layers spread corrupted attribute values to neighboring nodes, degrading representations more severely than sequence-based \ditto. Finally, in the missing-data experiments, \deepmatcher's RNN-based attention is brittle under MCAR/MAR/MNAR because removing key tokens disrupts temporal dependencies, while \ditto's self-attention can redistribute mass to remaining informative tokens (e.g., from a missing ``Brand'' to the ``Title''), preserving stable performance even at high missingness rates.}

Our analysis focuses primarily on \textit{semantic heterogeneity}. This decision stems from the observation that semantic variations are often the most subtle and challenging to detect, yet they are underexplored in empirical EM research. However, we acknowledge that this choice limits our coverage of representation heterogeneity (e.g., multimodal or schema format differences), which also plays a critical role in many EM scenarios. Future work should expand these experiments to cover diverse forms of representation heterogeneity, especially as multimodal and semi-structured data become more common.

Addressing HEM effectively requires a combination of deep semantic modeling, dataset-specific adaptation, generalization-focused learning strategies, and human-in-the-loop capabilities. Our findings serve as a guide for developing EM systems that are both resilient to heterogeneity and adaptable to real-world variability.

\section{Conclusion and Future Research}\label{sec:future}

This paper addresses the challenge of data heterogeneity in EM. We proposed a
taxonomy of heterogeneity, surveyed recent methods with a focus on semantic
variation, analyzed their relationship to the FAIR principles, and conducted
extensive experiments that evaluate model robustness and generalizability.
Our results show that heterogeneity remains a major barrier to reliable EM,
even for state-of-the-art models.

Several key directions can guide future work on HEM. Below we focus on areas
that remain underexplored even after adding dedicated sections for LLMs,
multimodal EM, and benchmarking in the main body of the paper.

\begin{itemize}[leftmargin=0.35cm]

\item {\emph{EM in Data Lakes.} Data lakes produce extreme representation and structural heterogeneity due to
schema drift, sparse or unreliable metadata, and files spanning structured, semi-structured, and unstructured formats. Prior work on dataset discovery and ER in lakes~\cite{bogatu2020dataset,bouabdelli2025towards,fernandez2018datafusion} shows that mismatched or incomplete schemas make even simple alignment tasks difficult. Our experiments (Section~\ref{sec:exp}) confirm that neural models degrade significantly under such schema and granularity shifts. Future research should develop adaptive matchers that combine schema inference, metadata enrichment, and multimodal content-based signals. Useful building blocks include table-understanding models such as TURL and TaPas~\cite{yu2020turl,tapas2019}. Promising directions include:
(i) continual-learning EM models that update as lake schemas evolve; (ii) unified embeddings that reconcile structured, text, and image attributes; and (iii) pipelines that fuse metadata with content signals for robust matching at scale.}

\item {\emph{Human-in-the-Loop and Explainability.} HITL EM has been explored for resolving difficult or ambiguous matches~\cite{altowim2014regularizing,bellogin2023interactive}, and recent studies on
explainable ER \cite{meduri2020elt,esmaili2021explainable} show its relevance in practice. However, these systems rarely account for heterogeneity-driven errors such as context shifts or representation mismatches. Our experiments identify such cases as persistent failure modes. Future HEM research should combine uncertainty-aware active learning~\cite{qian2020activeER} with explanations tailored to our heterogeneity taxonomy—for example, highlighting when mismatches arise from missing attributes, conflicting context, or schema differences. HITL pipelines should
also support incremental updates of match rules and embeddings as users provide feedback, enabling more robust and interactive EM.}

\item {\emph{Privacy and Security.} Privacy-preserving linkage has a long history~\cite{vatsalan2017ppsurvey,christen2020book}, and federated or distributed EM techniques~\cite{karapiperis2020federated,ranbaduge2022dpER} are gaining attention. However, most current systems assume consistent schemas and data types across parties. Heterogeneous schemas, mixed modalities, and evolving attributes create new privacy challenges not addressed by existing work. Future directions include designing DP-aware blocking and matching methods
that work across heterogeneous attributes, building privacy-preserving multimodal embeddings, and developing secure multi-party protocols that handle schema drift. HEM can offer a structured way to reason about how different forms of heterogeneity interact with privacy risk.}

\item {\emph{Fairness and Inclusivity.} Recent work on fairness in EM~\cite{shahbazi2023through,moslemi2024evaluating,nilforoushan2022entity,efthymiou2021fairer,moslemi2024threshold} has shown that real-world EM pipelines can amplify disparities across subgroups. Our empirical results indicate that heterogeneity—such as differing
levels of attribute completeness or domain-specific terminology—intensifies these fairness issues. Future research should develop fairness metrics and mitigation strategies that explicitly account for semantic, contextual, and structural heterogeneity. Promising directions include causal analysis to trace how heterogeneous attributes propagate bias, dynamic re-weighting or adversarial
debiasing to maintain fairness as data evolves, and schema-informed balancing that adjusts for subgroup-specific representation gaps.}
\item {\emph{Robustness to Temporal and Schema Drift.} Temporal evolution creates new forms of heterogeneity even within a single source. Prior work on temporal ER~\cite{christen2013adaptive}
shows that entity relationships and attribute semantics can change substantially over time. Future work should design drift-aware EM pipelines that detect and localize semantic and schema changes, maintain cross-version attribute alignment, and update matchers via continual or online learning. 
The heterogeneity taxonomy offers a natural framework for identifying which aspects of drift, including semantic, contextual and structural drift, are most impactful and for guiding how systems should adapt.}
\end{itemize}

Future advances in HEM require methods that explicitly handle the forms of
heterogeneity outlined in our taxonomy and adapt as these conditions change.
Such developments are essential for building EM systems that remain robust
and reliable in real, evolving data ecosystems.

\section{Declaration of generative AI and AI-assisted technologies in the writing process}
During the preparation of this work, the authors used ChatGPT (OpenAI) to
improve the readability and language of the manuscript. After using this
tool, the authors reviewed and edited the content as needed and took full
responsibility for the content of the published article.





\bibliographystyle{elsarticle-num} 
\bibliography{ref}






\end{document}